\documentclass[preprint,3p,times,onecolumn]{elsarticle}
\usepackage{amsmath}
\usepackage{epsfig}
\usepackage{amssymb,amsthm,graphicx}
\usepackage{tabularx}
\usepackage{multirow}
\usepackage{tabularx}
\usepackage{float}

\newcommand{\bea}{\begin{eqnarray}}
\newcommand{\eea}{\end{eqnarray}}
\newcommand{\bes}{\begin{subequations}}
\newcommand{\ees}{\end{subequations}}

\newcommand{\sn}{{\rm sn}}

\newcommand{\dn}{{\rm dn}}
\newcommand{\cn}{{\rm cn}}
\newcommand{\sech}{{\rm sech}}
\journal{Elsevier}
\begin{document}
\begin{frontmatter}
\title{Nonparaxial elliptic waves and solitary waves in coupled nonlinear Helmholtz equations}
\author[bhc]{K. Tamilselvan}\ead{tamsel.physics@gmail.com}
\author[bhc]{T. Kanna\corref{tk}}\cortext[tk]{Corresponding author}\ead{kanna\_phy@bhc.edu.in}
\address[bhc]{Post Graduate and Research Department of Physics, Bishop Heber College, Tiruchirapalli 620 017, Tamil Nadu, India}
\author[iiser]{Avinash Khare} \ead{khare@physics.unipune.ac.in}
\address[iiser]{Raja Ramanna Fellow, Department of Physics, Savitribai Phule Pune University, Pune -- 411007, India}

\begin{abstract}
 We  obtain a class of  elliptic  wave solutions of coupled nonlinear Helmholtz (CNLH) equations describing nonparaxial ultra-broad beam propagation in nonlinear Kerr-like media, in terms of the Jacobi elliptic functions and also discuss their limiting forms (hyperbolic solutions). Especially, we show the existence of non-trivial solitary wave profiles in the CNLH system. The  effect of nonparaxiality on the speed, pulse width  and amplitude of the nonlinear waves is analysed in detail.  Particularly a mechanism for tuning the speed by altering the nonparaxial parameter is proposed. We also identify a novel phase-unlocking behaviour due to the presence of nonparaxial parameter.
\end{abstract}

\end{frontmatter}
\section{Introduction}
Study of nonlinear waves has time honoured history \cite{whith}. In recent years the interest on nonlinear waves is exponentially increasing among researchers of various disciplines due to their ubiquitous appearance in different physical systems. There exist several types of nonlinear waves such as elliptic waves, solitons, rogue waves, shock waves, compactons etc. The underlying physical systems in which these nonlinear waves appear are in general described by different interesting nonlinear evolution equations. All these systems appear in diverse areas of physics like fluid dynamics \cite{whith}, nonlinear optics \cite{kivs}, Bose-Einstein condensates \cite{pethic}, bio-physics \cite{scott}, plasma physics \cite{david}, etc. Especially, in  nonlinear optics the coupled nonlinear Schr\"odinger (CNLS) system is an important mathematical model with various potential applications \cite{kivs}. From a physical perspective, this CNLS system arises in the context of partially coherent beam propagation in photorefractive media and also appears as governing equation for pico second pulse propagation in multimode optical fiber \cite{cros}-\cite{men}. The CNLS system becomes integrable for specific choice of parameters \cite{tk1, mana}. The integrable CNLS (so called Manakov model) system admits bright soliton solutions for focusing  nonlinearity \cite{mana, radhak}  and the dark soliton as well as dark-bright soliton solutions for defocusing nonlinearity \cite{shep, vm}. Such type of multicomponent solitons have also been realized in left handed metamaterials \cite{laza}.

 In the framework of nonlinear optics the derivation of CNLS equations stems naturally from the Maxwell's equations by employing the paraxial wave approximation or slowly varying envelope approximation (SVEA) \cite{kivs}. Under this approximation the second order derivative of normalized wave with respect to propagation direction is ignored. As a result of this the CNLS model excludes the diffraction length of the beam in the transverse direction of the physical system \cite{ciat}. Here a beam broader than its carrier wavelength, with moderate intensity and is propagating in (or at a negligible angle ) with respect to the reference direction is said to be a paraxial beam. If the beam fails to satisfy at least any one of the aforementioned properties is said to be a nonparaxial beam. The ultra-narrow beam nonparaxiality violates only the property of beam broader than its carrier wavelength. Such type of nonparaxiality has been studied in detail in Ref. \cite{chi}. The notion of nonparaxiality was first introduced by Lax \textit{et al}. \cite{lax}, in the investigation of ultra-narrow beam, for which the transverse waist $w_0$ and the carrier wavelength $\lambda$ are comparable, propagating in an inhomogeneous isotropic nonlinear media. The corrections to paraxial approximation have been obtained in Ref. \cite{lax} by expanding the fields as a power series in the ratio of beam diameter to diffraction length. Ultimately, it has been shown that the resulting governing equation will be a nonlinear Schr\"odinger (NLS) equation with higher order diffractive correction terms. After that, the question posed in Ref. \cite{akh} on the reliability of standard self-focusing NLS system in describing beam propagation  lead to the inclusion of additional longitudinal field oscillations. Going one step further, the two dimensional NLS equation in self-focusing Kerr medium with nonparaxial term has been numerically studied in Ref. \cite{fibi}.

 On the other hand, the nonparaxial ultra-broad beam  possesses a broad beam having moderate intensity but is propagating at an arbitrary angle with respect to the reference direction \cite{cham0}. To overcome the above inadequacy of the paraxial approximation, we modify the CNLS system by including the nonparaxial effect following the nonparaxial theory developed for beam propagation in Kerr like nonlinear media \cite{blair}. The evolution of broad optical beams in Kerr like nonlinear media can be well described by the coupled nonlinear Helmholtz (CNLH) type equations. The general dimensionless CNLH equations describing interaction of two obliquely propagating incoherently coupled optical fields in Kerr type nonlinear media can be casted as \cite{christ3}-\cite{cham1},
\bes\label{1}\bea
iq_{1,\xi}+\kappa q_{1,\xi\xi}+\frac{1}{2}q_{1,\tau\tau}+(\sigma_{1}|q_1|^2+\sigma_{2}|q_2|^2)q_{1}=0\,,\\
iq_{2,\xi}+\kappa q_{2,\xi\xi}+\frac{1}{2}q_{2,\tau\tau}+(\sigma_{1}|q_1|^2+\sigma_{2}|q_2|^2)q_{2}=0\,,
\eea
\ees
where $q_{j}, j=1,2,$ are the envelope fields of first and second components, subscripts $\xi$ and $\tau$ represent the longitudinal and transverse coordinates respectively. In Eq.~(\ref{1}), the second term corresponds to Helmholtz nonparaxiality and the co-efficient $\kappa (> 0)$ is the nonparaxial parameter. The nonparaxial parameter $\kappa$ value has been estimated of the order of $10^{-3}$ to $10^{-4}$  \cite{christ3,cham1}.   Normally, in the paraxial approximation (the beam wavelength is much smaller than the width of the beam) the terms $\kappa q_{j,\xi\xi}, j=1,2,$ are neglected. However, these terms will come into picture when the width of the beam is of the order of wavelength \cite{fibi} and influence the dynamics of the nonlinear waves. Physically, here we have included the diffraction in both transverse and longitudinal directions.

In Eq.~(\ref{1}), we have explicitly introduced the real nonlinearity coefficients $\sigma_{l}, l=1,2$, that can be absorbed into the equation itself with a simple transformation, in order to deal with a broader class of nonlinearities, namely focusing nonlinearity ($\sigma_{1}>0$, $\sigma_{2}>0$), defocusing nonlinearity ($\sigma_{1}<0$, $\sigma_{2}<0$ and  mixed (focusing-defocusing) nonlinearity ($\sigma_{1}>0$, $\sigma_{2}<0$ or $\sigma_{1}<0$, $\sigma_{2}>0$) within a single framework. For $\sigma_{1}=\sigma_{2}=\pm1$, the system (\ref{1}) reduces to the Helmholtz-Manakov system discussed in Ref. \cite{christ3}. In real situations one require the cross-phase modulation coefficients (coefficients of $|q_2|^2$ in Eq.(1a) and $|q_1|^2$ in Eq.(1b)) to be the same. This can be achieved for the nonlinearity $\sigma_{1}>0$, $\sigma_{2}<0$, by considering the complex conjugate of Eq.~(1a) along with Eq. (1b) as it is and making the transformation $\psi_{1}\rightarrow\sqrt{\sigma_{1}}q_{1}^{*}$, $\psi_{2}\rightarrow\sqrt{|\sigma_{2}|}q_{2}$ will result in a coupled equation in which $\psi_{1}$ is in the anomalous dispersion regime and $\psi_{2}$ is in the normal dispersion region. Note that now in spite of having opposite signs for  $\sigma_{1}$ and $\sigma_{2}$, the cross-phase modulation coefficients are equal to 1. Similar analysis can be carried out for the other choice ($\sigma_{1}<0$, $\sigma_{2}>0$) too. The mixed (focusing-defocusing) nonlinearity has been considered earlier in the absence of nonparaxial terms ($\kappa q_{j,\xi\xi}$, $j=1,2$) in Ref. \cite{soy}. Here we include the nonparaxial term and discuss the mixed type nonlinearity. This type of mixed (focusing-defocusing) nonlinearity can be realized in left-handed materials, couplers with left-handed and right-handed composite materials, optical materials with quadratic nonlinearities and also in two component Bose-Einstein condensates using Feshbach mechanism.

There exist several analytical studies on the scalar (single component) nonlinear Helmholtz (NLH) system. Particularly, it has been shown that the one dimensional scalar NLH system with focusing nonlinearity admits bright soliton solutions \cite{cros1} and  scalar NLH system with defocusing nonlinearity supports dark soliton solutions \cite{ciat1} while the two dimensional scalar NLH system supports the vortex solitons \cite{wang}. However, results are scarce for the above CNLH (vector NLH) system. Recently, Eq. (1) with $\sigma_{1}=\sigma_{2}=\pm1$ has been studied in \cite{christ3} Kerr medium and bright and dark soliton solutions have been reported for focusing and defocusing nonlinearities respectively. Collision studies of solitons in CNLH system revealed the fact that the interaction angle between two solitons is varied by altering the nonparaxial parameter \cite{cham1}.

Moreover, studying periodic nonlinear waves and their hyperbolic limiting forms in nonlinear evolution equations is one of the frontier topics. Such type of elliptic waves can be realized in nonlinear optics, atomic condensates, plasma physics, etc. For describing real situations one may need special type of periodic solutions. Several periodic solutions of integrable and non-integrable multicomponent NLS equations with focusing, defocusing and mixed type nonlinear interactions have been obtained in Refs. \cite{chow1}, in terms of Jacobi elliptic functions. Apart from this NLS framework, recently periodic solutions in several other nonlinear evolution equations have been studied due to their physical importance \cite{anjan}. In this paper, for the first time, we present several types of elliptic and solitary wave solutions of the CNLH system.  This work is mainly aimed at bringing out the effect of nonparaxial parameter on the elliptic as well as solitary waves in the CNLH equations.

This paper is organised as follows: In Sec. 2, we briefly outline the mathematical procedure to construct the elliptic solutions of the CNLH system. In Secs. 3 and 4, respectively we obtain the elliptic wave solutions in terms of Lam\'e polynomials of order one and two, and also present a complete discussion on the effect of the nonparaxial term. We conclude our results in Sec. 5.
\section{The Method}
 In this section, we briefly outline our procedure to construct a rich variety of doubly periodic elliptic waves as well as localized solitary wave solutions of the CNLH system (1). This will also provide the criteria for the existence of nonlinear elliptic waves in the CNLH system. For constructing the Jacobi elliptic solutions of system (\ref{1}), we introduce the traveling wave ansatz
\bes\label{2}\bea
q_{j} = f_{j}(u)e^{i\alpha_{j}}
\,,
\eea
where
\bea
\quad u=\beta(\tau-v\xi+\delta_0),\,\quad \alpha_{j}=k_{j}\xi-\omega_{j}\tau +\delta_{j}\,,  \quad \quad j=1,2,
\eea
in Eq. (1). Here, $f_{j}$, $j=1,2,$ are real functions of $\xi$ and $\tau$ while $\beta$, $\delta_{0}$ and $\delta_{j}$ $j=1,2$ are real constants, $\omega_{j}$s are frequencies of the two components of the CNLH system, $v$ is the velocity and $k_{j}$ is the wave number of the $q_j^{th}$ component. On equating the real and imaginary parts of the resulting equations, we obtain the following equations:
\bea\label{2c}
&&\frac{d^{2} f_{j}}{du^{2}}-\frac{1}{(1+2\kappa v^{2})\beta^{2}}\left[2k_{j}(1+\kappa k_{j})+\omega_{j}^{2}-2(\sigma_{1}f_{1}^{2}+\sigma_{2}f_{2}^{2})\right]f_{j}=0,\,\,\,\,\,\, j=1,2,\nonumber\\
\eea
where
\bea\label{2d}
&&\qquad \qquad v=-\left(\frac{ \omega_{2}}{1+2\kappa k_{2}}\right),\,\,\,\, \omega_{1}=\omega_{2}\left(\frac{1+2\kappa k_{1}}{1+2 \kappa k_{2}}\right).
\eea\ees
Note that Eq.~(\ref{2c}) is different from the earlier systems studied in Refs. \cite{hioe} and \cite{avin} due to the explicit appearance of nonparaxial term $\kappa$ in Eq.~(\ref{2c}). We assume the Lam\'e function ansatz for $f_{j}$, that is,
\bea
f_j=\rho_j \psi_{j}^{(l)},   \quad \quad l,j=1,2,
\eea
where $\psi_{j}^{(l)}$ can be any one of the three first order Lam\'e polynomials for $ l=1$ and for $l=2$ it can be any of the five second order Lam\'e polynomials and with this assumption Eq.~(\ref{2c}) reduces to the Lam\'e equation \cite{whit},
\bea
\frac{d^{2}\psi_{j}^{(l)}}{du^{2}}+[\lambda_{j}^{(l)} - l(l+1)m \sn^{2}(u,m)]\psi_{j}^{(l)}=0,
\eea
where $m$ ($0\leq m \leq 1$) is the modulus parameter of the Jacobi elliptic function  $\sn(u,m)$,~~$l~(=1,2)$ represents the order of the Lam\'e polynomial $\psi_{j}^{(l)}$ and $\lambda_{j}^{(l)}$ is the corresponding eigenvalue. Thus we will have two distinct families of solutions corresponding to the Lam\'e polynomials of order 1 ($l=1$) and of order 2 ($l=2$). The resulting first order and second order elliptic solutions expressed in terms of the Lam\'e polynomials of order one and two are explicitly obtained and discussed in the following sections.

\section{First order elliptic and solitary wave solutions}
The first order solutions of the CNLH system (\ref{1}) consist of six distinct solutions. Additionally, one novel superposed elliptic solution can be constructed in terms of $\dn$ and $\cn$ elliptic functions. These elliptic solutions reduce to two broad types of hyperbolic solitary wave solutions namely bright solitary wave and dark solitary wave in the hyperbolic limit.  Each of the  six elliptic solutions is characterized by twelve real parameters with five conditions, thereby admitting  seven arbitrary real parameters. In the following, we present these first order solutions one by one and clearly demonstrate how the amplitude, speed and width of the nonlinear waves are influenced by the nonparaxial parameter $\kappa$.

\subsection*{\bf Solution 1}
\subsubsection*{\bf (a) Cnoidal waves }

The first elliptic solution of the CNLH system (\ref{1}) is given below.
\bes \label{4}\bea
\left(\begin{array}{l}
q_{1} \\
q_{2} \\
\end{array}\right)=\left(\begin{array}{l}
A~e^{i\alpha_{1}} \\
B~e^{i\alpha_{2}} \\
\end{array}\right)\sqrt{m}~\textit{\cn (u,m)}\,,
\eea
where
\bea\label{4a}
&&v=-\left(\frac{\omega_{2}}{1+2\kappa k_{2}}\right),\, \omega_{1}=\omega_{2}\delta\,,~~\delta=\left(\frac{1+2\kappa k_{1}}{1+2\kappa k_{2}}\right),\, \beta^{2}=\frac{2k_{1}(1+\kappa k_{1})+ \omega_{2}^{2}\delta^{2}}{(2m-1)(2\kappa v^{2}+1)}\,, \nonumber\\
&&A^{2}=\frac{1}{\sigma_{1}}\left[(2\kappa v^{2}+1)\beta^{2}-\sigma_{2}B^{2}\right],\,
\omega_{2}^{2}=\frac{2}{\left[\delta^{2}-1\right]} \left[k_{2}(1+\kappa k_{2})-k_{1}(1+\kappa k_{1})\right]\,.
\eea
Note that all the first order and second order solutions discussed in this paper admit the same constraint conditions (\ref{2d}) for $v$ and $\omega_{1}$ . The above solution 1(a) is possible for the focusing nonlinearity ($\sigma_{1}$ and $\sigma_{2}$ are positive) but for the defocusing nonlinearity ($\sigma_{1}$ and $\sigma_{2}$ are negative) the above solution does not exist. However for the mixed nonlinearity physically admissible solution is possible for appropriate choice of the solution parameters.

\subsubsection*{\bf (b)  Bright solitary waves }

 The solution 1(a) reduces to the following hyperbolic bright solitary wave solution in the limit $m=1$.
\bea\label{5}
\left(\begin{array}{l}
q_{1} \\
q_{2} \\
\end{array}\right)=\left(\begin{array}{l}
A~e^{i\alpha_{1}} \\
B~e^{i\alpha_{2}} \\
\end{array}\right)\sech(u)\,,
\eea
provided
\bea\label{6}
 \beta^{2}=\frac{2k_{1}(1+\kappa k_{1})+ \omega_{2}^{2}\delta^{2}}{(2\kappa v^{2}+1)}\,.
\eea\ees
Here all other constraint parameters are same as given in Eq.~(\ref{4a}).

\subsection*{\bf Solution 2}

\subsubsection*{\bf (a) Snoidal waves }

The second elliptic solution of the CNLH system (\ref{1}) can be expressed as.
\bes \label{7}\bea
\left(\begin{array}{l}
q_{1} \\
q_{2} \\
\end{array}\right)=\left(\begin{array}{l}
A~e^{i\alpha_{1}} \\
B~e^{i\alpha_{2}} \\
\end{array}\right)\sqrt{m}~\textit{\sn (u,m)}\,,
\eea
where
\bea\label{7a}
&&\beta^{2}=-\frac{2k_{1}(1+\kappa k_{1})+\omega_{2}^{2}\delta^{2}}{(1+m)(2\kappa v^{2}+1)}\,,\nonumber\\
&&A^{2}=-\frac{1}{\sigma_{1}}\left[(2\kappa v^{2}+1)\beta^{2}+\sigma_{2}B^{2}\right],\,
\omega_{2}^{2}=\frac{2}{\left[\delta^{2}-1\right]} \left[k_{2}(1+\kappa k_{2})-k_{1}(1+\kappa k_{1})\right]\,.
\eea
The solution parameters and the system parameters have to be properly chosen to ensure that $\beta^{2}$, $A^{2}$ and $\omega_{2}^{2}$ are indeed positive. It can be inferred that for the focusing nonlinearity ($\sigma_{1}>0$ and $\sigma_{2}>0$), $A^{2}$ becomes negative and the above solution does not exist. However, the solution exists for the defocusing and mixed nonlinearities with appropriate choice of parameters.

\subsubsection*{\bf (b) Dark solitary waves }

In the limit $m=1$ the solution 2(a) becomes as the following hyperbolic solution
\bea\label{8}
\left(\begin{array}{l}
q_{1} \\
q_{2} \\
\end{array}\right)=\left(\begin{array}{l}
A e^{i\alpha_1} \\
B e^{i\alpha_2} \\
\end{array}\right)\tanh(u)\,.
\eea
Now, all the constraints parameters are same as given in Eq. (\ref{7a}) except for $\beta^{2}$ which now becomes as
\bea\label{9}
\beta^{2}=-\frac{2k_{1}(1+\kappa k_{1})+\omega_{2}^{2}\delta^{2}}{2(2\kappa v^{2}+1)}\,.
\eea\ees

\subsection*{\bf Solution 3~:~ Dnoidal waves }

Another solution for the CNLH system is given below:
\bes \label{10}\bea
\left(\begin{array}{l}
q_{1} \\
q_{2} \\
\end{array}\right)=\left(\begin{array}{l}
A~e^{i\alpha_{1}} \\
B~e^{i\alpha_{2}} \\
\end{array}\right)\textit{\dn (u,m)}\,.
\eea
The constraint conditions turn out to be the same as given in Eq. (3b) except for $\beta^{2}$ which is now given as
\bea
\beta^{2}=\frac{2k_{1}(1+\kappa k_{1})+\omega_{2}^{2}\delta^{2}}{(2-m)(2\kappa v^{2}+1)}\,.
\eea\ees
This solution also exists for focusing and mixed nonlinearities. The solution 3 reduces to the hyperbolic solution 1(b) for the limiting case $m=1$.

\subsection*{\bf Solution 4}

\subsubsection*{\bf (a)  Dnoidal-Snoidal waves }

Yet another elliptic solution of the CNLH system (\ref{1}) is,
\bes \label{11}\bea
\left(\begin{array}{l}
q_{1} \\
q_{2} \\
\end{array}\right)=\left(\begin{array}{l}
A~\textit{\dn (u,m)}~e^{i\alpha_{1}} \\
B~\sqrt{m}~\textit{\sn (u,m)}~e^{i\alpha_{2}} \\
\end{array}\right)\,,
\eea
provided
\bea\label{11a}
&&A^{2}=\frac{1}{\sigma_{1}}\left[(2\kappa v^{2}+1)\beta^{2}+\sigma_{2}B^{2}\right]\,,\nonumber\\
&&B^{2}=\frac{1}{2\sigma_{2}}[2k_{1}(1+\kappa k_{1})+
\omega_{2}^{2}\delta^{2}+(m-2)(2\kappa v^{2}+1)\beta^{2}]\,,\nonumber\\
&&\omega_{2}^{2}=\frac{1}{\left[\delta^{2}-\left(1+\frac{2\kappa \beta^{2}}{(1+2\kappa k_{2})^{2}}\right)\right]}\left[2k_{2}(1+\kappa k_{2})-2k_{1}(1+\kappa k_{1})+\beta^{2}\right]\,.
\eea

The above solution exists in case $\sigma_1 > 0, \sigma_2 < 0$ and it
definitely does not exist in case $\sigma_1 < 0, \sigma_2 > 0$. On the
other hand, for focusing and defocusing nonlinearities, the existence
of the solution requires the parameters to satisfy some conditions.
Particularly, for the defocusing case we require
$(2\kappa v^{2}+1)\beta^{2}<|\sigma_{2}B^{2}|$ and
$|(m-2)(2\kappa v^{2}+1)\beta^{2}|>2k_{1}(1+\kappa k_{1})
+\omega_{2}^{2}\delta^{2}$. One should also choose the solution as well as
the system parameters appropriately to ensure that $\omega_{2}^{2}$ is indeed
positive definite.

\subsubsection*{\bf (b)  Bright-Dark solitary waves }

The above solution 4(a) reduces  to the following hyperbolic solution in the limiting case $m=1$.
\bea\label{11}
\left(\begin{array}{l}
q_{1} \\
q_{2} \\
\end{array}\right)=\left(\begin{array}{l}
A~\sech(u)~e^{i\alpha_{1}}\\
B~\tanh(u)~e^{i\alpha_{2}}\\
\end{array}\right)\,,
\eea
where
\bea\label{12}
B^{2}=\frac{1}{2\sigma_{2}}[2k_{1}(1+\kappa k_{1})+
\omega_{2}^{2}\delta^{2}-(2\kappa v^{2}+s)\beta^{2}]\,.
\eea\ees
Here the first component $q_{1}$ and the second component $q_{2}$ comprise of bright and dark solitary waves respectively. Such type of coexistence of bright and dark solitary waves even for the focusing nonlinearity is a special feature of the present system.

\subsection*{\bf Solution 5:~ Cnoidal-Snoidal waves }

Next elliptic wave solution of the CNLH system (\ref{1}) is obtained as
\bes \label{13}\bea
\left(\begin{array}{l}
q_{1} \\
q_{2} \\
\end{array}\right)=\left(\begin{array}{l}
A~\sqrt{m}~\textit{\cn (u,m)}~e^{i\alpha_{1}} \\
B~\sqrt{m}~\textit{\sn (u,m)}~e^{i\alpha_{2}} \\
\end{array}\right)\,.
\eea
The constraint relations for this solution are same as those of solution 4(a) (see Eq.(\ref{11a})) except for $B^{2}$ and $\omega^{2}$ which are now given by
\bea
&&m B^{2}=\frac{1}{2 \sigma_{2}}\left[2k_{1}(1+\kappa k_{1})+\omega_{2}^{2}\delta^{2}-(2m-1)(2\kappa v^{2}+1)\beta^{2}\right]\,, \nonumber\\
&&\omega_{2}^{2}=\frac{1}{\left[\delta^{2}-(1+\frac{2 m \kappa \beta^{2}}{(1+2 \kappa k_{2})^{2}})\right]}\left[2 k_{2}(1+ \kappa k_{2})-2 k_{1}(1+\kappa k_{1})+m  \beta^{2}\right]\,.
\eea\ees
The above solution 5 reduces to the hyperbolic solution 4(b) in the limit $m=1$.

\subsection*{\bf Solution 6:~ Dnoidal-Cnoidal waves }

The explicit form of the sixth solution of the CNLH system (\ref{1}) is given below.
\bes \label{14}\bea
\left(\begin{array}{l}
q_{1} \\
q_{2} \\
\end{array}\right)=\left(\begin{array}{l}
A~\textit{\dn (u,m)}~e^{i\alpha_{1}} \\
B~\sqrt{m}~\textit{\cn (u,m)}~e^{i\alpha_{2}} \\
\end{array}\right)\,.
\eea
The constraint conditions associated with the above solution are
\bea
&&A^{2}=\frac{1}{\sigma_{1}}\left[(2\kappa v^{2}+1)\beta^{2}-\sigma_{2}B^{2}\right]\,,\nonumber\\
&&(m-1) B^{2}=\frac{1}{2 \sigma_{2}}\left[2k_{1}(1+\kappa k_{1})+\omega_{2}^{2}\delta^{2}-(2-m)(2\kappa v^{2}+1)\beta^{2}\right]\,, \nonumber\\
&&\omega_{2}^{2}=\frac{1}{\left[\delta^{2}-\left(1-\frac{2  \kappa \beta^{2}}{(1+2\kappa k_{2})^{2}}\right)\right]}\left[2k_{2}(1+\kappa k_{2})-(\beta^{2}+2k_{1}(1+\kappa k_{1}))\right]\,.
\eea\ees
The above mentioned solution  6  can exist for focusing nonlinearity with the
condition $|2\kappa v^{2}+1)\beta^{2}|>|\sigma_{2}B^{2}|$ but does not
exist for the defocusing nonlinearity. For the mixed nonlinearity case, it
demands the parameters to satisfy some conditions: $|(2-m)(2\kappa v^{2}+1)\beta^{2}|>2k_{1}(1+\kappa k_{1})+\omega_{2}^{2}\delta^{2}$ for the mixed case with $\sigma_{1}>0, \sigma_{2}<0$. Also the other choice of the mixed case $\sigma_{1}<0,\sigma_{2}>0$, requires $(2\kappa v^{2}+1)\beta^{2}<\sigma_{2}B^{2}$. To make $\omega_{2}^{2}$  positive, we should appropriately choose the solution and system parameters.

As expected, in the limit $m=1$, this solution goes over to the hyperbolic
solution 1(b).

\subsection*{\bf Solution 7:~Superposed first order elliptic waves }

Finally, we find that the linear superposition of two different elliptic solutions is also a solution of the CNLH system (1) \cite{avin1}. The corresponding superposed solution is
\bes\label{15}\bea
\left(\begin{array}{l}
q_{1} \\
q_{2} \\
\end{array}\right)=\left(\begin{array}{l}
 \frac{1}{2} \left(A \dn(u,m)
+D \sqrt{m} \cn(u,m) \right)
~e^{i\alpha_{1}}\,\\
\frac{1}{2} \left(B \dn(u,m)
+E \sqrt{m} \cn(u,m) \right)
~e^{i\alpha_{2}}\,
\end{array}\right)\,,
\eea
provided
\bea
&& A^{2}=\frac{1}{\sigma_{1}}\left[(2\kappa v^{2}+1)\beta^{2}-\sigma_{2}B^{2}\right]\,,\nonumber \\
&&\beta^{2}=2\frac{\left[2k_{1}(1+\kappa k_{1})+\omega_{2}^{2}\delta^{2}\right]}{(1+m)(2\kappa v^{2}+1)}\,,
\omega_{2}^{2}=\frac{2}{\left[\delta^{2}-1\right]} \left[k_{2}(1+\kappa k_{2})-k_{1}(1+\kappa k_{1})\right]\,.
\eea\ees
Here the signs of $D = \pm A$ and $E = \pm B$ are correlated. This solution exists for the focusing nonlinearity but for the defocusing nonlinearity  solution 7 can not exist. The mixed type nonlinearity also allows this solution only for the choice $(2\kappa v^{2}+1)\beta^{2}<|\sigma_{2}B^{2}|$. One can note that the solution 7 reduces to the hyperbolic solution 1(b) in the limit $m=1$.

In our above discussion on different elliptic waves we have pointed out the type of nonlinearity admitting those solutions. We tabulate below the solutions along with nonlinearities which support the solutions for a better understanding.
\begin{table}[H]
\centering
\caption {Types of nonlinearities and their corresponding solutions }
\begin{tabular}{|c|c|}
  \hline
 Solutions &Types of nonlinearity supporting the solutions\\\hline
 \multirow{3}{2cm}{1, 3, 6, 7} & Focusing nonlinearity ($\sigma_{1}>0$ and $\sigma_{2}>0$)\\
            & and mixed type nonlinearity [($\sigma_{1}>0$ and $\sigma_{2}<0$) \\
            &or ($\sigma_{1}<0$ and $\sigma_{2}>0$)] \\\hline
  \multirow{3}{2cm}{2} & Defocusing nonlinearity ($\sigma_{1}<0$ and $\sigma_{2}<0$)\\
            & and mixed type nonlinearity [($\sigma_{1}>0$ and $\sigma_{2}<0$) \\
            &or ($\sigma_{1}<0$ and $\sigma_{2}>0$)] \\\hline
 \multirow{3}{2cm}{4, 5} & Focusing nonlinearity ($\sigma_1 > 0$ and $\sigma_2 > 0$)\\
      &  Defocusing nonlinearity ($\sigma_1 < 0$ and $\sigma_2 < 0$)\\
      & and mixed type nonlinearity ($\sigma_1 > 0$ and $\sigma_2 < 0$)\\ \hline
\end{tabular}
\end{table}
Having obtained the solutions, the next natural step is to examine the physical significance of the obtained new solutions. From a physical perspective, it is of interest to identify the role of the nonparaxial parameter on the above different nonlinear elliptic wave solutions. This is an important aspect of the present study. For illustrative purpose, we consider solution 1(a), superposed solution [solution 7] and the bright soliton/solitary wave solution 1(b). One can infer from the first order solutions in the absence of nonparaxial term $(\kappa=0)$ (note that for this choice the CNLH system reduces to the standard integrable CNLS system) the  frequencies as well as the wave numbers in both components become equal. Ultimately the solutions become phase-locked \cite{Akhm1}. This phase-locking property will be removed when the nonparaxial parameter is introduced and now the two components will possess distinct phases. Further in order to interpret the role of the nonparaxial parameter, one should carefully analyse the constraint conditions on the physical parameters like speed (modulus of velocity), pulse width $(1/\beta)$ and the amplitudes.

Figs.~1(a-c) show the variation of speed, pulse width and amplitude of the first component of solutions 1(a), 7 and the hyperbolic solution 1(b) respectively, with respect to $\kappa$. The nature of the curve depends upon the sign of $k_{2}$ in particular the denominator term $(1+2\kappa k_{2})$ in the expression for $v$ (see Eq.~(\ref{4a})). In Fig.~1(a) the speed decreases gradually as $\kappa$ increases and reaches a saturation for the parametric choice $\omega_{2}=0.8,~B=1.6,~k_{2}=1.0$ and $\kappa=0.0001$. Then the pulse width and amplitude $A$ of the first component behave in an opposite  manner as $\kappa$ increases. It can be inferred from the above three solutions (1(a), 1(b), 7) that the amplitude of the second component is independent of $\kappa$. In Figs.~1(b-c), also we notice the speed decreases rapidly as $\kappa$ increases and reaches a saturation. But the pulse width initially increases as $\kappa$ increases and after some nonparaxial value $\kappa$, it decreases. Finally, we observe that the first component amplitude $A$ also initially increases as $\kappa$ increases and after some nonparaxial value, it attains a constant value. In all the three figures we notice that the speed is decreasing as $\kappa$ is increased. This shows that by tuning the nonparaxial parameter we can decelerate the elliptic waves as well as the solitary waves.

Another important observation from the Figs.~1(b-c) corresponding to superposed solution and hyperbolic solution is that the amplitude of the first component attains a saturation after a particular value of $\kappa$. However the speed continues to change. Thus in this saturation region by altering $\kappa$ we can change the speed without affecting the amplitude of the first component.

\renewcommand{\floatpagefraction}{0.7}
\begin{figure}[H]
\centering~\includegraphics[width=0.3\linewidth]{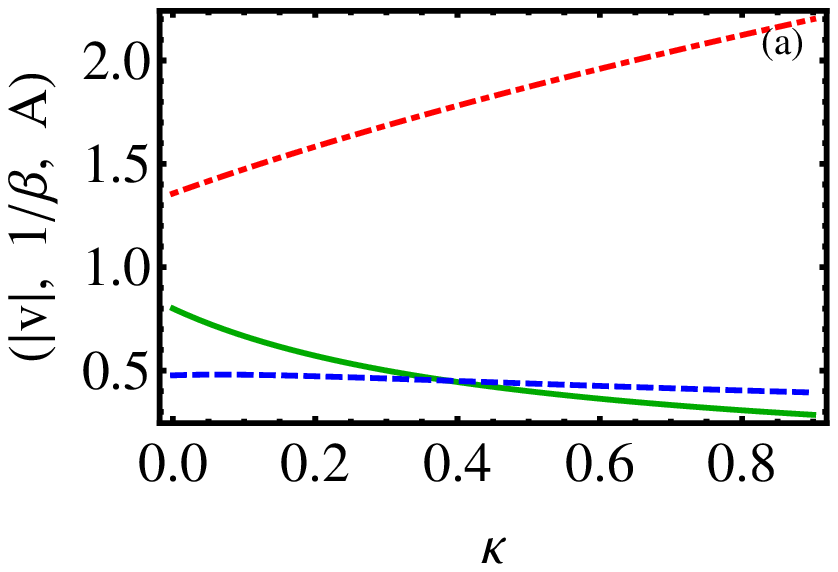}~
\includegraphics[width=0.3\linewidth]{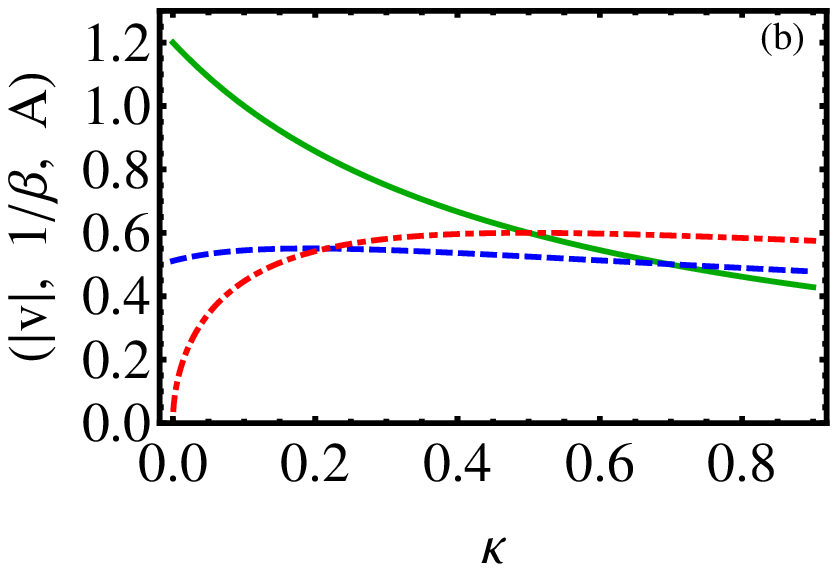}~
\includegraphics[width=0.3\linewidth]{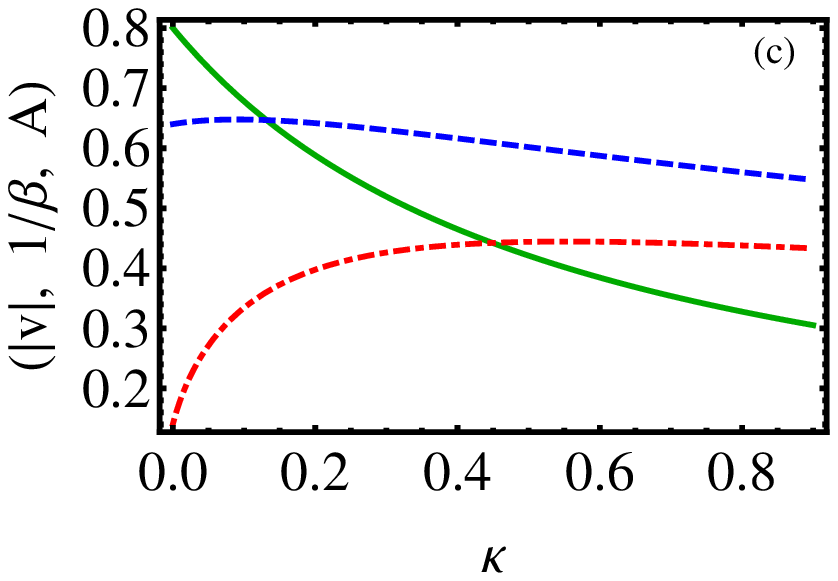}~
\caption{Plots of speed (solid dark green), pulse width (dotted blue) and amplitude of the first component (dotdashed red) versus nonparaxial parameter $\kappa$. (a) solution 1(a) with $\omega_{2}=0.8$, $k_{2}=1.0$, $B= 1.6$, (b) solution 7 with $\omega_{2}=1.2$, $k_{2}=1.0$, $B=E=1$ and (c) solution 1(b) with $\omega_{2}=0.8$, $k_{2}=0.9$, $B=0.99$, $m=1$. In all the figures $\sigma_{1}=1$, $\sigma_{2}=1$, $\delta_{0}=\delta_{1}=\delta_{2}=1$, $\xi=1.5$, $\tau=1.5$. Except in Fig.~1(c) $m=0.8$.} \label{Figure1}
\end{figure}
Next we consider the solutions 4(a), 5 and 6. A careful look at these solutions and their constraint relations shows that these solutions behave distinctly from the solutions 1(a), 7 and 1(b). For these solutions, the pulse width becomes arbitrary while the  amplitude of second component is not arbitrary. Also the pulse width affects the frequency and ultimately influences the speed. To get further insight we present the plots of amplitude of the first and second components and speed for these three solutions 4(a), 5 and 6 in Fig.~2.

In Figs.~2(a-c) we display the plots of speed (modulus of velocity), amplitude of the first and second components of solutions 4(a), 5 and 6 respectively, with respect to $\kappa$. In Figs.~2(a-c) the speed decreases gradually as $\kappa$ increases. But the amplitudes of the first (A) and second (B) components are increasing as $\kappa$ increases. One can infer from the above three solutions that the pulse width of the  solution is independent of $\kappa$. Fig.~2 also shows that by changing the nonparaxial parameter we can decelerate the elliptic waves as well as the solitary waves.

\begin{figure}[H]
\centering\includegraphics[width=0.3\linewidth]{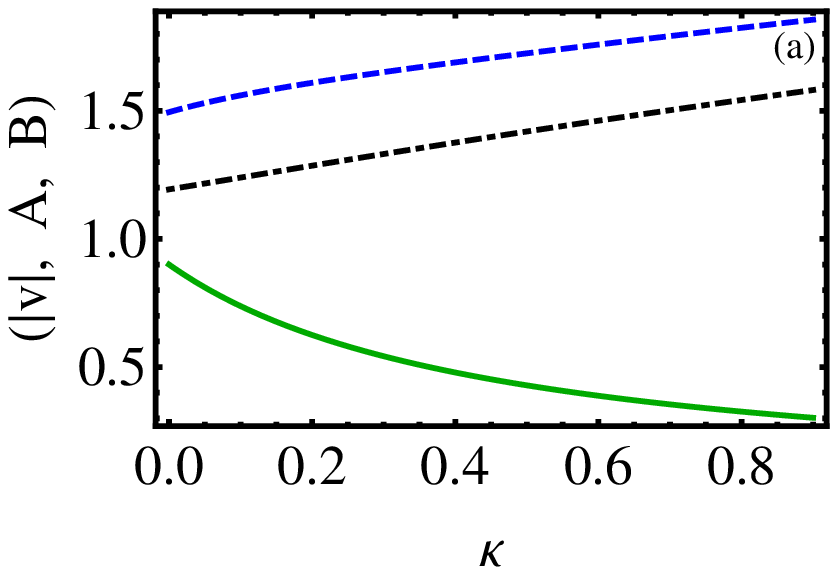}~~~
\includegraphics[width=0.3\linewidth]{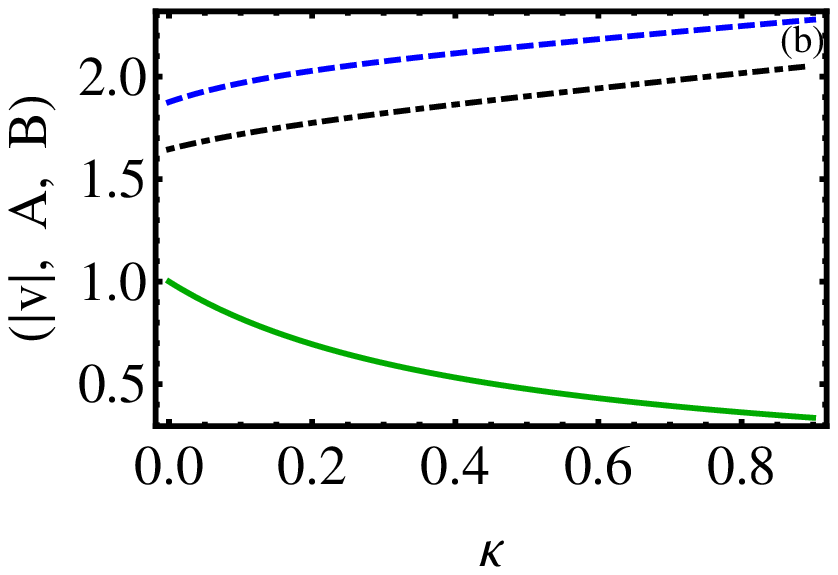}~~
\includegraphics[width=0.3\linewidth]{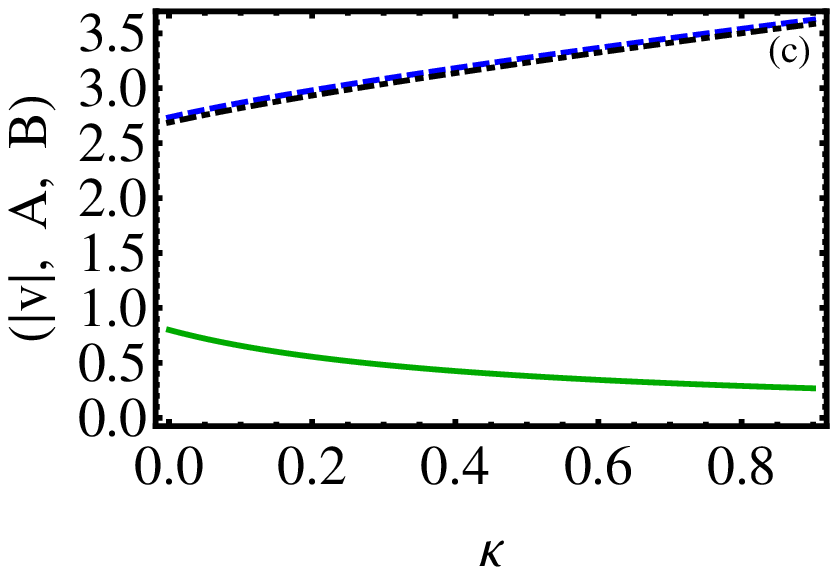}
\caption{Plots of speed (solid dark green),\, amplitude of the first component (dotted blue) and amplitude of the second component (dotdashed black) versus nonparaxial parameter. (a) solution 4(a) with $\omega_{2}=0.9$, $k_{2}=1.1$, $\beta=0.9$, (b) solution 5 with $\omega_{2}=1$, $k_{2}=1.1$, $\beta=0.9$ and (c) solution 6 with $\omega_{2}=0.8$, $k_{2}=1.1$, $\beta=0.5$, $\sigma_{2}=-1$. In all the figures $\sigma_{1}=1$, $\sigma_{2}=1$ (except in Fig.~2(c)), $\delta_{0}=\delta_{1}=\delta_{2}=1$, $\tau=1.5$, $m=0.8$.}
\end{figure}
We show the propagation of the elliptic waves supported by the solutions 1(a), 2(a) and 4(a) respectively in Figs.~3(a-c). Here, the top and middle panels of Fig.~3(a) show the intensity profiles of the solution 1(a). The intensity profiles of the first and second components are in-phase and this is demonstrated in the bottom panel of Fig.~3(a). Next, the top and middle panels in Figs .~3(b-c) show the intensity profiles for the  solutions 2(a) and 4(a) respectively. The bottom panel of Fig.~3(b) shows that  both components are in-phase while that of Fig.~3(c) shows both components are out-of-phase.
\begin{figure}[H]
\centering
\includegraphics[width=0.32\linewidth]{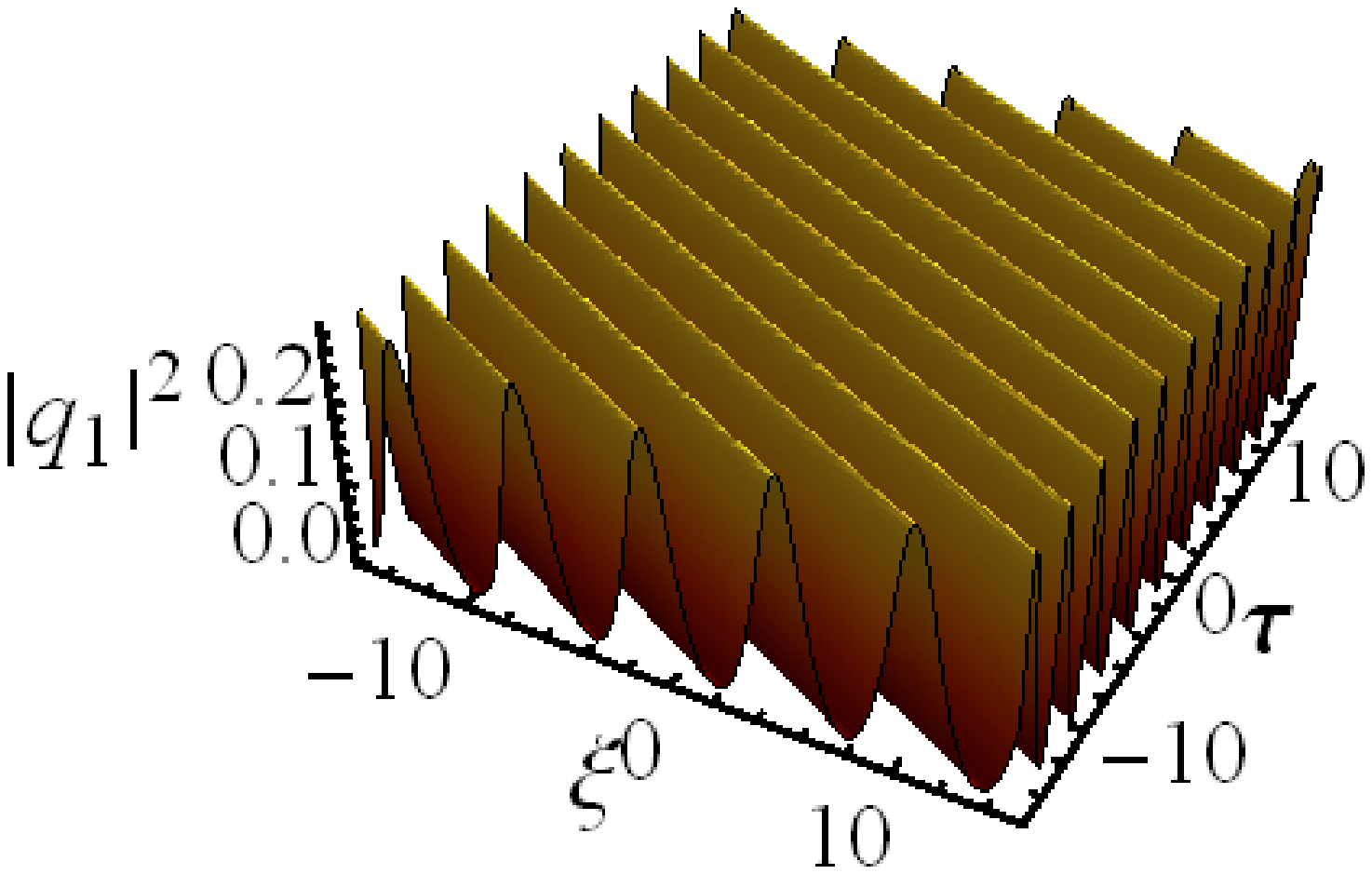}
\includegraphics[width=0.32\linewidth]{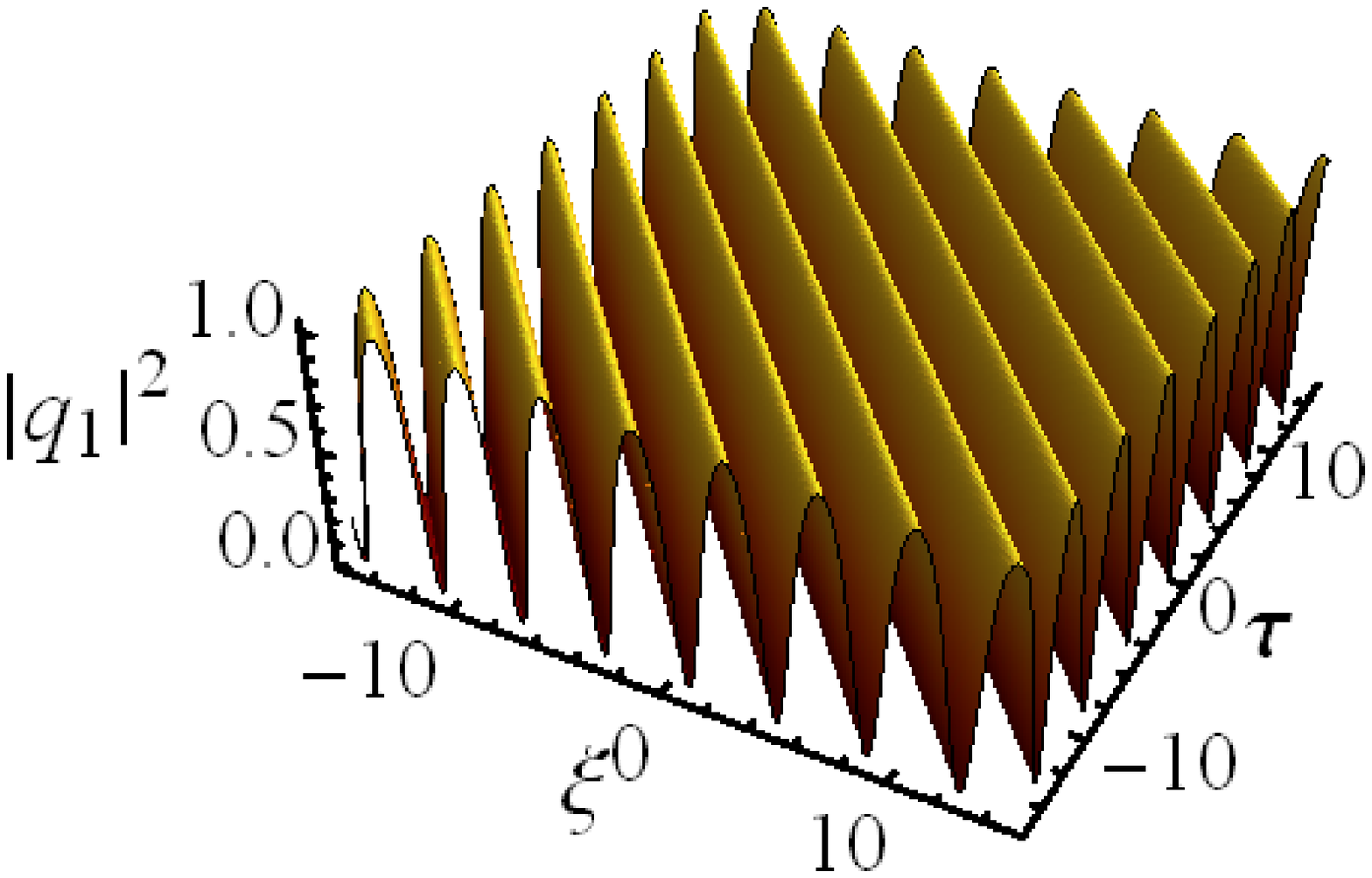}
\includegraphics[width=0.32\linewidth]{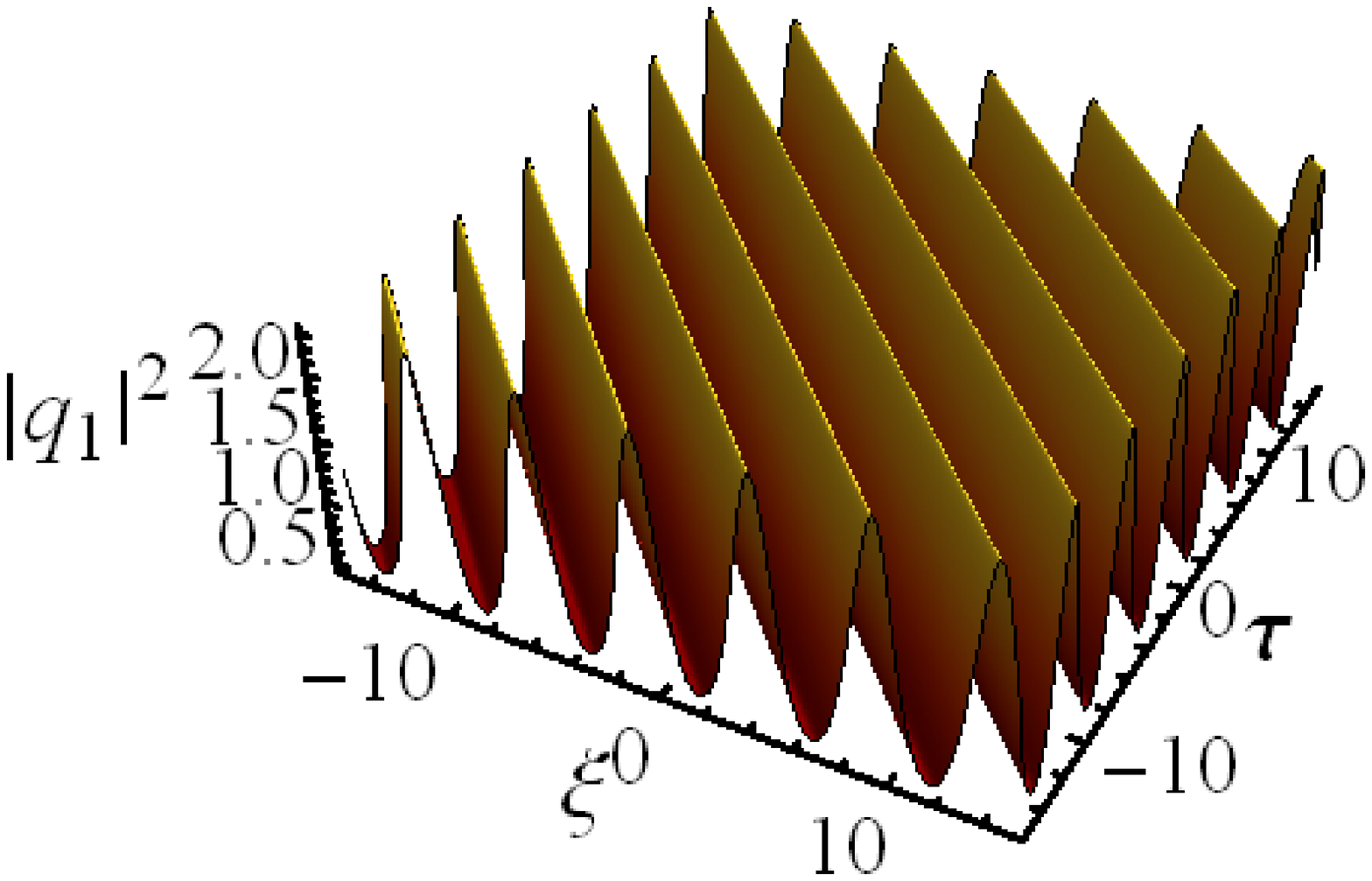}\\
\includegraphics[width=0.32\linewidth]{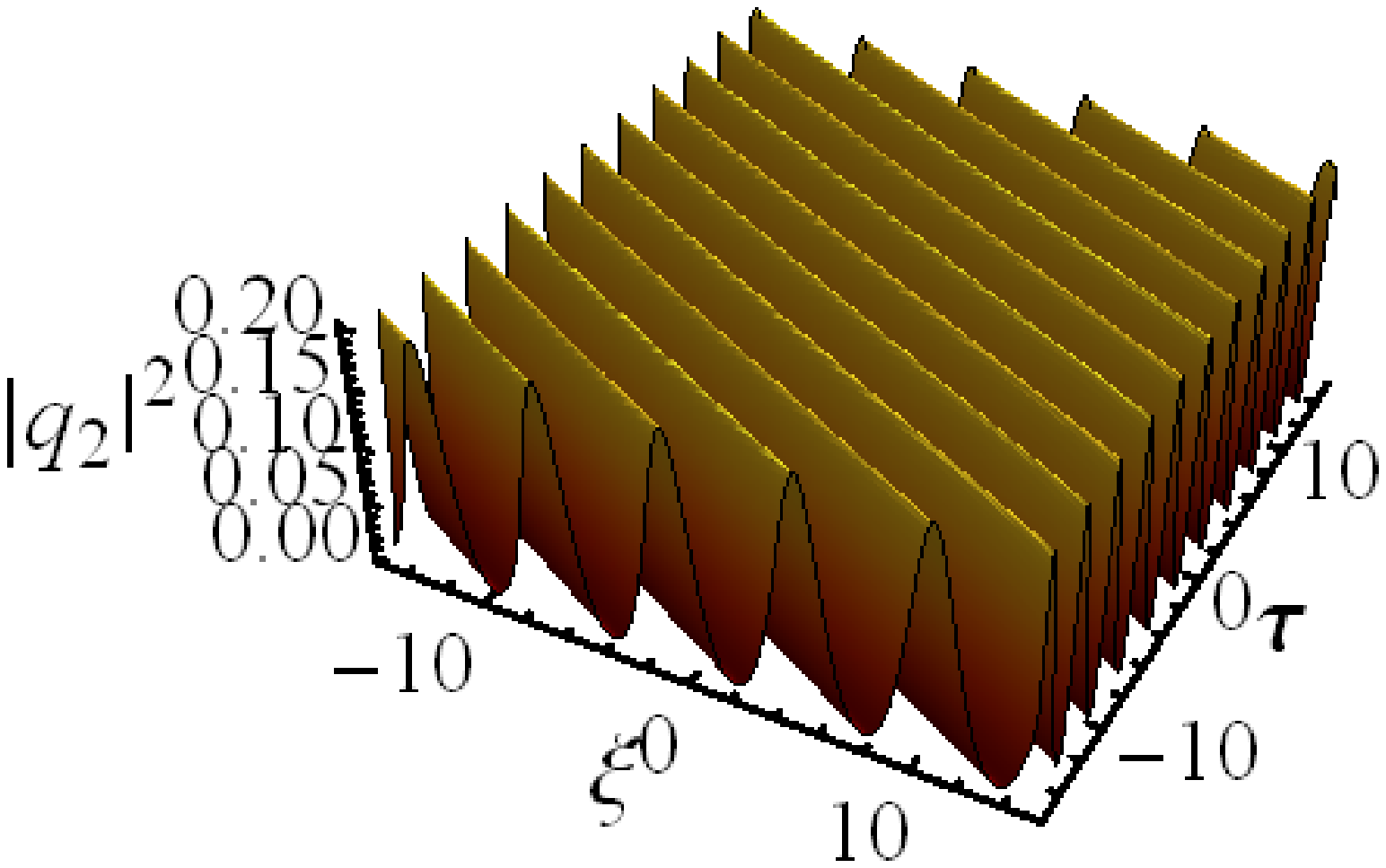}
\includegraphics[width=0.32\linewidth]{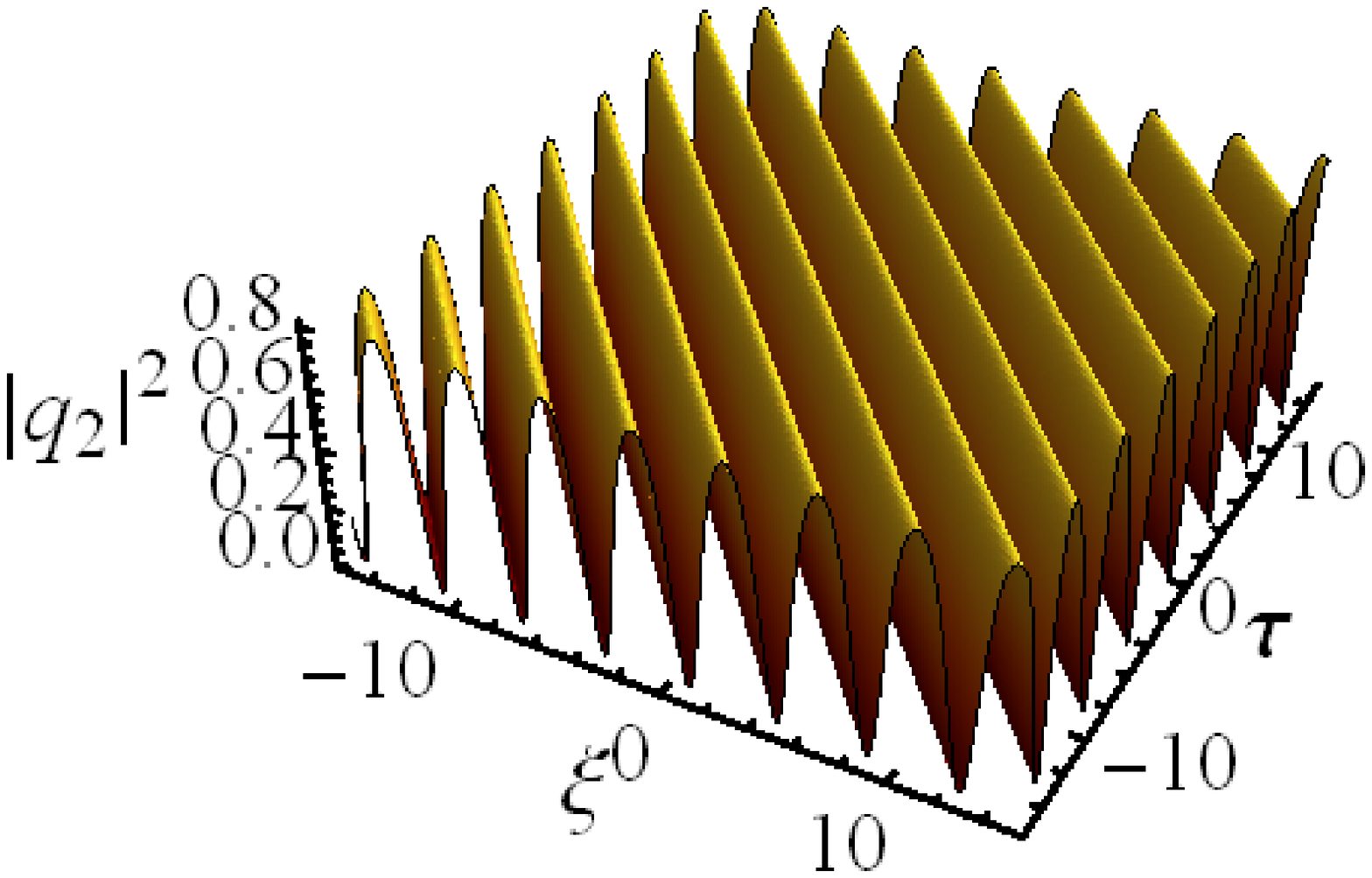}
\includegraphics[width=0.32\linewidth]{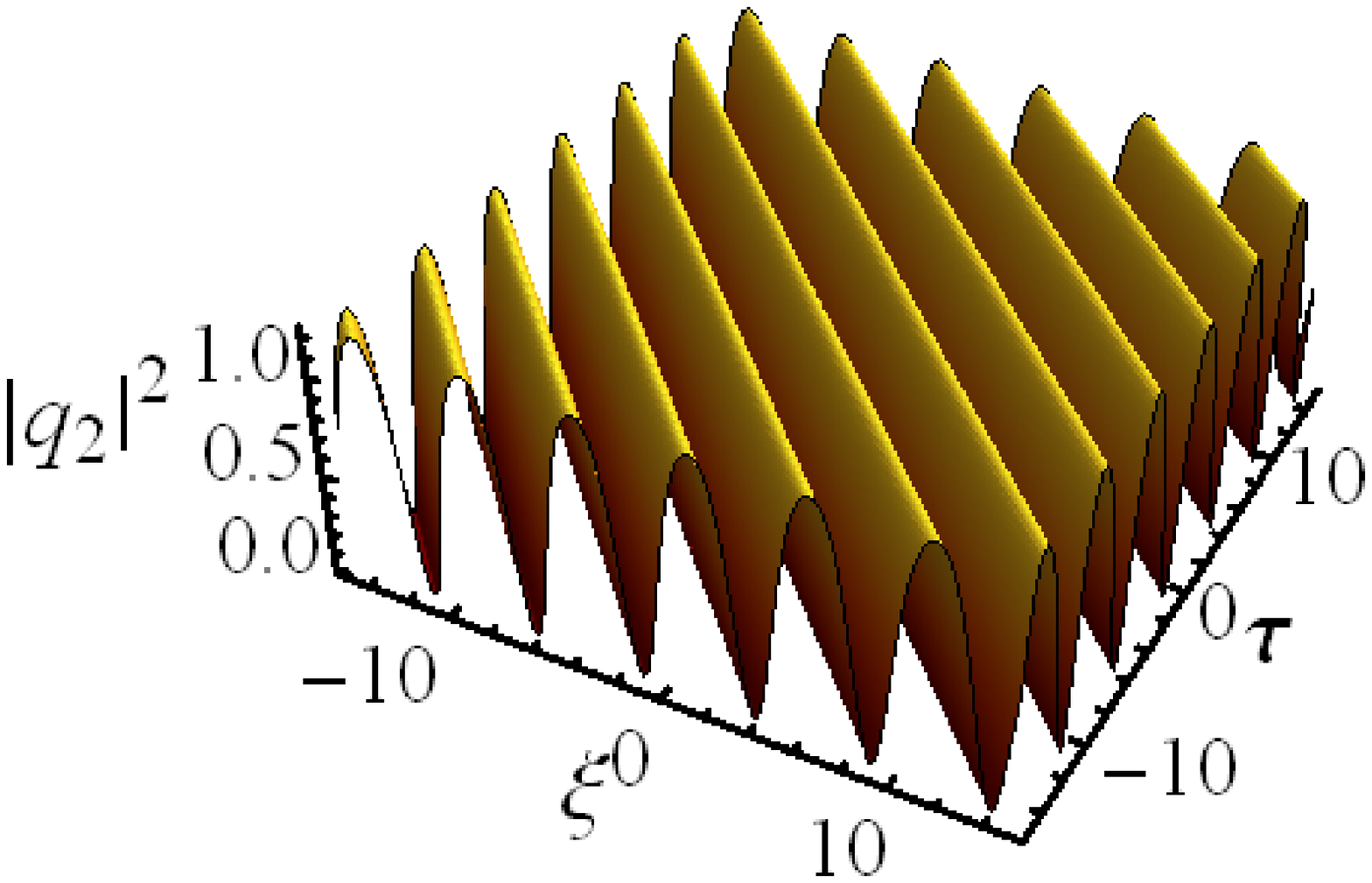}\\
\includegraphics[width=0.32\linewidth]{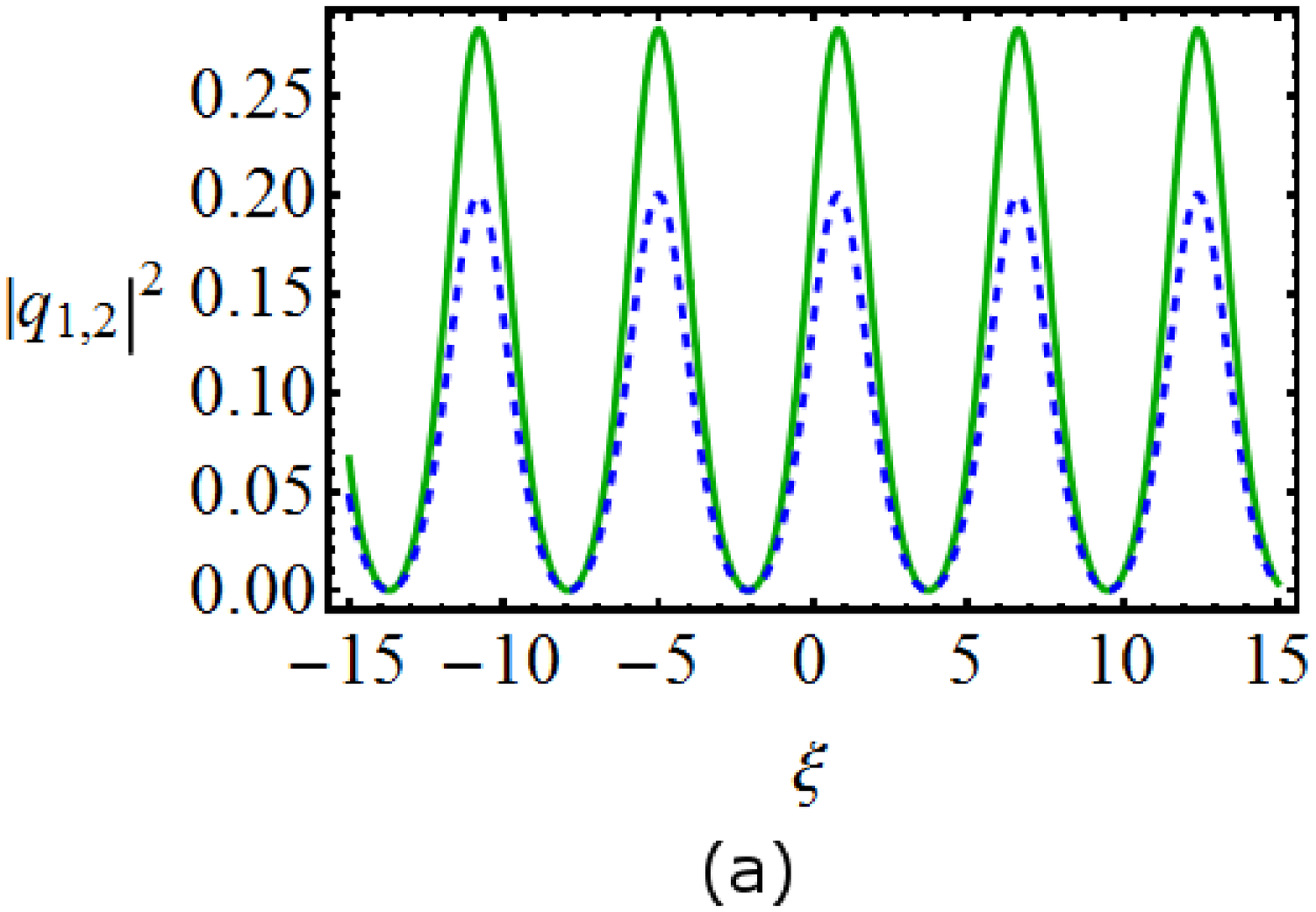}
\includegraphics[width=0.32\linewidth]{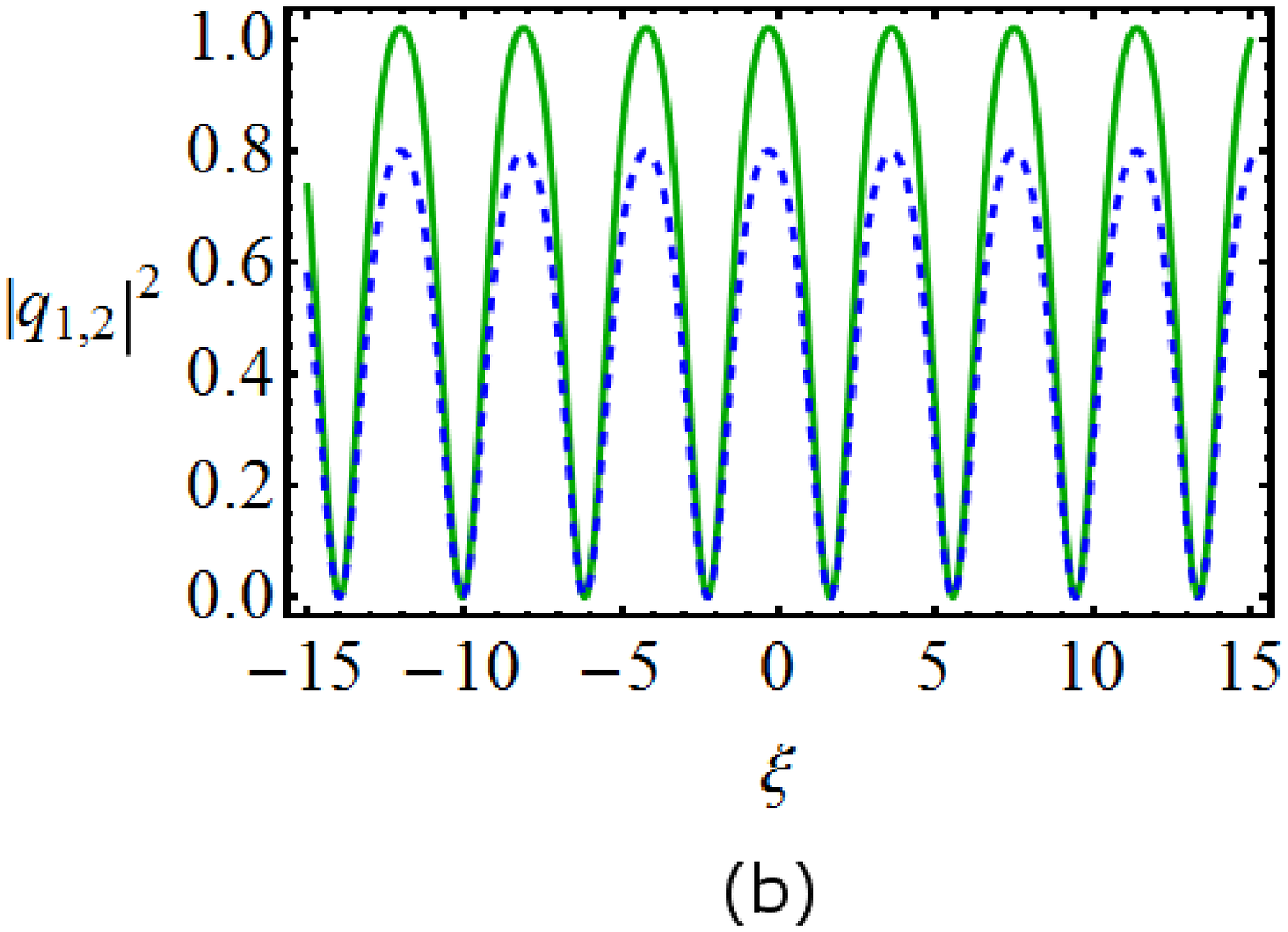}
\includegraphics[width=0.32\linewidth]{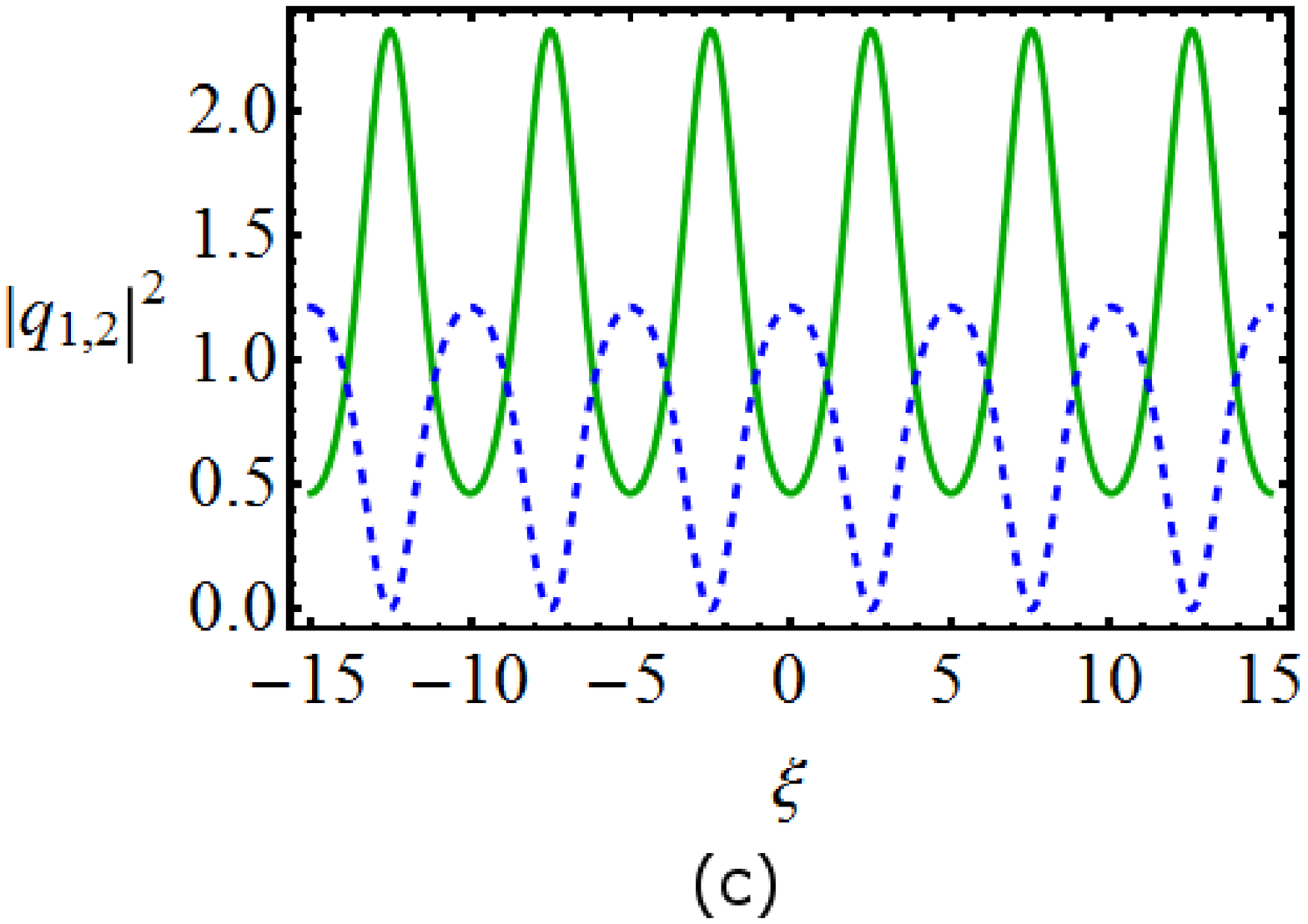}
\end{figure}
\begin{figure}[H]
\caption{The intensity plots of solution 1(a) with $\omega_{2}=0.5$, $k_{2}=0.6$, $B=1$, solution 2(a) with $\omega_{2}=1.1$, $k_{2}=-1.6$, $B=2$, $\sigma_{1}=-1$ and solution 4(a) for the choice $\omega_{2}=1$, $k_{2}=1.1$, $\beta=0.9$ are shown in the first two rows of Figs.~3(a-c) respectively. The third row shows the corresponding two dimensional plots of $q_1$ (dark green) and $q_2$  (dashed blue) components with $\tau=1.5$.  In all the figures $\sigma_{1}=1$ (except in Fig.~3(b)), $\kappa=0.0001$, $\sigma_{2}=1$, $\delta_{0}=\delta_{1}=\delta_{2}=1$, $m=0.8$.}
\end{figure}

Further, we illustrate all first order hyperbolic solutions 1(b), 2(b) and 4(b) in Fig.~4. Particularly,   Fig.~4(a) shows the bright soliton/solitary wave profiles exhibited by solution 1(b). Next, we show in Fig.~4(b) that in solution 2(b) both components admit dark soliton/solitary wave profiles.  Finally, we depict solution 4(b) in Fig.~4(c) showing the co-existence of  bright and dark solitary waves respectively in the first and second components.

\begin{figure}[H]
\centering
\includegraphics[width=0.32\linewidth]{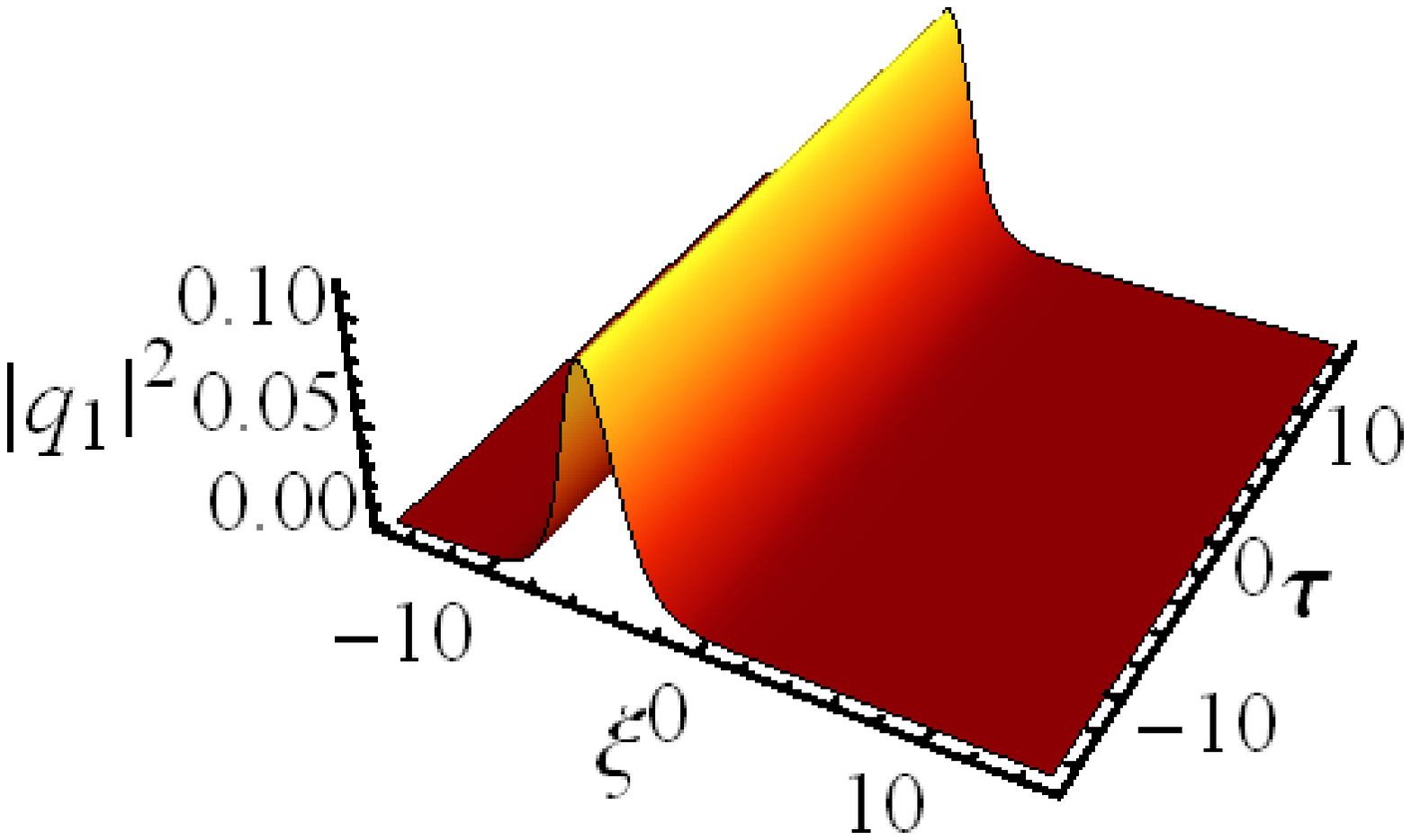}
\includegraphics[width=0.32\linewidth]{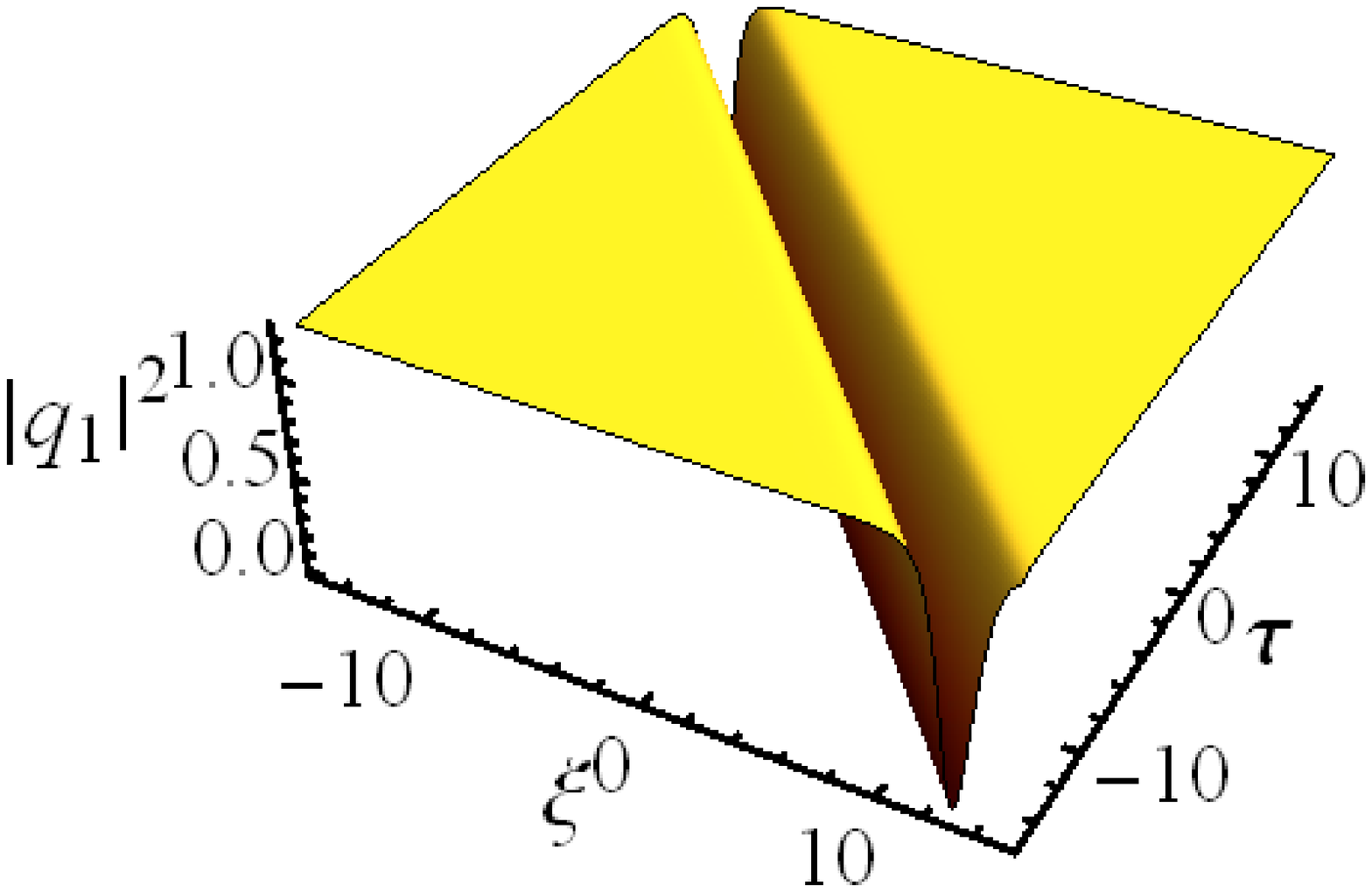}
\includegraphics[width=0.32\linewidth]{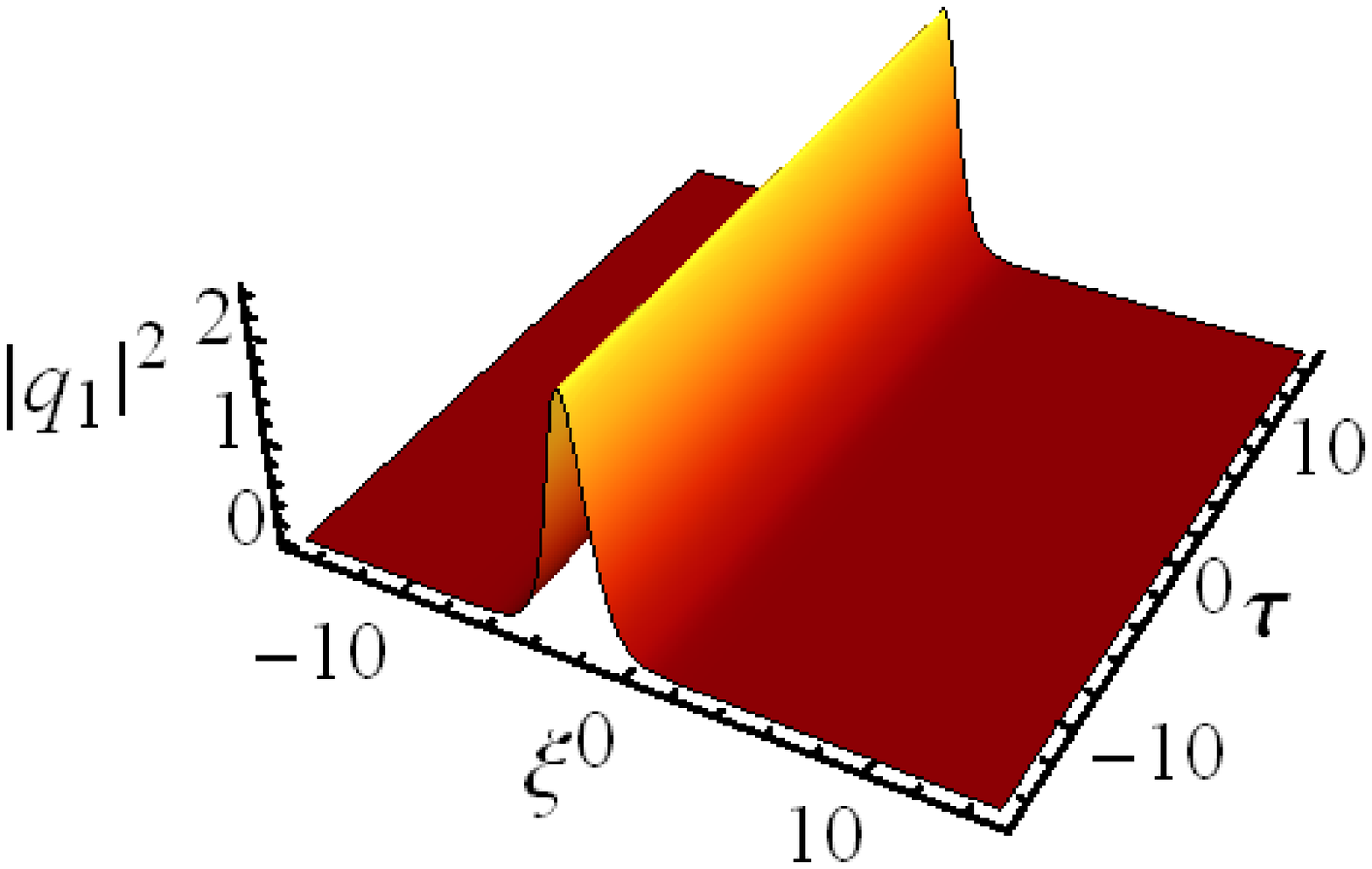}\\
\includegraphics[width=0.32\linewidth]{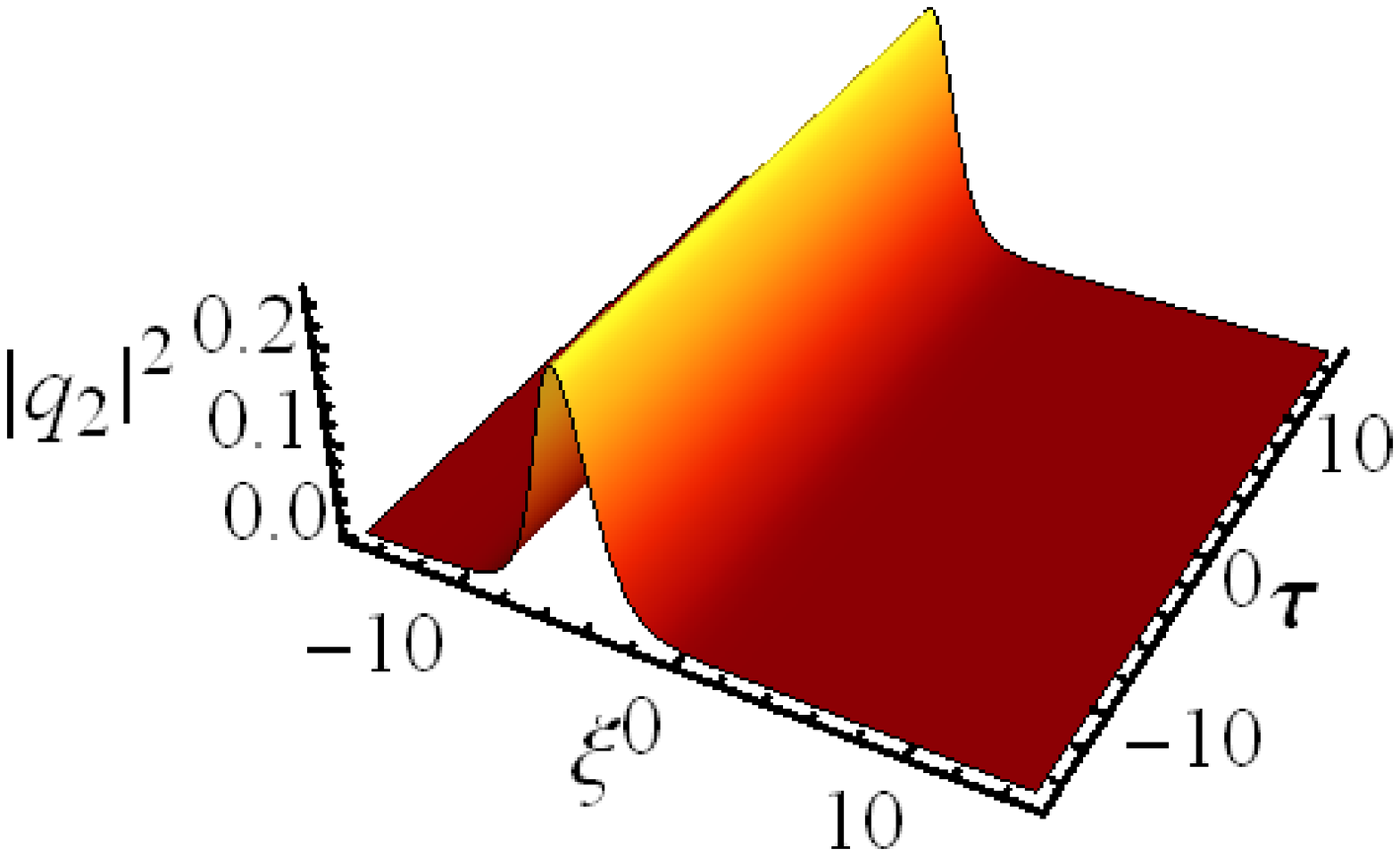}
\includegraphics[width=0.32\linewidth]{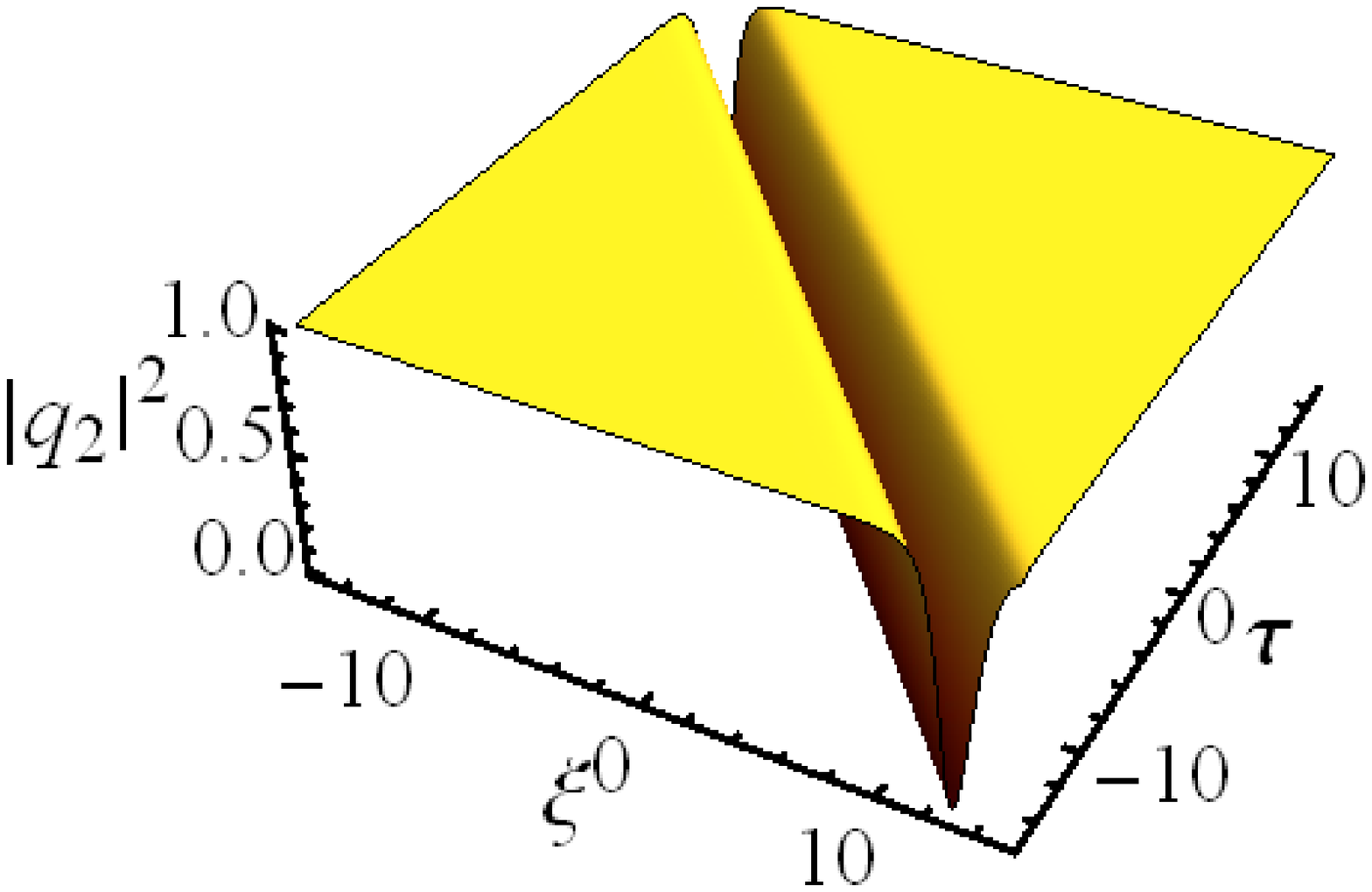}
\includegraphics[width=0.32\linewidth]{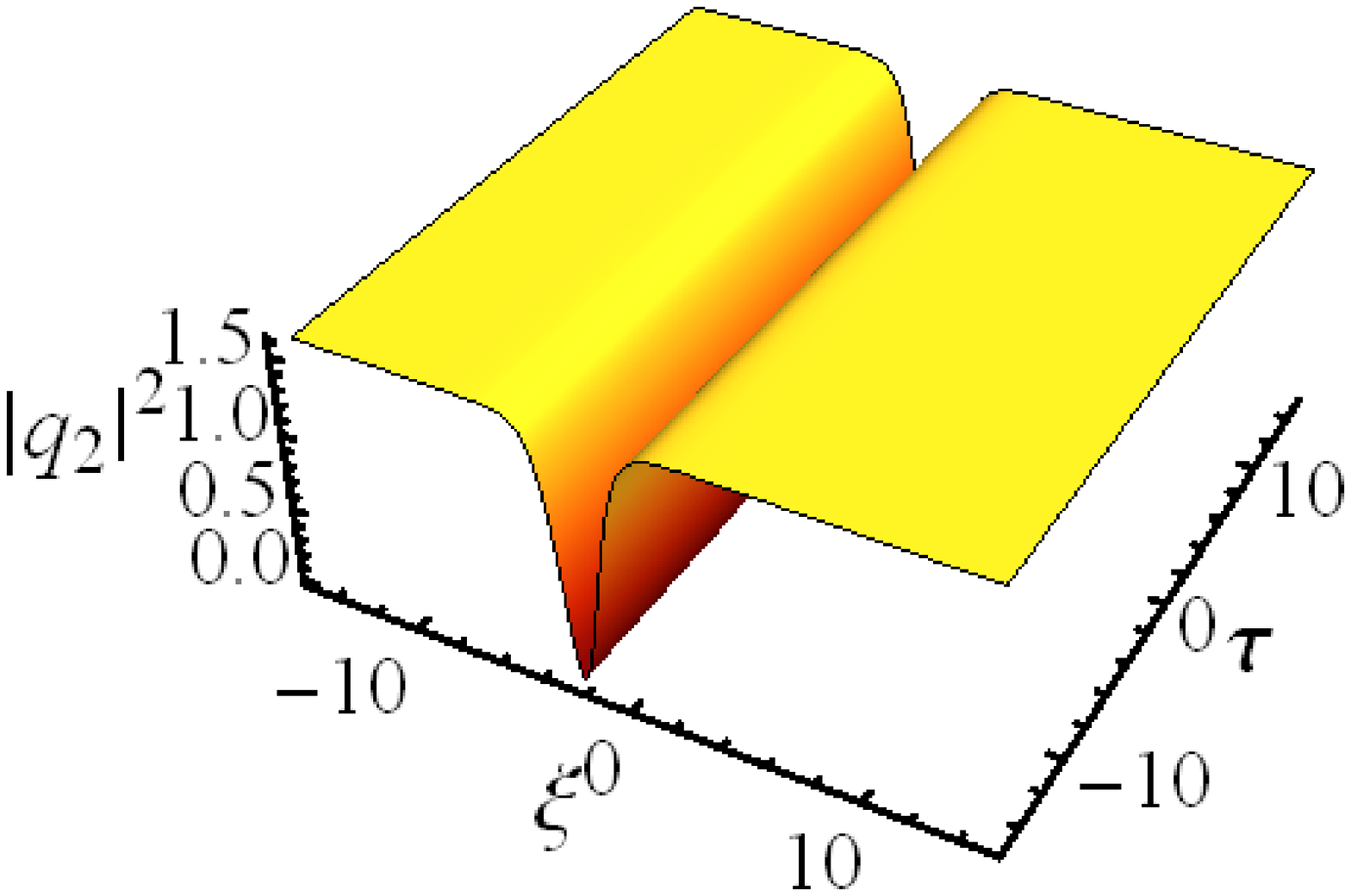}\\
\includegraphics[width=0.32\linewidth]{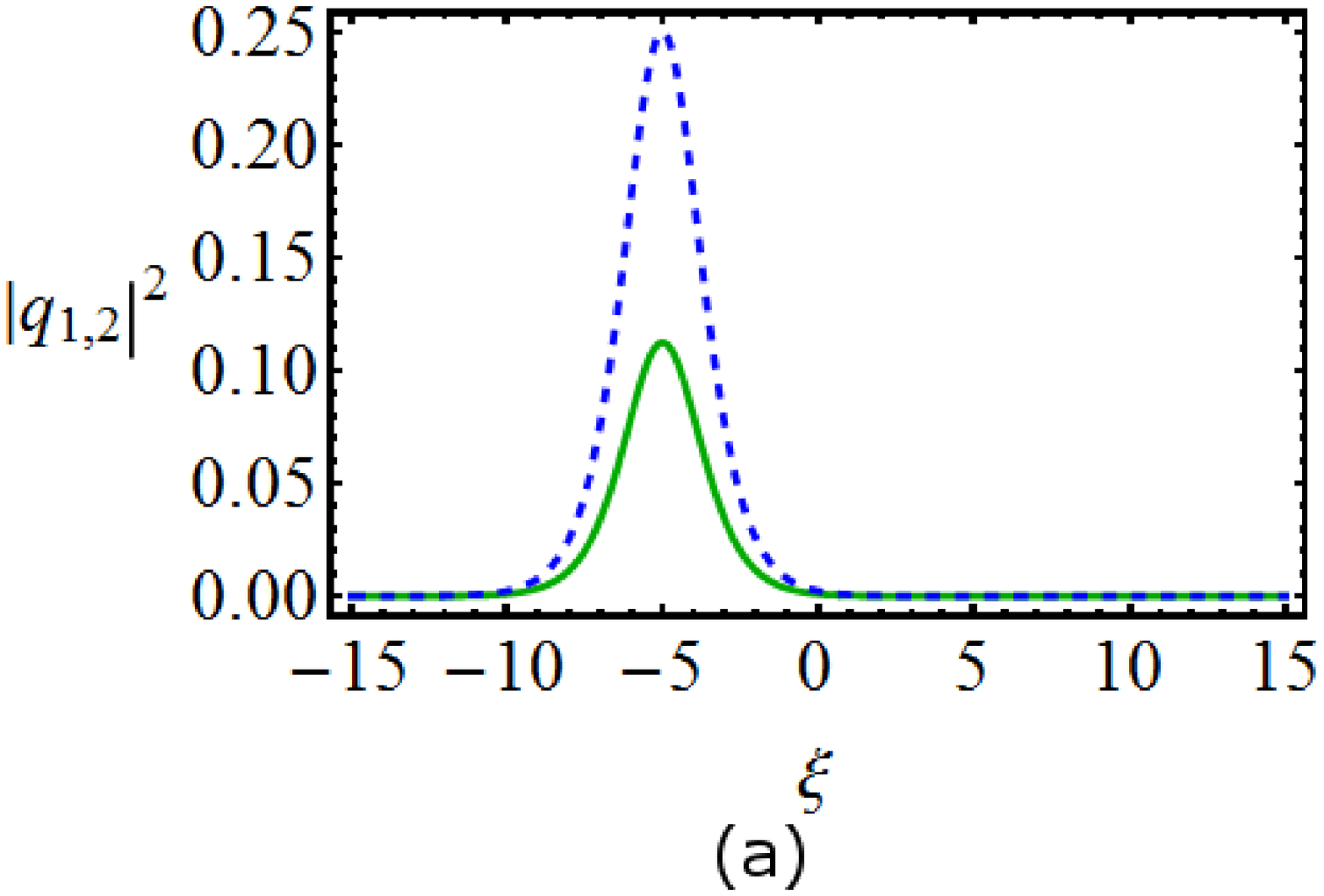}
\includegraphics[width=0.32\linewidth]{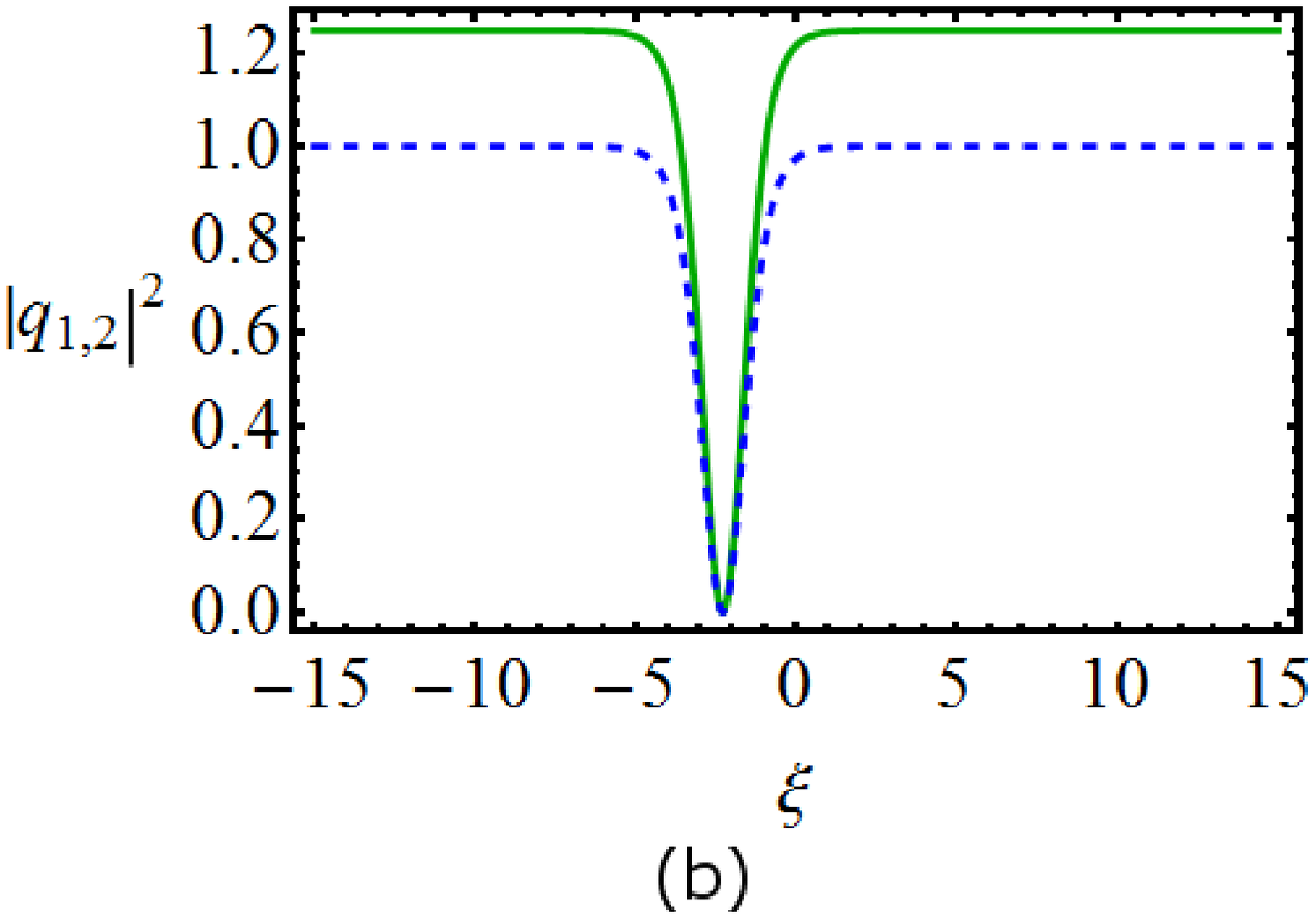}
\includegraphics[width=0.32\linewidth]{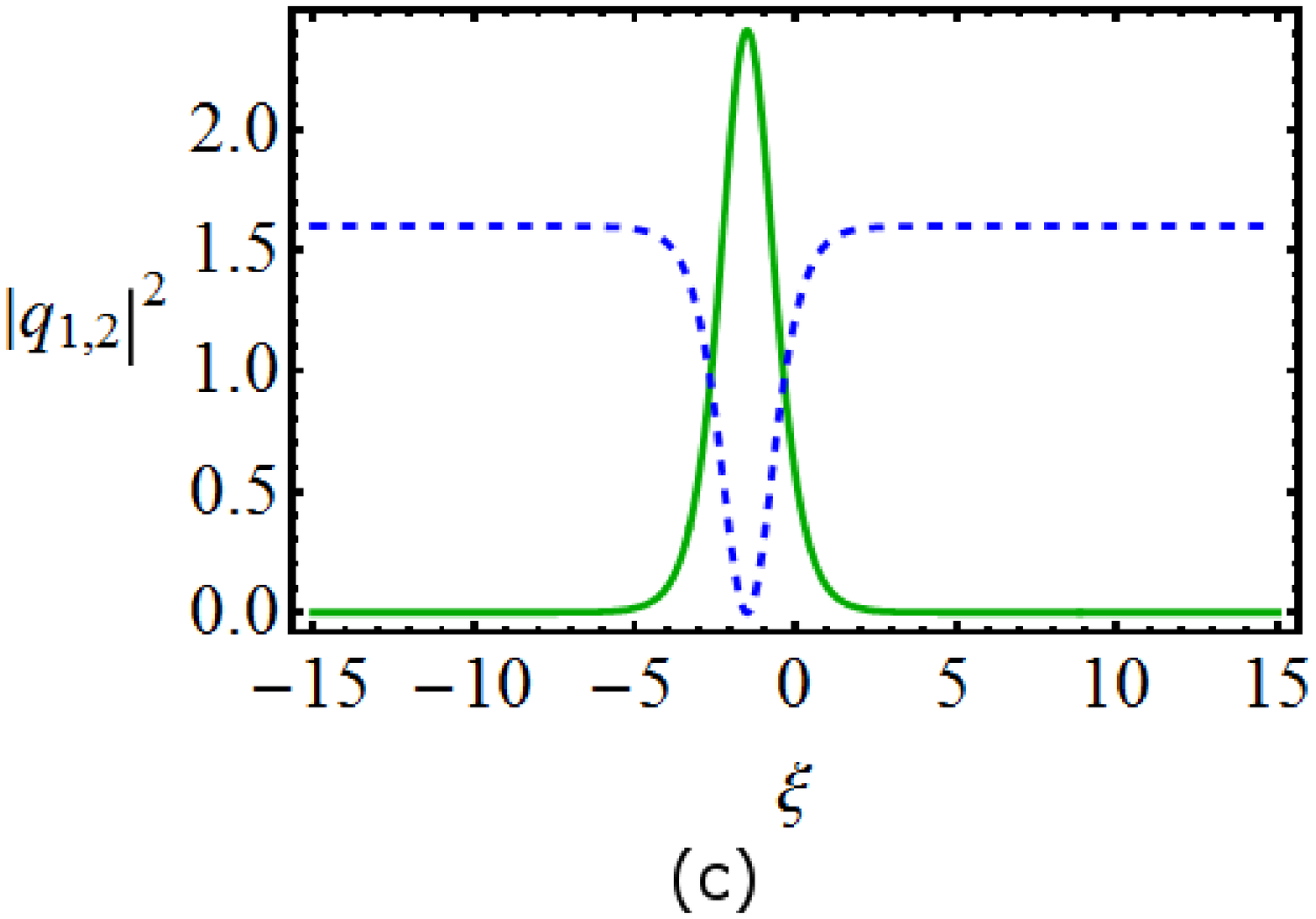}
\caption{The intensity plots of solution 1(b) with $\omega_{2}=0.5$, $k_{2}=0.6$, $B=1$, solution 2(b) with $\omega_{2}=1.1$, $k_{2}=-1.6$, $B=2$, $\sigma_{1}=-1$ and solution 4(b) for the choice $\omega_{2}=1$, $k_{2}=1.1$, $\beta=0.9$ are shown in the first two rows respectively. The third row shows the corresponding two dimensional plots $q_1$ (dark green) and $q_2$  (dashed blue) components with $\tau=1.5$.  In all the figures $\sigma_{1}=1$ (except in Fig.~4(b)), $\kappa=0.0001$, $\sigma_{2}=1$, $\delta_{0}=\delta_{1}=\delta_{2}=1$, $m=1$.}
\end{figure}

\section{Second order Elliptic and solitary wave solutions }
In this section, we construct seven distinct elliptic solutions of the CNLH system (\ref{1}) in terms of Lam\'e polynomials of order 2 and also obtain a novel superposed elliptic solution. We also discuss the associated hyperbolic solutions when the modulus parameter $m$ becomes unity. Each of the elliptic solutions given below has six arbitrary parameters. We refer to these solutions as second order elliptic solutions as they are stemming from the second order Lam\'e polynomials.

\subsection*{\bf Solution 8}

\subsubsection*{\bf (a) \boldmath{$\cn~\dn-\sn~\dn$} waves }

The first second order elliptic solution of the CNLH system (\ref{1}) is obtained as,
\bes \label{15a}\bea
\left(\begin{array}{l}
q_{1} \\
q_{2} \\
\end{array}\right)=\left(\begin{array}{l}
A~\sqrt{m}~\textit{\cn (u,m)}~\textit{\dn (u,m)}~e^{i\alpha_{1}} \\
B~\sqrt{m}~\textit{\sn (u,m)}~\textit{\dn (u,m)}~e^{i\alpha_{2}} \\
\end{array}\right)\,,
\eea
where $\alpha_{1,2}$ and $u$ are as defined in Eq. (\ref{2}).\\
The above solution requires,
\bea\label{15b}
&&\sigma_{1}A^{2}=\sigma_{2}B^{2}\,,\nonumber\\
&&m A^{2}=\frac{1}{\sigma_{1}}\left[3(2\kappa v^{2}+1)\beta^{2}\right],\,\beta^{2}=\left[\frac{2k_{1}(1+\kappa k_{1})+\omega_{2}^{2}\delta^{2}}{(2\kappa v^{2}+1)(5-m)}\right]\,, \nonumber\\
&&\omega_{2}^{2}=\frac{2}{\left((5m-4)\delta^{2}-(5-m)\right)}
\left[k_{2}(5-m)(1+\kappa k_{2})-k_{1} (5m-4)(1+\kappa k_{1})\right]\,.
\eea
A careful look at the expression for $A^{2}$ shows that the choice $\sigma_{1}<0$ is ruled out. Ultimately, from $A^{2}=\frac{\sigma_{2}}{\sigma_{1}}B^{2}$, we find that the solution  will exist only for the focusing nonlinearity. Also, the parameters $k_{1}$, $k_{2}$ and $m$ have to be chosen appropriately so that  $\omega_{2}$ is indeed positive.

\subsubsection*{\bf (b) Blue-White solitary waves}

In the hyperbolic limit ($m=1$) the above solution results in a specially coupled solitary wave solution. Here the first component admits  symmetric profile whereas that of the second component is antisymmetric. Such type of  $q_{1}$ and $q_{2}$ components can also be viewed as blue and white solitary waves respectively \cite{hioe1} and can be expressed as

\bea \label{16}
\left(\begin{array}{l}
q_{1} \\
q_{2} \\
\end{array}\right)=\left(\begin{array}{l}
A \sech^2(u)~e^{i\alpha_{1}} \\
B \sech(u)~\tanh(u)~e^{i\alpha_{2}} \\
\end{array}\right)\,.
\eea
The solution parameters satisfy the following conditions:
\bea\label{17}
A^{2}=\frac{1}{\sigma_{1}}\left[3(2\kappa v^{2}+1)\beta^{2}\right],\,
\beta^{2}=\left[\frac{2k_{1}(1+\kappa k_{1})+\omega_{2}^{2}\delta^{2}}{4(2\kappa v^{2}+1)}\right],\,
\omega_{2}^{2}=\frac{2}{\left(\delta^{2}-4\right)}
\left[4 k_{2}(1+\kappa k_{2})-k_{1} (1+\kappa k_{1})\right]\,
\eea\ees
and all other constraints parameters are as given in Eq. (\ref{15b}).

\subsection*{\bf Solution 9:~\boldmath{$\cn~\dn-\sn~\cn$} waves }

Another second order elliptic solution of the CNLH system (\ref{1}) is
\bes \label{18}\bea\label{18a}
\left(\begin{array}{l}
q_{1} \\
q_{2} \\
\end{array}\right)=\left(\begin{array}{l}
A~\sqrt{m}~\textit{\cn (u,m)}~\textit{\dn (u,m)}~e^{i\alpha_{1}} \\
B~m~\textit{\sn (u,m)}~\textit{\cn (u,m)}~e^{i\alpha_{2}} \\
\end{array}\right)\,,
\eea
in which the solution parameters satisfy the relations,
\bea
&&\sigma_{1}A^{2}=\sigma_{2}B^{2}\,,\nonumber\\
&&A^{2}=\frac{1}{\sigma_{1}}\left[3(2\kappa v^{2}+1)\beta^{2}\right],\,
\beta^{2}=\left[\frac{2k_{1}(1+\kappa k_{1})+\omega_{2}^{2}\delta^{2}}{(2\kappa v^{2}+1)(5m-1)}\right]\,,\nonumber\\
&&\omega_{2}^{2}=\frac{2}{\left((5m-4)\delta^{2}-(5m-1)\right)}
\left[k_{2}(5m-1)(1+\kappa k_{2})-k_{1}(5m-4)(1+\kappa k_{1})\right]\,.
\eea\ees
As in the previous solution, here too the solution is ruled out for the choice $\sigma_{1}<0$ and the solution exists only for the focusing type nonlinearity. Here also, the same blue and white solitary waves given by solution 8(b) result for $q_{1}$ and $q_{2}$, respectively in the hyperbolic limit $m=1$ along with same constraint conditions.

\subsection*{\bf Solution 10:~\boldmath{$\sn~\dn-\sn~\cn$} waves }

Yet another second order elliptic solution for the CNLH system (\ref{1}) is
\bes \label{19}\bea
\left(\begin{array}{l}
q_{1} \\
q_{2} \\
\end{array}\right)=\left(\begin{array}{l}
A~\sqrt{m}~\textit{\sn (u,m)}~\textit{\dn (u,m)}~~e^{i\alpha_{1}} \\
B~m~\textit{\sn (u,m)}~\textit{\cn (u,m)}~e^{i\alpha_{2}} \\
\end{array}\right)\,,
\eea
with the conditions
\bea
&&\sigma_{1}A^{2}=-\sigma_{2}B^{2}\,,\nonumber\\
&&(1-m) A^{2}=-\frac{1}{\sigma_{1}}\left[3(2\kappa v^{2}+1)\beta^{2}\right],\,
\beta^{2}=-\frac{1}{(1+4m)}\left[\frac{2k_{1}(1+\kappa k_{1})+\omega_{2}^{2}\delta^{2}}{(2\kappa v^{2}+1)}\right]\,,\nonumber\\
&&\omega_{2}^{2}=\frac{2}{\left((4+m)\delta^{2}-(4m+1)\right)}
\left[k_{2}(4m+1)(1+\kappa k_{2})-k_{1}(4+m)(1+\kappa k_{1})\right]\,.
\eea\ees
One can infer from the expression for $A^{2}$ and the condition $\sigma_{1} A^{2}=-\sigma_{2} B^{2}$ that the above solution is possible only for the mixed type nonlinearity. This solution becomes singular in the hyperbolic limit $(m=1)$.

\subsection*{\bf Solution 11}

\subsubsection*{\bf (a) \boldmath{$\dn^{2}- \sn~\cn$} waves}

We also identify the following second order solution for the CNLH system (\ref{1})
\bes \label{20}\bea
\left(\begin{array}{l}
q_{1} \\
q_{2} \\
\end{array}\right)=\left(\begin{array}{l}
\left(A~\textit{\dn}^2 \textit{(u,m)}+D\right)~e^{i\alpha_{1}} \\
B~m~\textit{\sn (u,m)}~\textit{\cn (u,m)}~e^{i\alpha_{2}} \\
\end{array}\right)\,,
\eea
provided
\bea
&&\sigma_{1}A^{2}=\sigma_{2}B^{2},\,p=\frac{D}{A}\,,\nonumber\\
&&p=-\frac{1}{3}\left((2-m)\pm\sqrt{1-m+m^{2}}\right),\,
(2+2p-m) A^{2}=\frac{1}{\sigma_{1}}[3(2\kappa v^{2}+1)\beta^{2}]\,,~~
\beta^{2}=\left[\frac{2k_{1}(1+\kappa k_{1})+\omega_{2}^{2}\delta^{2}}{(2\kappa v^{2}+1)b_{1}}\right]\,,\nonumber\\
&&\omega_{2}^{2}=\frac{2}{(a_{1}\delta^{2}-b_{1})}\left[b_{1}k_{2}(1+\kappa k_{2})-a_{1}k_{1}(1+\kappa k_{1})\right]\,,
(2+2p-m) a_{1}= [6(1+p)^{2}-(4+m)(2+2p-m)]\,,\nonumber\\
&&(2+2p-m) b_{1}=2[3(1+p)^{2}-(2+2p-m)((1+m)+\sqrt{1-m+m^{2}}].
\eea
It can be inferred from the expression for $p$ that it admits two values corresponding to the $\pm$ sign before the term $\sqrt{1-m+m^{2}}$. Ultimately, for the positive sign the above solution 11(a) exists only for the defocusing nonlinearity but for the negative sign this solution exists only for the focusing nonlinearity.

\subsubsection*{\bf  (b) Red-White solitary waves}

In the hyperbolic limit this solution 11(a) with $p=-\frac{1}{3}\left((2-m)-\sqrt{1-m+m^{2}}\right)$  reduces to the solution  8(b). The same solution 11(a) with $p=-\frac{1}{3}\left((2-m)+\sqrt{1-m+m^{2}}\right)$ reduces to the following special hyperbolic solitary wave solution in the limit $m=1$:
\bea\label{21}
\left(\begin{array}{l}
q_{1} \\
q_{2} \\
\end{array}\right)=\left(\begin{array}{l}
A \left(\sech^2(u)-\frac{2}{3}\right)~e^{i\alpha_{1}} \\
B\,\sech(u)~\tanh(u)~e^{i\alpha_{2}} \\
\end{array}\right)\,,
\eea
provided
\bea\label{22}
&& \sigma_1 A^2 = \sigma_2 B^2\,,~~A^{2}=-\frac{1}{\sigma_{1}}
\left[9(2\kappa v^{2}+1)\beta^{2}\right]\,, \nonumber \\
&&\beta^{2}=-\left[\frac{2k_{1}(1+\kappa k_{1})+\omega_{2}^{2}\delta^{2}}{8(2\kappa v^{2}+1)}\right]\,,
\omega_{2}^{2}=\frac{2}{(7\delta^{2}-8)}\left[8k_{2}(1+\kappa k_{2})-7k_{1}(1+\kappa k_{1})\right]\,.
\eea\ees
Here all other constraints parameters are same as that of solution 11(a). One can also view the first component $q_{1}$ and the second component $q_{2}$ as red and white solitary waves respectively \cite{hioe1}. In fact, $q_{1}$ admits like W-shape solitary wave profile.

\subsection*{\bf Solution 12}

\subsubsection*{\bf (a) \boldmath{$\dn^{2}-\sn~\dn$} waves}

One more second order elliptic solution for the CNLH system (\ref{1}) is
\bes \label{23}\bea
\left(\begin{array}{l}
q_{1} \\
q_{2} \\
\end{array}\right)=\left(\begin{array}{l}
\left(A~\textit{\dn}^2 \textit{(u,m)}+D\right)~e^{i\alpha_{1}} \\
B~\sqrt{m}~\textit{\sn (u,m)}~\textit{\dn (u,m)}~e^{i\alpha_{2}} \\
\end{array}\right)\,,
\eea
where the constraint conditions read as
\bea
&&\sigma_{1}A^{2}=\sigma_{2}B^{2},\,p=\frac{D}{A}\,,\nonumber\\
&&p=-\frac{1}{3}\left((2-m)\pm\sqrt{1-m+m^{2}}\right)\,,
A^{2}=\frac{1}{(2p+1)\sigma_{1}}\left[3(2\kappa v^{2}+1)\beta^{2}\right],\,
\beta^{2}=\left[\frac{2k_{1}(1+\kappa k_{1})+\omega_{2}^{2}\delta^{2}}{2(2\kappa v^{2}+1)b_{2}}\right],\nonumber\\
&&\omega_{2}^{2}=\frac{2}{(a_{2}\delta^{2}-b_{2})}
\left[b_{2}k_{2}(1+\kappa k_{2})-a_{2}k_{1}(1+\kappa k_{1})\right],\,
a_{2}=\left(\frac{6(1+p)^{2}}{2p+1}-(1+4m)\right)\,,\nonumber\\
&&b_{2}=2\left(\frac{3(1+p)^{2}}{2p+1}
-\left((1+m)+\sqrt{1-m+m^{2}}\right)\right).\,
\eea\ees

The $\pm$ sign appearing in the expression for $p$ results in two different
values for $2p+1$. Particularly, for all values of $m$, the $+ (-)$ sign yields
negative (positive) values for $2p+1$. For $2p+1>0$, the solution is
allowed only
for focusing nonlinearity while for $2p+1<0$, the solution will exist if the
nonlinearity is of defocusing type. However to get regular solutions the
parameters have to be chosen appropriately. Here too the solution 12 in the
limit $m=1$ with $p=0$ reproduces the solution 8(b) while in the limit $m=1$ with $p=-\frac{2}{3}$ it reduces
to solution 11(b).

\subsection*{\bf Solution 13}

\subsubsection*{\bf (a) \boldmath{$\dn^{2}-\cn~\dn$} waves}

We have obtained the following elliptic solution for the CNLH system (\ref{1}):

\bes \label{24}\bea
\left(\begin{array}{l}
q_{1} \\
q_{2} \\
\end{array}\right)=\left(\begin{array}{l}
\left(A~\textit{\dn}^2 \textit{(u,m)}+D\right) ~e^{i\alpha_{1}} \\
B~\sqrt{m}~\textit{\cn (u,m)}~\textit{\dn (u,m)}~e^{i\alpha_{2}} \\
\end{array}\right)\,,
\eea
along with the constraints
\bea
&&\sigma_{1}A^{2}=-\sigma_{2}B^{2},\,p=\frac{D}{A}\,,\nonumber\\
&&p=-\frac{1}{3}\left((2-m)\pm\sqrt{1-m+m^{2}}\right),\,
(2p+1-m) A^{2}=\frac{1}{\sigma_{1}}[3(2\kappa v^{2}+1)\beta^{2}]\,,
\beta^{2}=\left[\frac{2k_{1}(1+\kappa k_{1})+\omega_{2}^{2}\delta^{2}}{2(2\kappa v^{2}+1)b_{3}}\right]\,,\nonumber\\
&&\omega_{2}^{2}=\frac{2}{(a_{3}\delta^{2}-b_{3})}
\left[b_{3}k_{2}(1+\kappa k_{2})-a_{3} k_{1}(1+\kappa k_{1})\right],\,
(2p+1-m) a_{3}=((1+m)-6((1+p)^{2}-m))\,,\nonumber\\
&&(2p+1-m) b_{3}=2\left(3((1+p)^{2}-m)-(2p+1-m)\left((1+m)+\sqrt{1-m+m^{2}}\right)\right)\,.
\eea
The condition $\sigma_{1} A^{2}=-\sigma_{2} B^{2}$ restricts the solution to be allowed only for the mixed type nonlinearities. For the condition $p=-\frac{1}{3}\left((2-m)+\sqrt{1-m+m^{2}}\right)$, we should take the value of $\sigma_{1}$ as negative to make $A^{2}$ as positive definite. But the same solution with $p=-\frac{1}{3}\left((2-m)-\sqrt{1-m+m^{2}}\right)$, requires $\sigma_{1}$ to be positive for getting positive definite value for $A^{2}$.

\subsubsection*{\bf  (b) Red-Blue solitary waves}

Here, in the limit $m=1$ the solution 13(a) with $p=0$ does not exist.
However, this solution 13(a)  with $p=-\frac{2}{3}$,  yields the
following hyperbolic solution comprising is W-shape bright solitary wave and a
standard solitary wave.
\bea\label{24a}
\left(\begin{array}{l}
q_{1} \\
q_{2} \\
\end{array}\right)=\left(\begin{array}{l}
\big(A\sech^2(u)-\frac{2}{3}\big) ~e^{i\alpha_{1}} \\
B \sech^2(u)~e^{i\alpha_{2}} \\
\end{array}\right)\,.
\eea
Here the solution parameters satisfy the relations
\bea\label{24b}
&&\sigma_1 A^2 = - \sigma_2 B^2\,,~~\sigma_1 < 0,~~\sigma_2 > 0\,,~~
A^{2}=-\frac{9}{\sigma_{1}}\left[(2\kappa v^{2}+1)\beta^{2}\right]\,,
p=-\frac{2}{3}\,, \nonumber \\
&&\beta^{2}=-\left[\frac{2k_{1}(1+\kappa k_{1})+\omega_{2}^{2}\delta^{2}}
{2(2\kappa v^{2}+1)}\right]\,,~~
\omega_{2}^{2}=-\frac{2}{(\delta^{2}+1)}
\left[k_{2}(1+\kappa k_{2})+k_{1}(1+\kappa k_{1})\right]\,.
\eea\ees
Here the first component $q_{1}$ and the second component $q_{2}$ are comprised of red and blue solitary waves respectively \cite{hioe1}.

\subsection*{\bf Solution 14}

\subsubsection*{\bf (a) \boldmath{$\dn^{2}-\dn^{2}$} waves}

A distinct second order elliptic solution for the CNLH system (\ref{1}) is
\bes \label{25}\bea
\left(\begin{array}{l}
q_{1} \\
q_{2} \\
\end{array}\right)=\left(\begin{array}{l}
A~(\textit{\dn}^2 \textit{(u,m)}+D)~e^{i\alpha_{1}} \\
B~(\textit{\dn}^2 \textit{(u,m)}+E)~e^{i\alpha_{2}} \\
\end{array}\right)\,,
\eea
provided
\bea
&&\sigma_{1}A^{2}=-\sigma_{2}B^{2}\,,\nonumber\\
&&p\neq q=-\frac{1}{3}\left((2-m)\pm\sqrt{1-m+m^{2}}\right)\,,
A^{2}=\frac{3}{2(p-q)\sigma_{1}}\left[(2\kappa v^{2}+1)\beta^{2}\right]\,,p=\frac{D}{A},\,q=\frac{E}{B},\,\nonumber\\
&&\beta^{2}=-\left[\frac{2k_{1}(1+\kappa k_{1})+\omega_{2}^{2}\delta^{2}}{2\sqrt{1-m+m^{2}}(2\kappa v^{2}+1)}\right],\,
\omega_{2}^{2}=-\frac{2}{(\delta^{2}+1)}
\left[k_{2}(1+\kappa k_{2})+k_{1}(1+\kappa k_{1})\right]\,.
\eea\ees
As in the previous solution, here also the condition $\sigma_{1} A^{2}=-\sigma_{2} B^{2}$ confines the system to admit the solution only for mixed type nonlinearities.

\subsubsection*{\bf  (b) Blue-Red solitary waves}

In the hyperbolic limit the solution 14(a) with $p=-\frac{2}{3}$ and $q=0$ reduces to the hyperbolic solution 13(b). The same solution 14(a) with $p=0$ and $q=-\frac{2}{3}$ reduces to the following hyperbolic solitary wave solution in the limit $m=1$.
\bes\label{25a}\bea
\left(\begin{array}{l}
q_{1} \\
q_{2} \\
\end{array}\right)=\left(\begin{array}{l}
A \sech^2(u)~e^{i\alpha_{1}} \\
B \left(\sech^2(u)-\frac{2}{3}\right)~e^{i\alpha_{2}} \\
\end{array}\right)\,,
\eea
where the solution parameters satisfy the relation
\bea\label{25b}
&&A^{2}=\frac{9}{\sigma_{1}}\left[(2\kappa v^{2}+1)\beta^{2}\right]\,.
\eea\ees
Here all other constraints parameters are same as that of solution 13(b). Then the first component $q_{1}$ and second component $q_{2}$ are comprised of blue and red solitary waves respectively \cite{hioe1}.

\subsection*{\bf Solution 15:~ Superposed second order elliptic waves}

Finally, we have also constructed a novel second order superposed elliptic solution that can be expressed as a combination of  $\dn^2$ and $\cn \dn$  elliptic functions. This special superposed elliptic solution is given below.
\bes\bea\label{26}
\left(\begin{array}{l}
q_{1} \\
q_{2} \\
\end{array}\right)=\left(\begin{array}{l}
\bigg (\frac{A}{2} \dn^{2}~(u,m) +D
+ \frac{F}{2} \sqrt{m}~\cn~(u,m)~\dn~(u,m) \bigg )~e^{i\alpha_{1}} \\
\bigg(\frac{B}{2} \dn^2~(u,m)+E
+ \frac{G}{2} \sqrt{m}~\cn~(u,m)
\dn~(u,m) \bigg )~e^{i\alpha_{2}} \\
\end{array}\right)\,.
\eea
Here the constraint conditions read as
\bea
&&\sigma_{1}A^{2}=-\sigma_{2}B^{2},\,
A^{2}=\frac{9}{2\sigma_{1}(p-q)}\left((2\kappa v^{2}+1)\beta^{2}\right),\nonumber \\
&&\beta^{2}=-\frac{2\left[2k_{1}(1+\kappa k_{1})+\omega_{2}^{2}\delta^{2}\right]}{2\sqrt{1+14m+m^{2}}(2\kappa v^{2}+1)}
,\,p_{1}=\frac{D}{A}\,,q_{1}=\frac{E}{B},\,p_{1}\neq q_{1}=-\frac{1}{12}\left((5-m)\pm \sqrt{1+14m+m^{2}}\right),\nonumber \\
&&\omega_{2}^{2}=-\frac{2}{(\delta^{2}+1)}
\left[k_{2}(1+\kappa k_{2})+k_{1}(1+\kappa k_{1})\right]
\,.\nonumber \\
\eea\ees
This solution exists exclusively for the mixed nonlinearity as follows from the constraint relation $\sigma_{1}A^{2}=-\sigma_{2}B^{2}$ and also from the expression for $A^{2}$. In solution 15, the signs of $F = \pm A$ and $G = \pm B$ are correlated. The hyperbolic solution is only possible for the case $F=A$ and $G=B$. Note that $p_{1}$ and $q_{1}$ admit two values corresponding to the $\pm$ sign before the term $\sqrt{1+14m+m^{2}}$ in their corresponding expression. Here, in the limit $m=1$ the solution 15 with $p_{1}=-\frac{2}{3}$ and $q_{1}=0$ reduces to the solution 11(b). But this same solution 15, in the limit $m=1$ with $p_{1}=0$ and $q_{1}=-\frac{2}{3}$ reduces to the solution 14(b).

In the following table, we tabulate the solutions and the types of nonlinearities admitting that solution.

\begin{table}[H]
\centering
\caption {Types of nonlinearities and their corresponding second order solutions}
\begin{tabular}{|c|c|}
  \hline
 Solutions & Types of nonlinearity supporting the solutions\\\hline
 \multirow{1}{2cm}{8, 9} & Focusing nonlinearity ($\sigma_{1}>0$ and $\sigma_{2}>0$)\\\hline
 \multirow{2}{2.5cm}{10, 13, 14, 15} & Mixed nonlinearity [($\sigma_{1}>0$ and $\sigma_{2}<0$) \\
                      &or ($\sigma_{1}<0$ and $\sigma_{2}>0$)] \\\hline
 \multirow{2}{2cm}{11, 12} & Defocusing nonlinearity ($\sigma_{1}<0$ and $\sigma_{2}<0$) when $p=-\frac{1}{3}\left((2-m)+\sqrt{1-m+m^{2}}\right)$ \\
                      &Focusing nonlinearity ($\sigma_{1} > 0$ and
$\sigma_{2} > 0$) when $p=-\frac{1}{3}\left((2-m)-\sqrt{1-m+m^{2}}\right)$  \\\hline
\end{tabular}
\end{table}
For illustrative purpose, we present the variation of speed (modulus of velocity), pulse width and amplitude of the first component of solutions 8(a), 10 and 13(a) respectively, with respect to $\kappa$ in Figs.~5(a-c). In Fig.~5(a) the speed and the pulse width increases slightly as $\kappa$ increases. Then the amplitude A of the first component behaves opposite to the speed and pulse width i.e., the amplitude A of the first component decreases slightly as $\kappa$ increases. In Figs.~5(b-c), also we notice the speed and pulse width increases as $\kappa$ increases. But the amplitude A decreases as $\kappa$ increases.
\renewcommand{\floatpagefraction}{0.7}
\begin{figure}[H]
\centering~\includegraphics[width=0.3\linewidth]{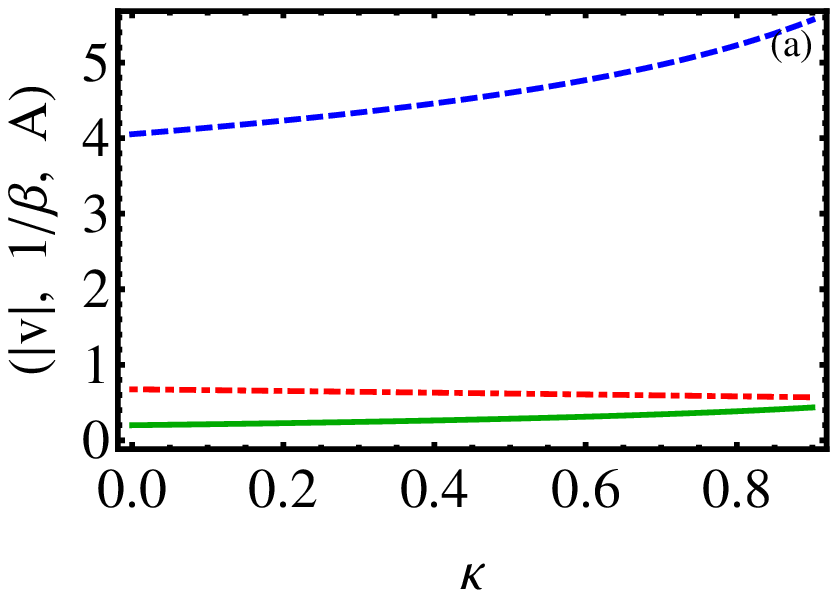}~
\includegraphics[width=0.3\linewidth]{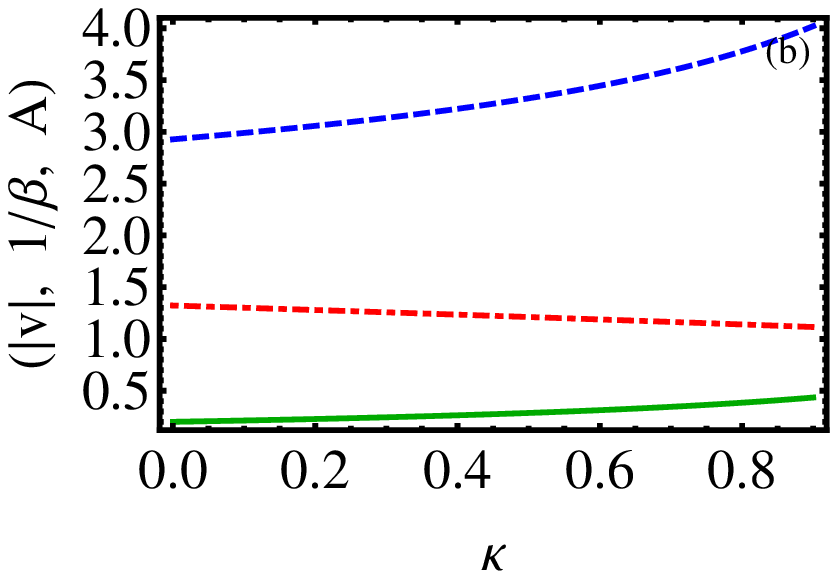}~
\includegraphics[width=0.3\linewidth]{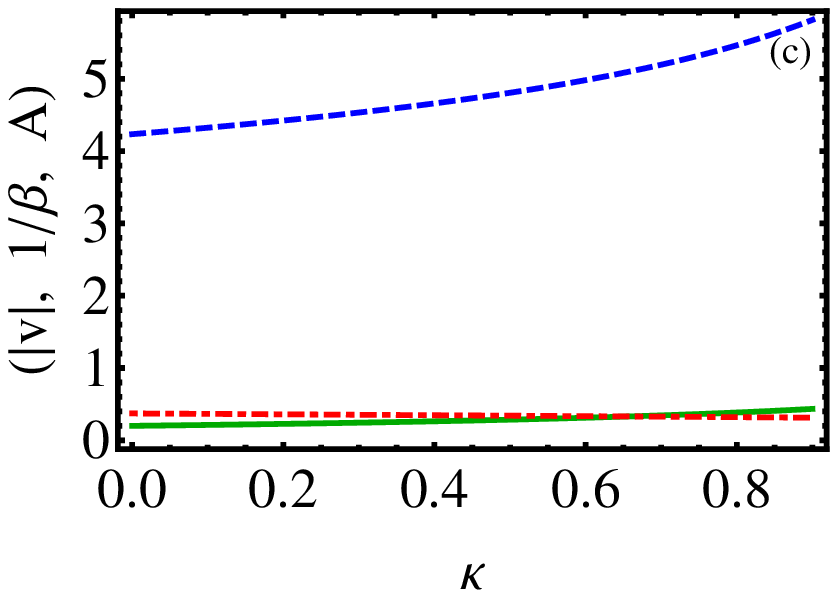}~
\caption{Plots of speed (solid dark green), pulse width (dotted blue) and amplitude of the first component (dotdashed red) versus nonparaxial parameter $\kappa$. (a) solution 8(a) with $\omega_{2}=0.2$, $k_{2}=-0.3$, $m=0.4$, $\sigma_{1}=1$, (b) solution 10 with $\omega_{2}=0.2$, $k_{2}=-0.3$ and (c) solution 13(a) with $\omega_{2}=0.2$, $k_{2}=-0.3$. In all the figures $\sigma_{1}=-1$ (except Fig.~5(a)), $\sigma_{2}=1$, $\delta_{0}=\delta_{1}=\delta_{2}=1$, $\xi=1.5$, $\tau=1.5$, $m=0.8$ (except in Fig.~5(a)).}
\end{figure}

However the intensity profiles of the second order solutions are not same as that of first order solutions. For illustrative  purpose, here we have plotted the intensity profiles of the elliptic solutions 8(a), 10 and 15 of the CNLH system (\ref{1}) with nonparaxial parameter $\kappa=0.0001$  in Fig.~6. The top and middle panels of Fig.~6(a) show the intensity plots of solution 8(a). The intensity profiles of first and second components of the elliptic waves are clearly out-of-phase as illustrated in the bottom panel of Fig.~6(a). Next, in the top and middle panels of Fig.~6(b), we plot the solution 10, where the intensity profile of first component is broader and admits a flat-top elliptic wave profile. To facilitate the understanding of this distinct wave profiles we have given the corresponding two dimensional plot in the bottom panel of Fig.~6(b). It is interesting to notice that in contrast to the first ($q_1$) component the width of the pulse train is compressed in the second ($q_2$) component. Finally, intensity profiles of the superposed solution 15, are shown  in the first and middle panels of Fig.~6(c) while the associated two dimensional plot is presented in the bottom panel. We also note that in the superposed solution the amplitude of the second ($q_2$) component gets enhanced and admit simply periodic pulse trains whereas the first ($q_1$) components admits doubly periodic pulse trains with slightly lesser amplitude. The above type of distinct behaviours of second order elliptic waves for different choice of solution parameters suggest the possibility of pulse shaping, a desirable property in nonlinear optics, in the CNLH system.

\renewcommand{\floatpagefraction}{0.7}
\begin{figure}[H]
\centering\includegraphics[width=0.32\linewidth]{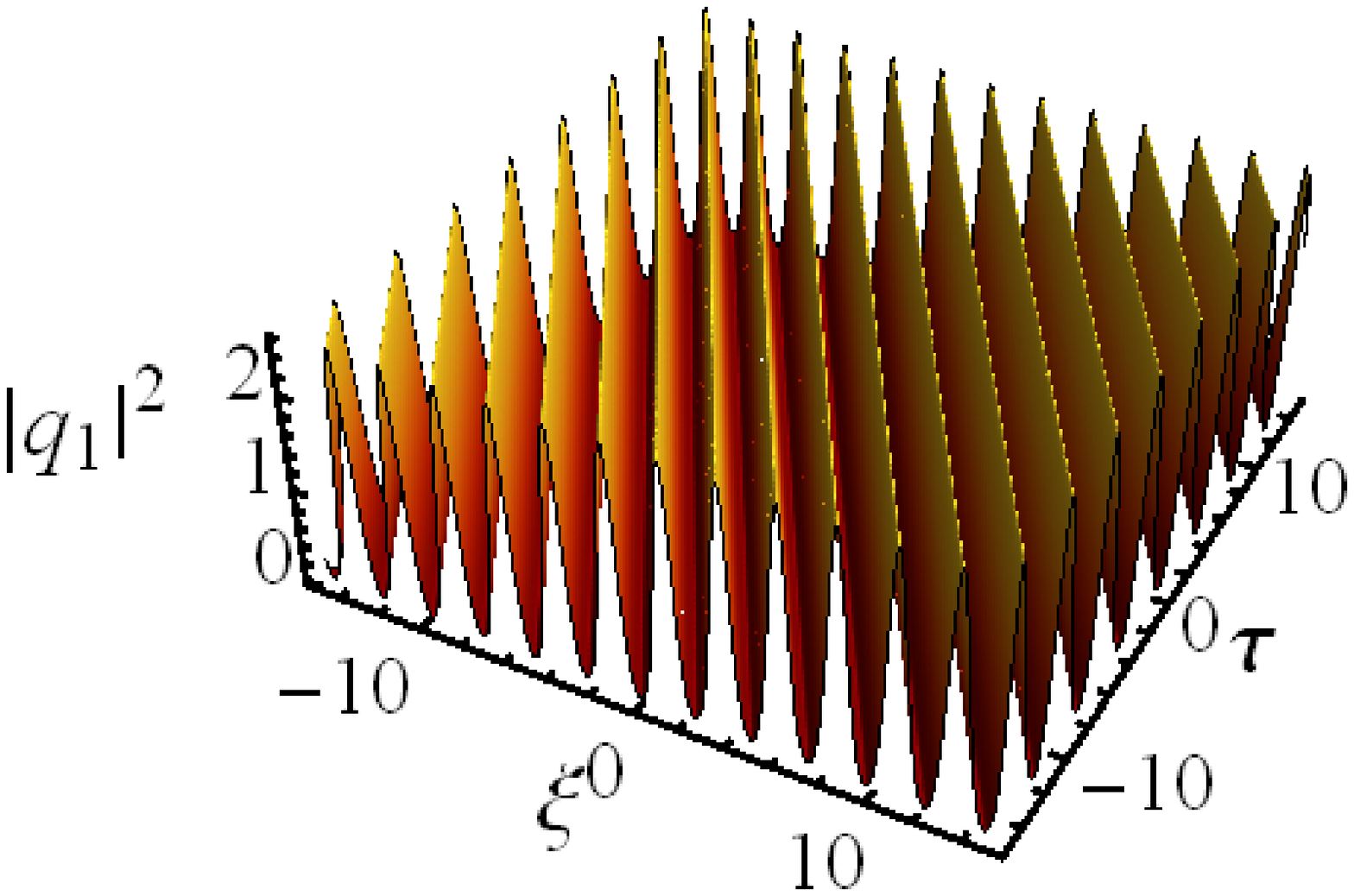}~~~
\includegraphics[width=0.32\linewidth]{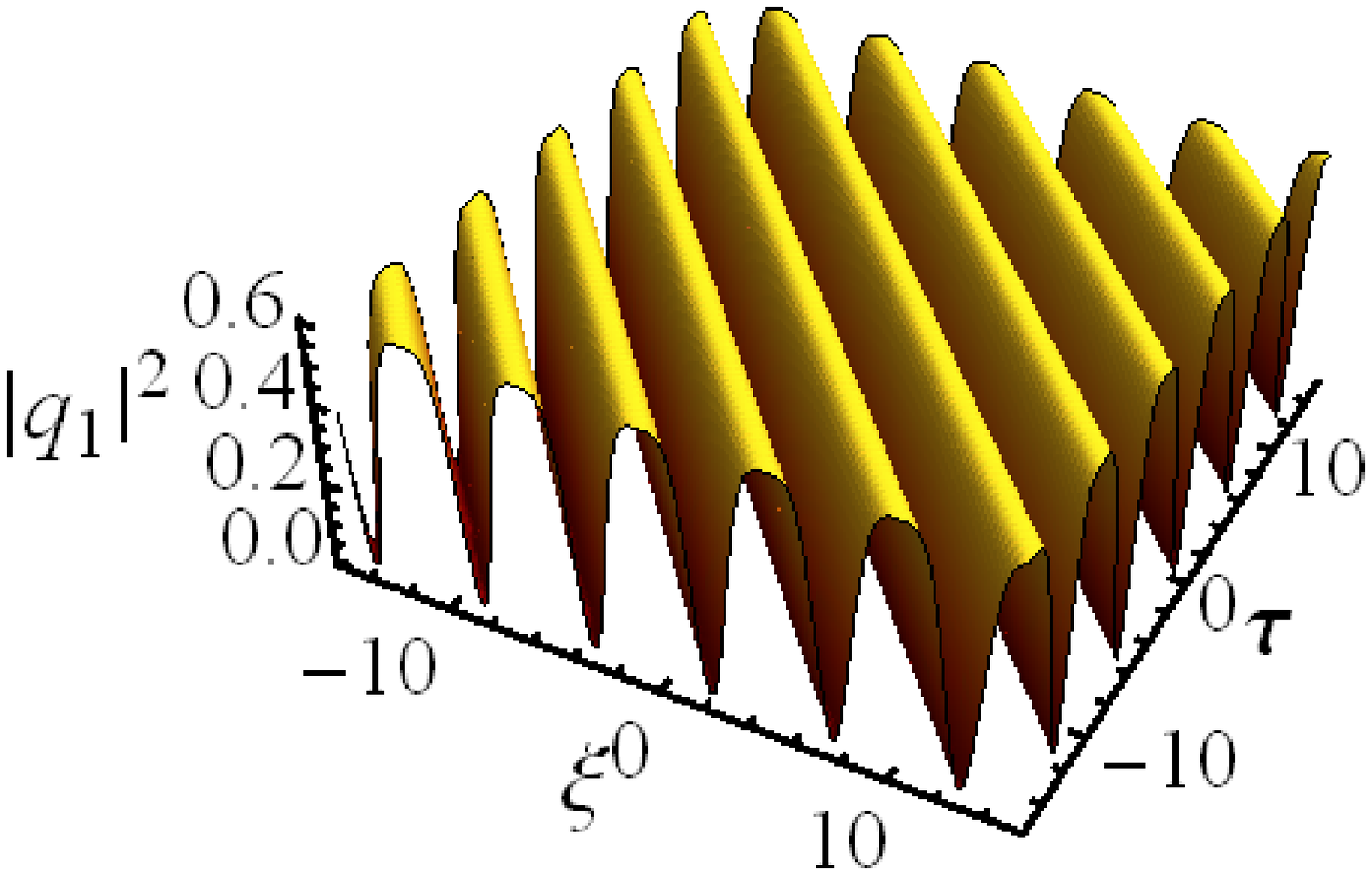}~~~
\includegraphics[width=0.32\linewidth]{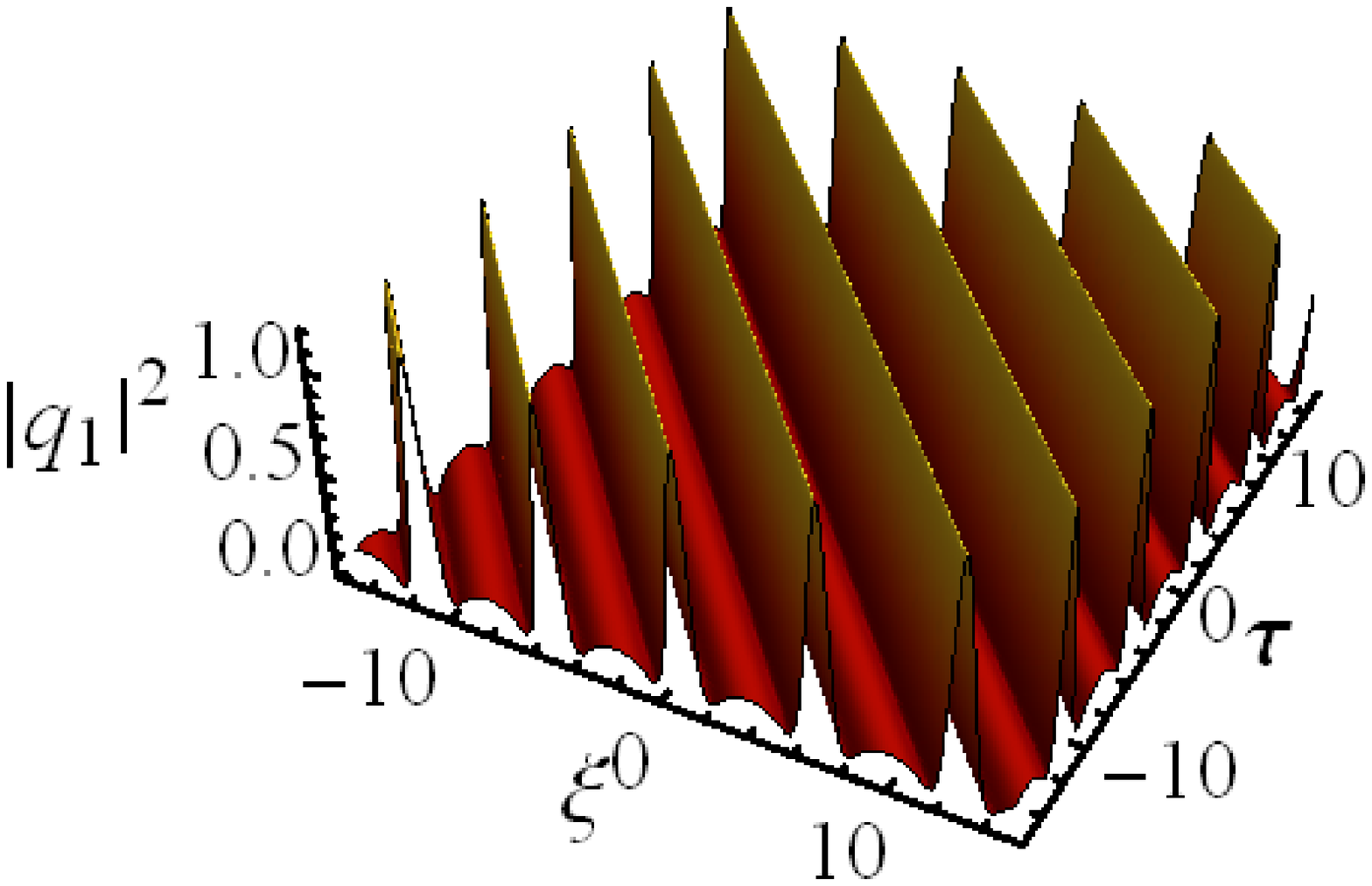}~~~\\
\includegraphics[width=0.32\linewidth]{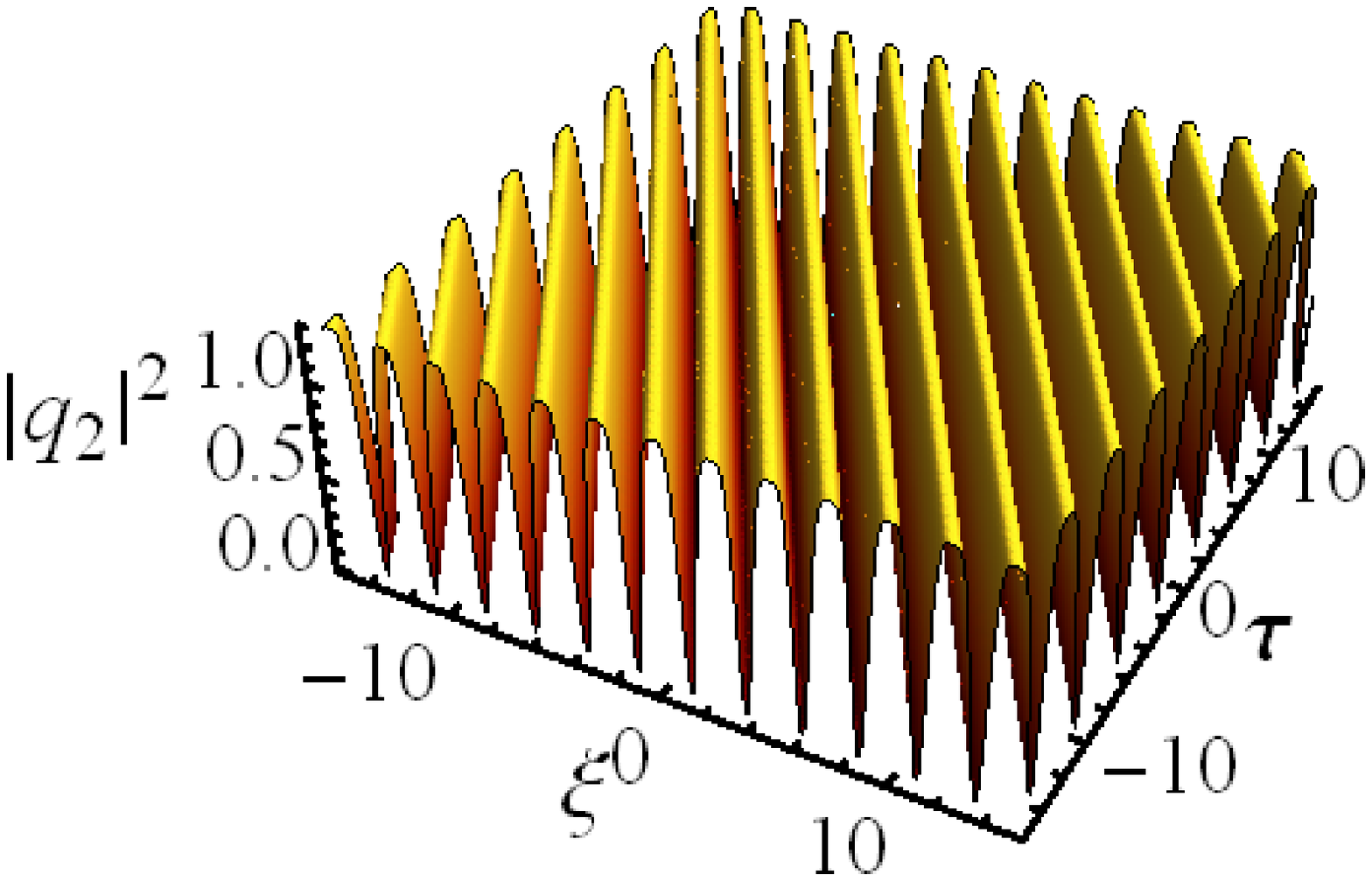}~~
\includegraphics[width=0.32\linewidth]{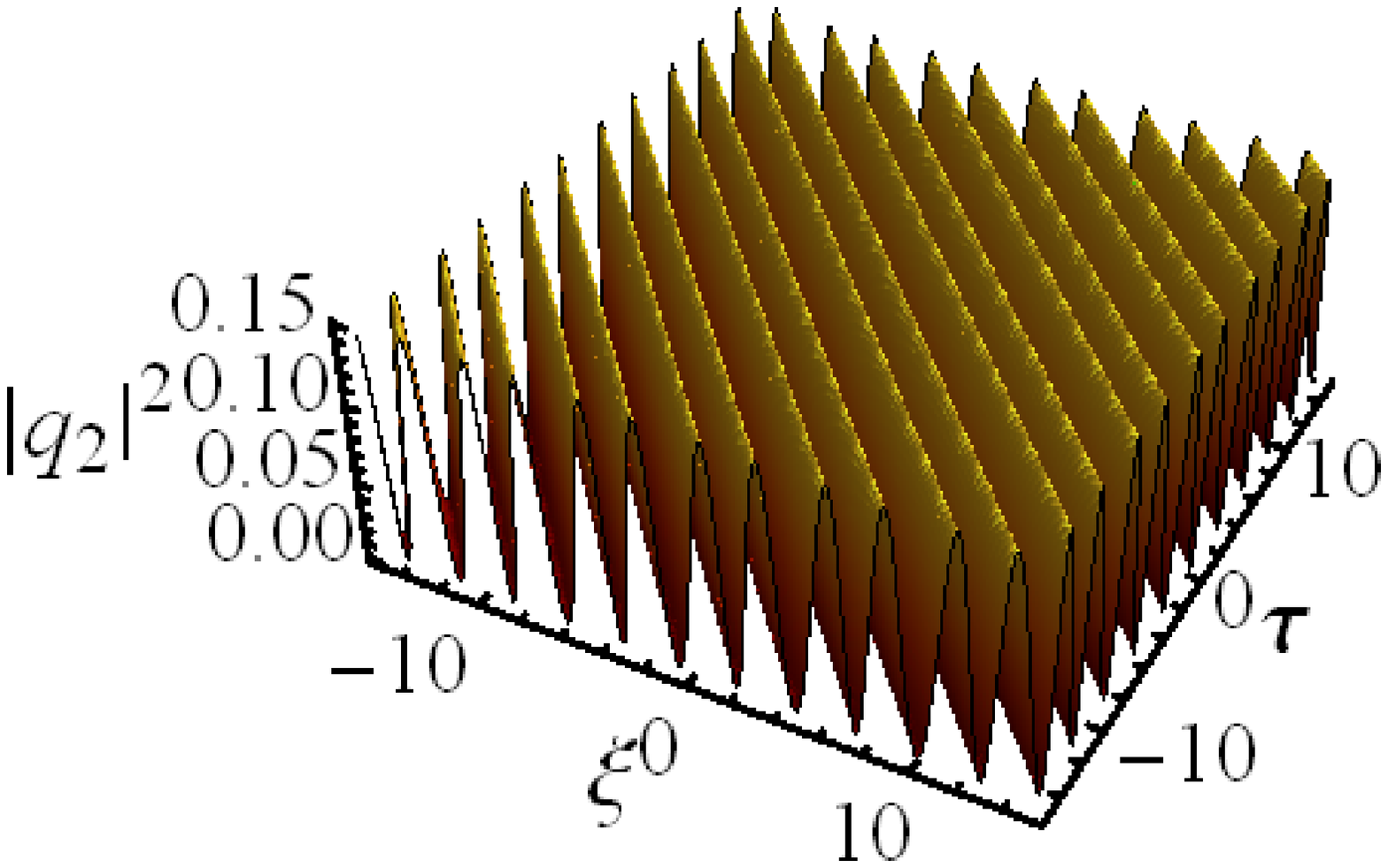}~~~
\includegraphics[width=0.32\linewidth]{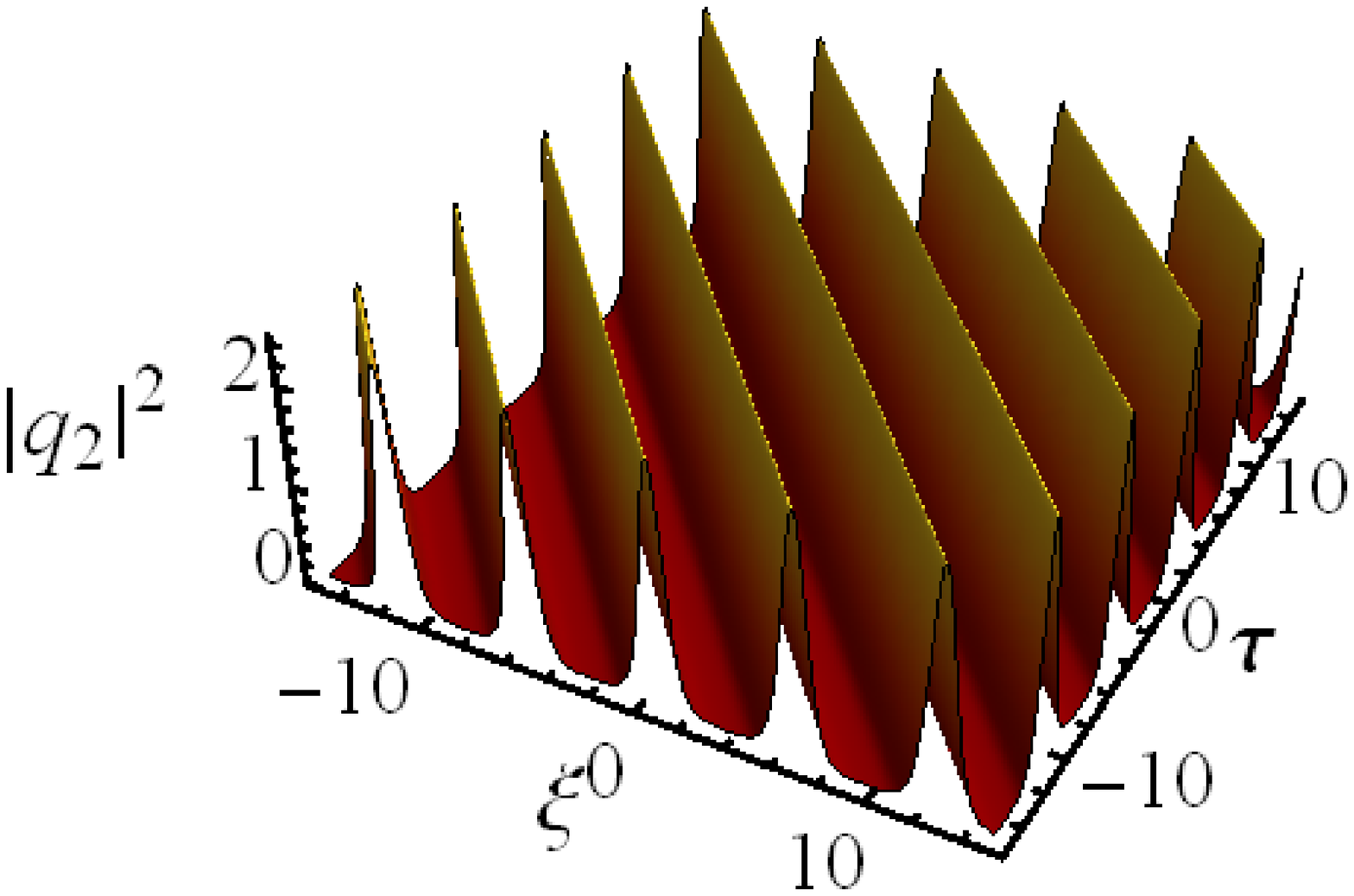}~~~\\
\includegraphics[width=0.32\linewidth]{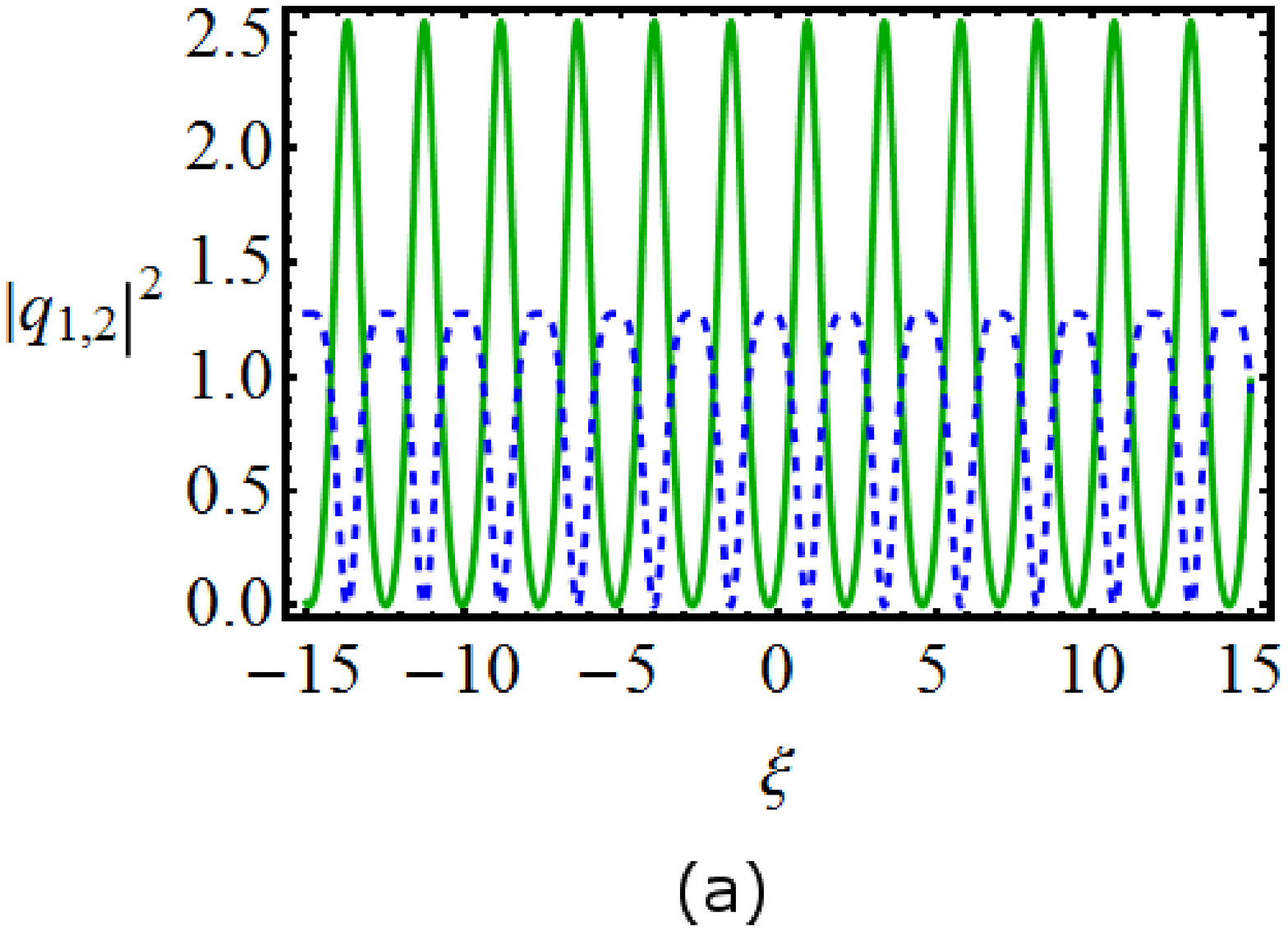}~~~
\includegraphics[width=0.32\linewidth]{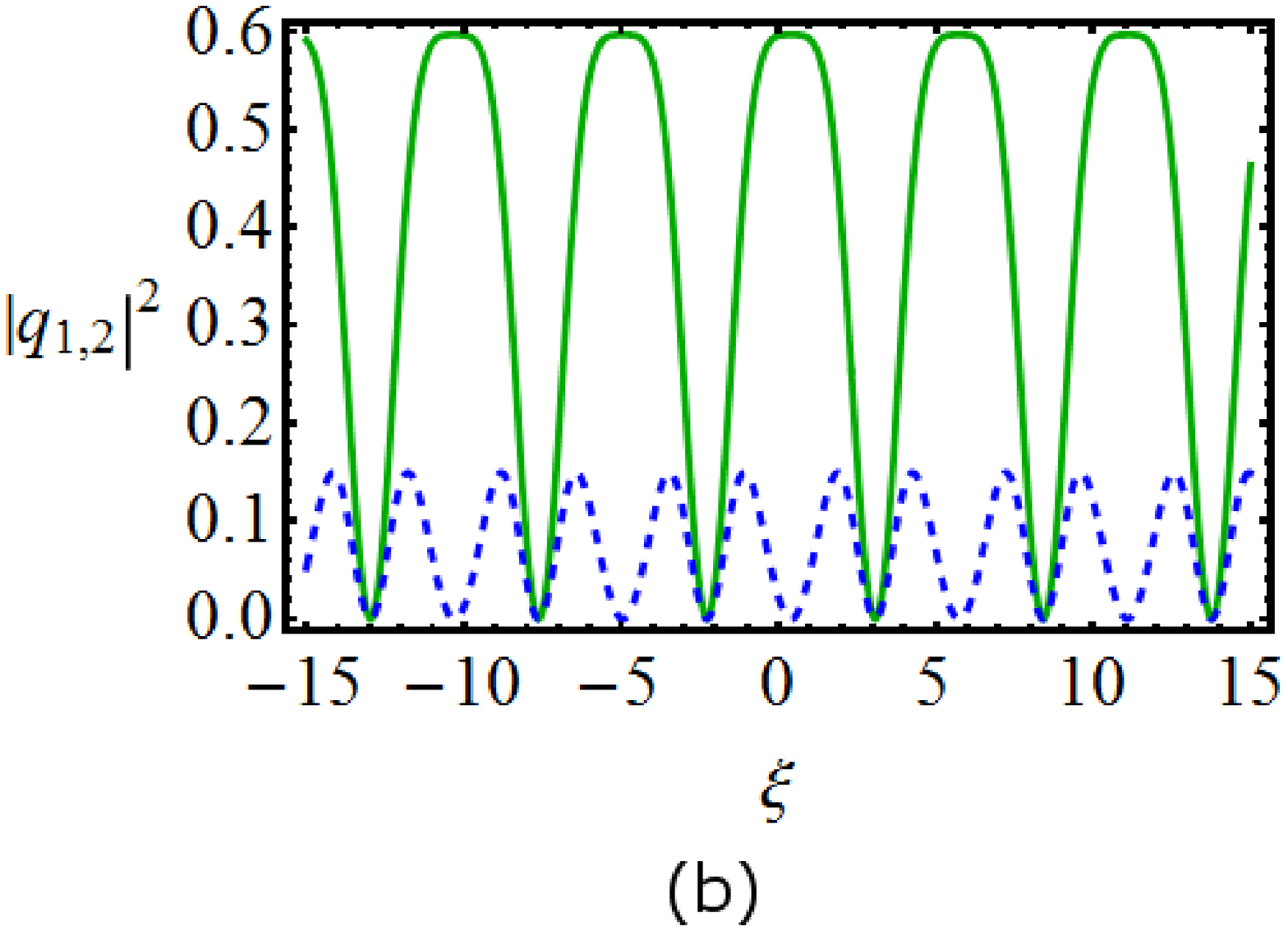}~~~
\includegraphics[width=0.32\linewidth]{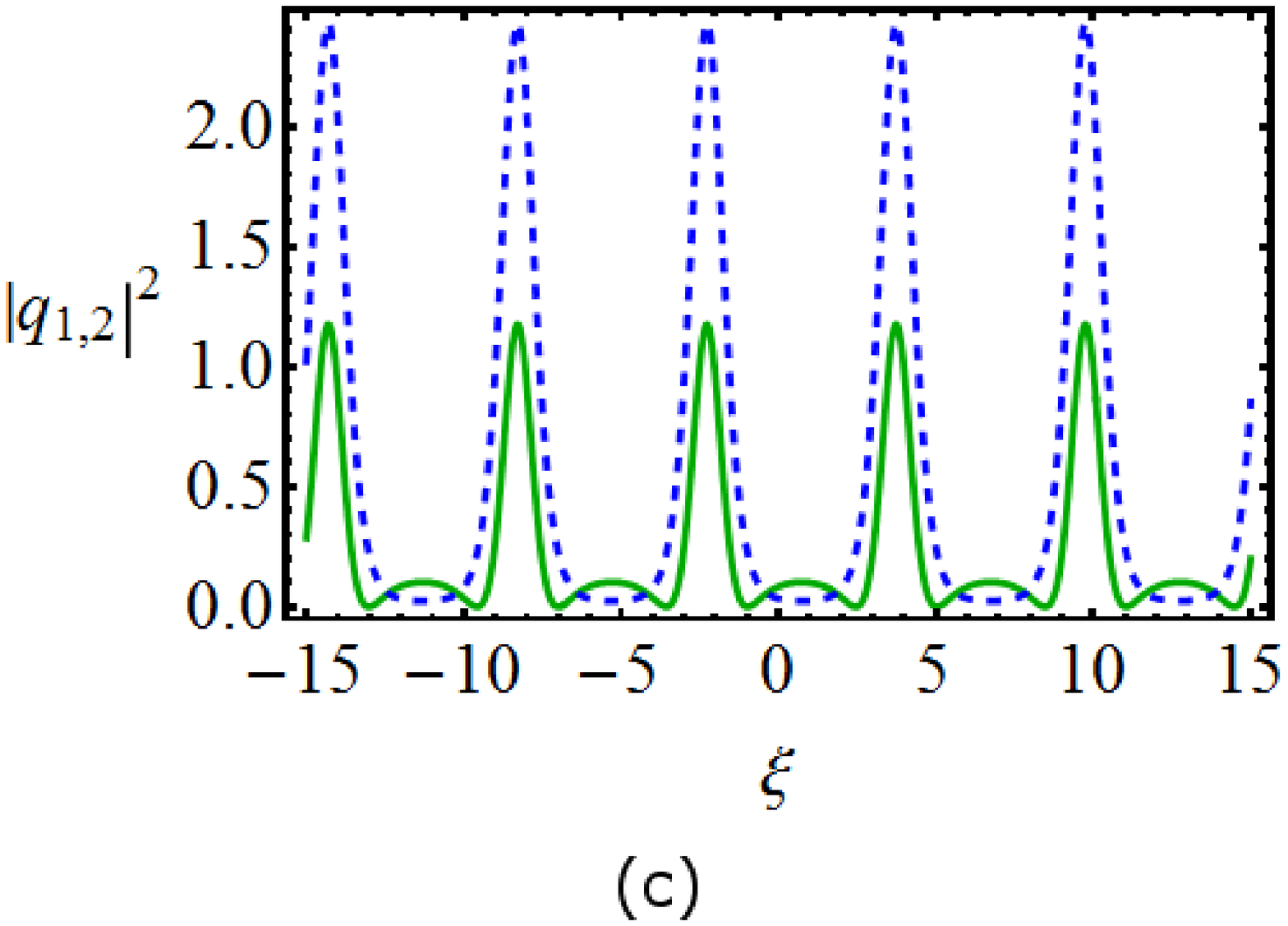}
\caption{The intensity plots of solution 8(a) with $\omega_{2}=1.6$, $k_{2}=-2$, $\sigma_{1}=1$, solution 10 with $\omega_{2}=1.1$, $k_{2}=-1.5$ and the superposed solutions 15 with $\omega_{2}=1.1$, $k_{2}=1.2$ are shown in first two rows of Figs.~6(a-c) respectively. The third row shows the corresponding two dimensional plots of the $q_1$ (dark green) and $q_2$  (dashed blue) components with $\tau=1.5$. In all the figures $\sigma_{1}=-1$ (except in Fig.~6(a)), $\kappa=0.0001$, $\sigma_{2}=1$, $\delta_{0}=\delta_{1}=\delta_{2}=1$, $m=0.5$.}
\end{figure}
The hyperbolic second order solutions display a rich variety of intensity profiles. To give further impetus in this aspect we have plotted all the second order hyperbolic solutions given above. In Fig.~7(a) we show the intensity plots of solution 8(b). Here the first component admits single-hump bright soliton/solitary wave and the second component bears a double-hump bright soliton/solitary wave intensity profile. Another interesting type of soliton/solitary wave is shown in Fig.~7(b), namely double-hump dark (W-shape) as well as bright soliton/solitary wave corresponds to solution 11(b). The last interesting co-existing bright-dark-structure comprising double-hump dark (W-shape) solitary wave in $q_{1}$ component and bright solitary wave in $q_{2}$ component are displayed in Fig.~7(c) (see the hyperbolic solution 13(b)). An important observation from these plots is that even though the parameters $k_{2}$ and $\omega_{2}$ are changed the intensity profiles remain unaltered. This property is a reminiscent of the degenerate property of solitary waves reported for coupled Sasa-Satsuma system in Ref.\cite{xu}.

\renewcommand{\floatpagefraction}{0.7}
\begin{figure}[H]
\centering~
\centering\includegraphics[width=0.32\linewidth]{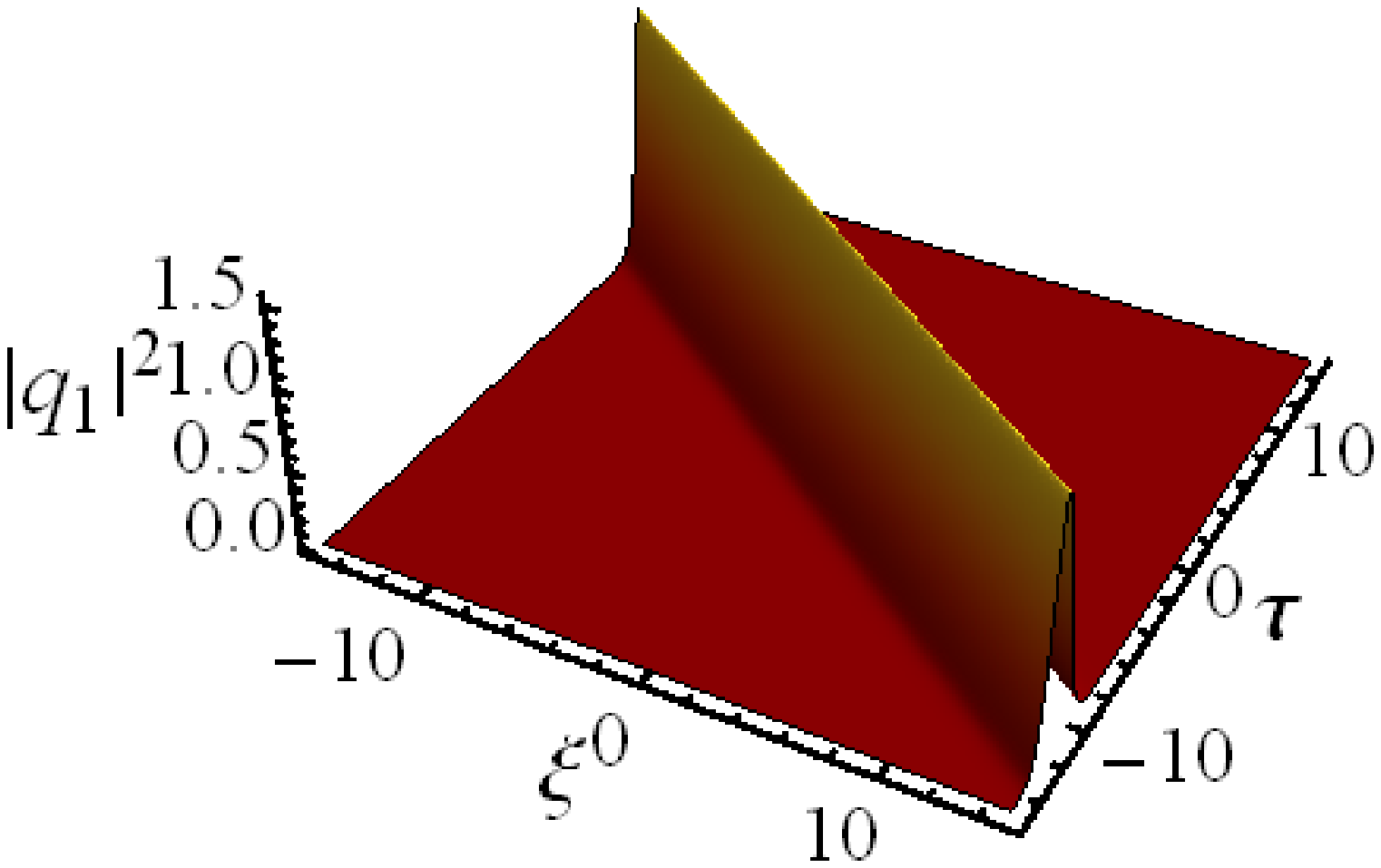}~~~
\includegraphics[width=0.32\linewidth]{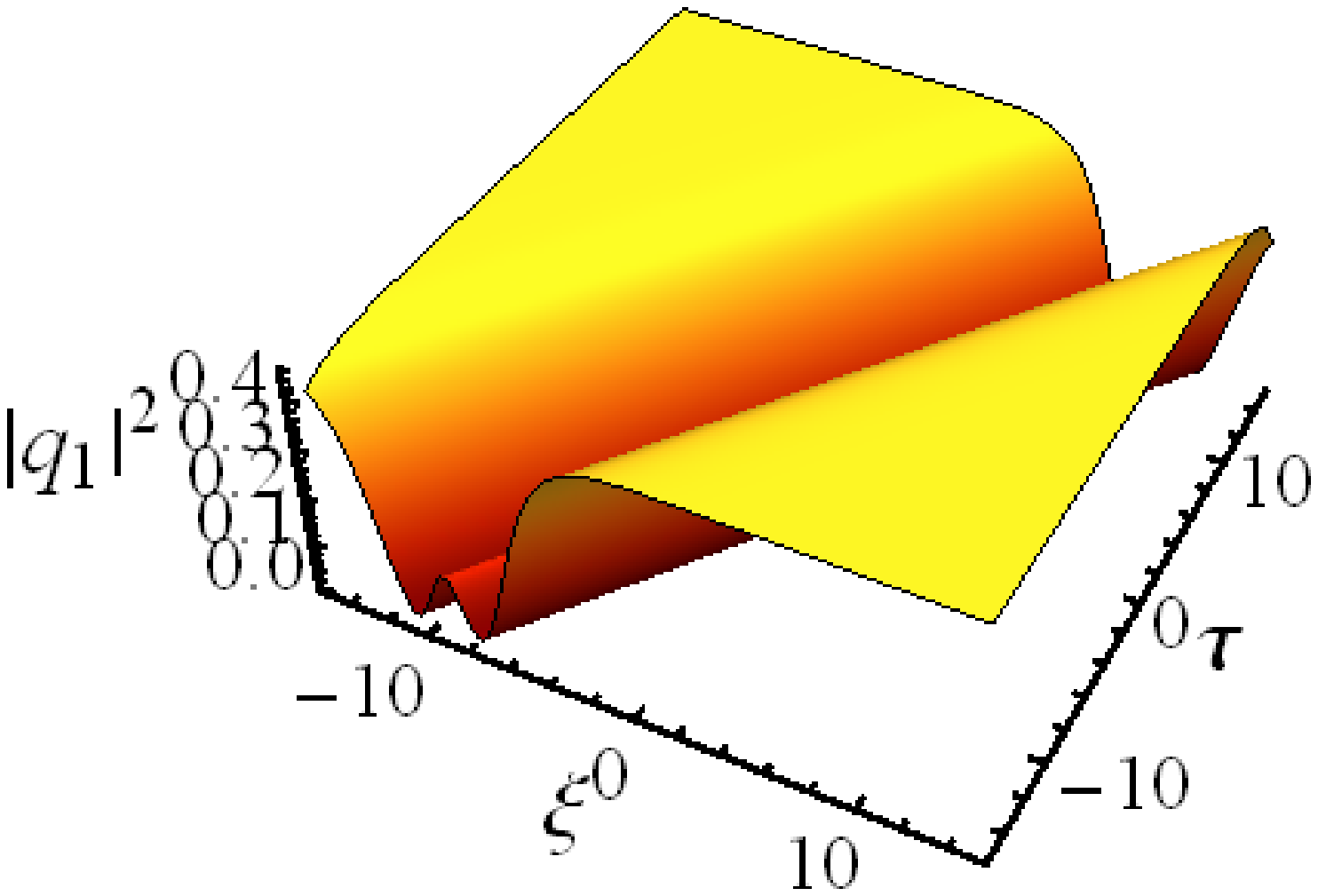}~~~
\includegraphics[width=0.32\linewidth]{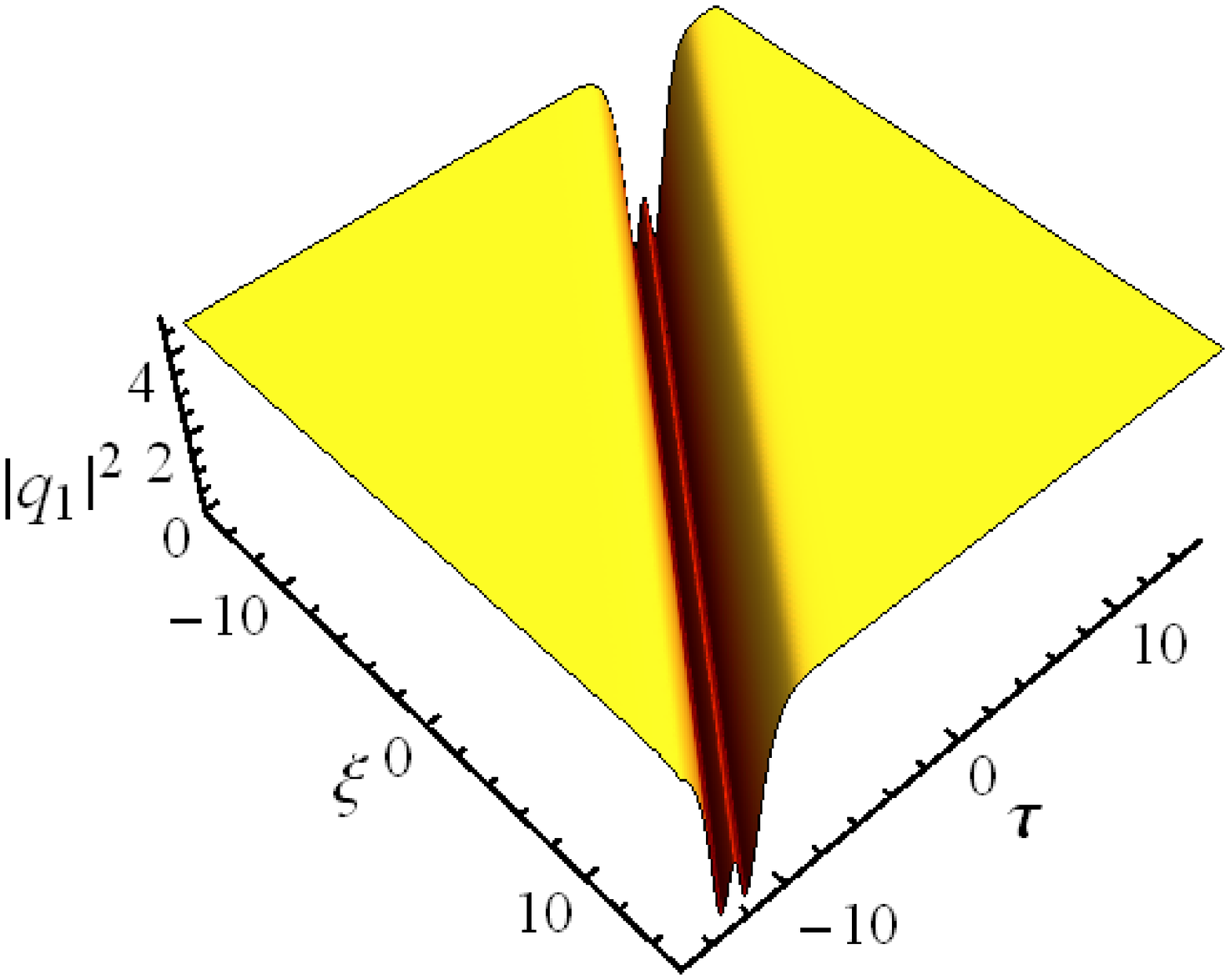}~~~\\
\includegraphics[width=0.32\linewidth]{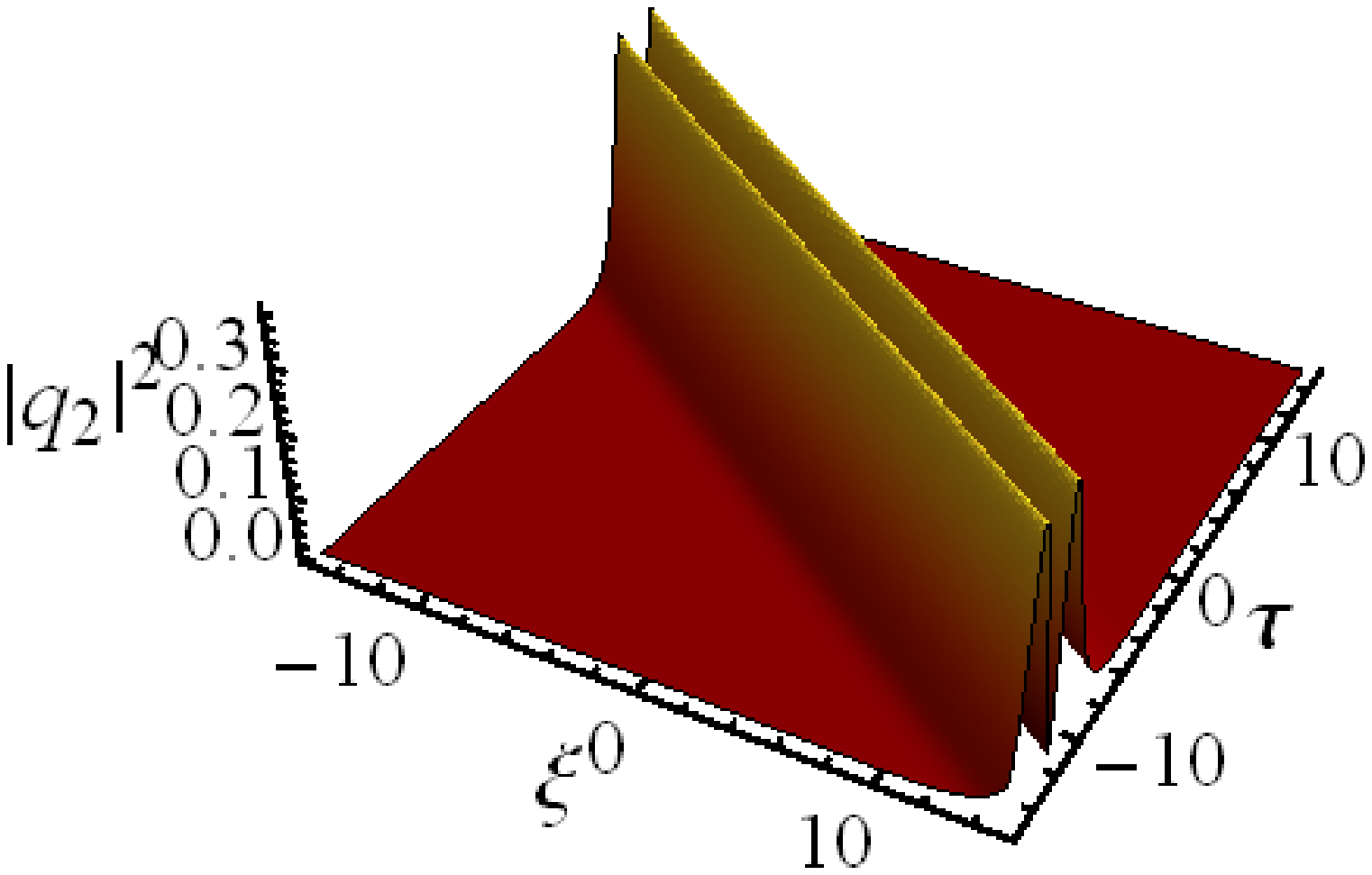}~~
\includegraphics[width=0.32\linewidth]{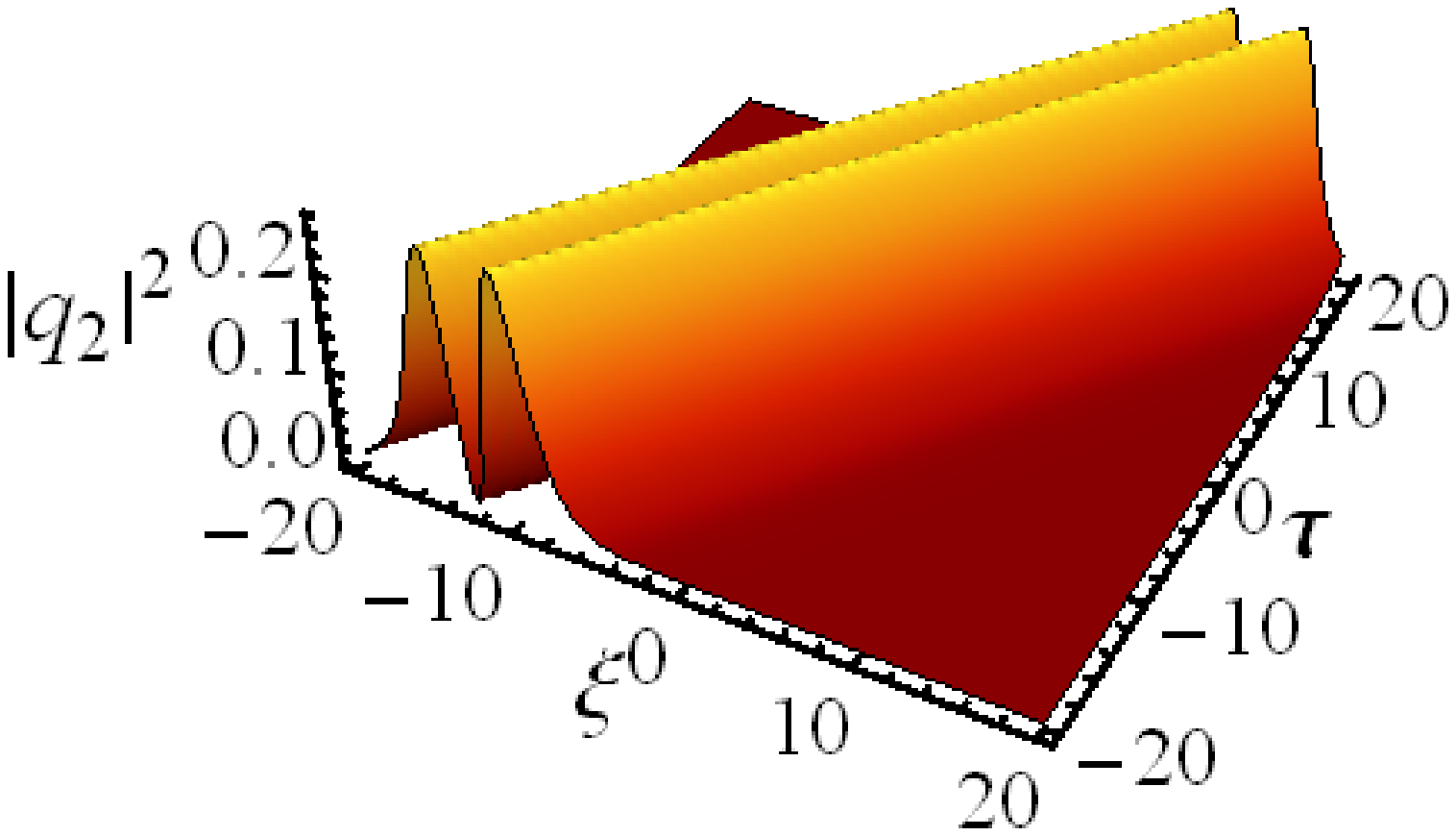}~~~
\includegraphics[width=0.32\linewidth]{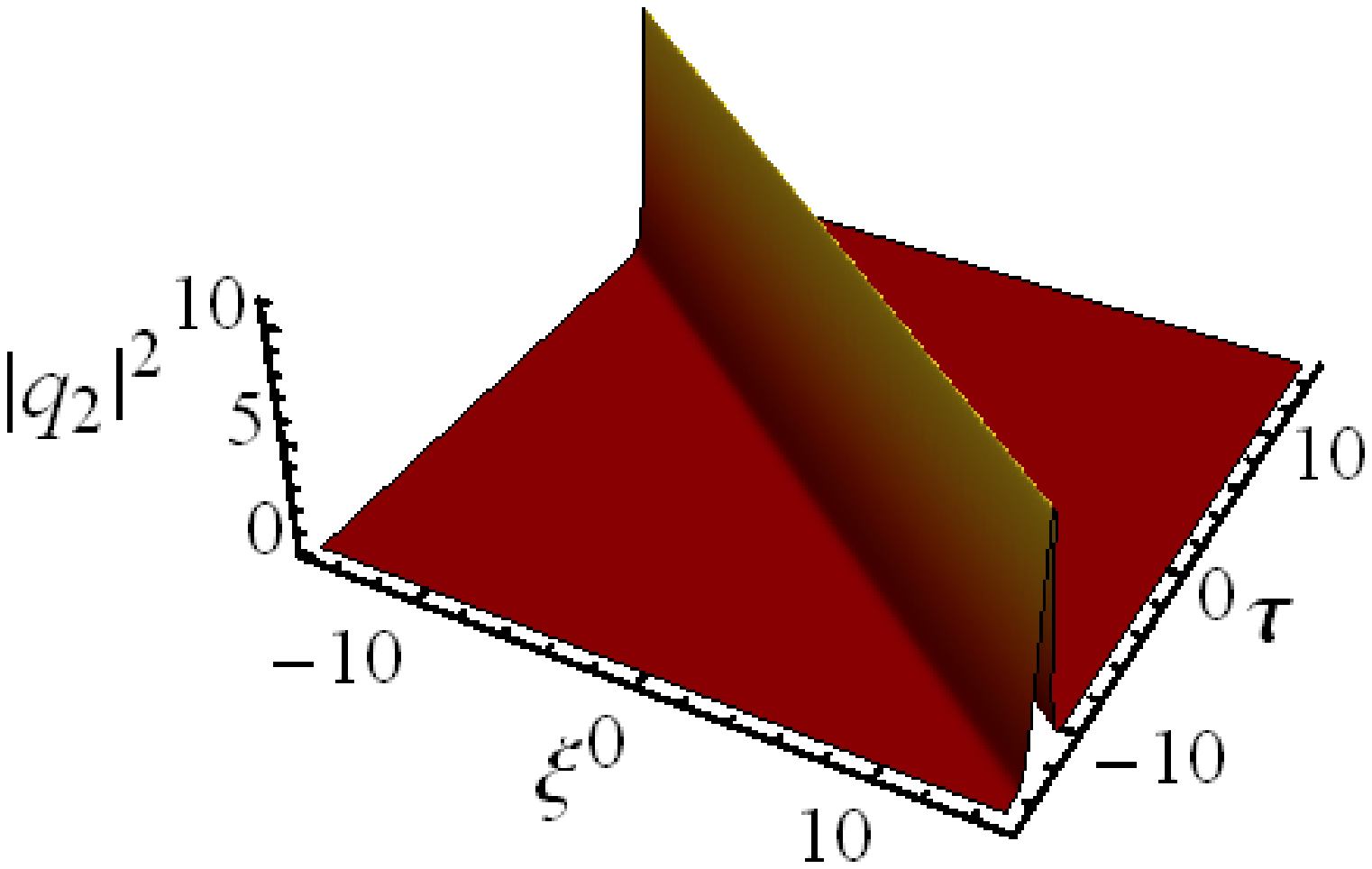}~~~\\
\includegraphics[width=0.32\linewidth]{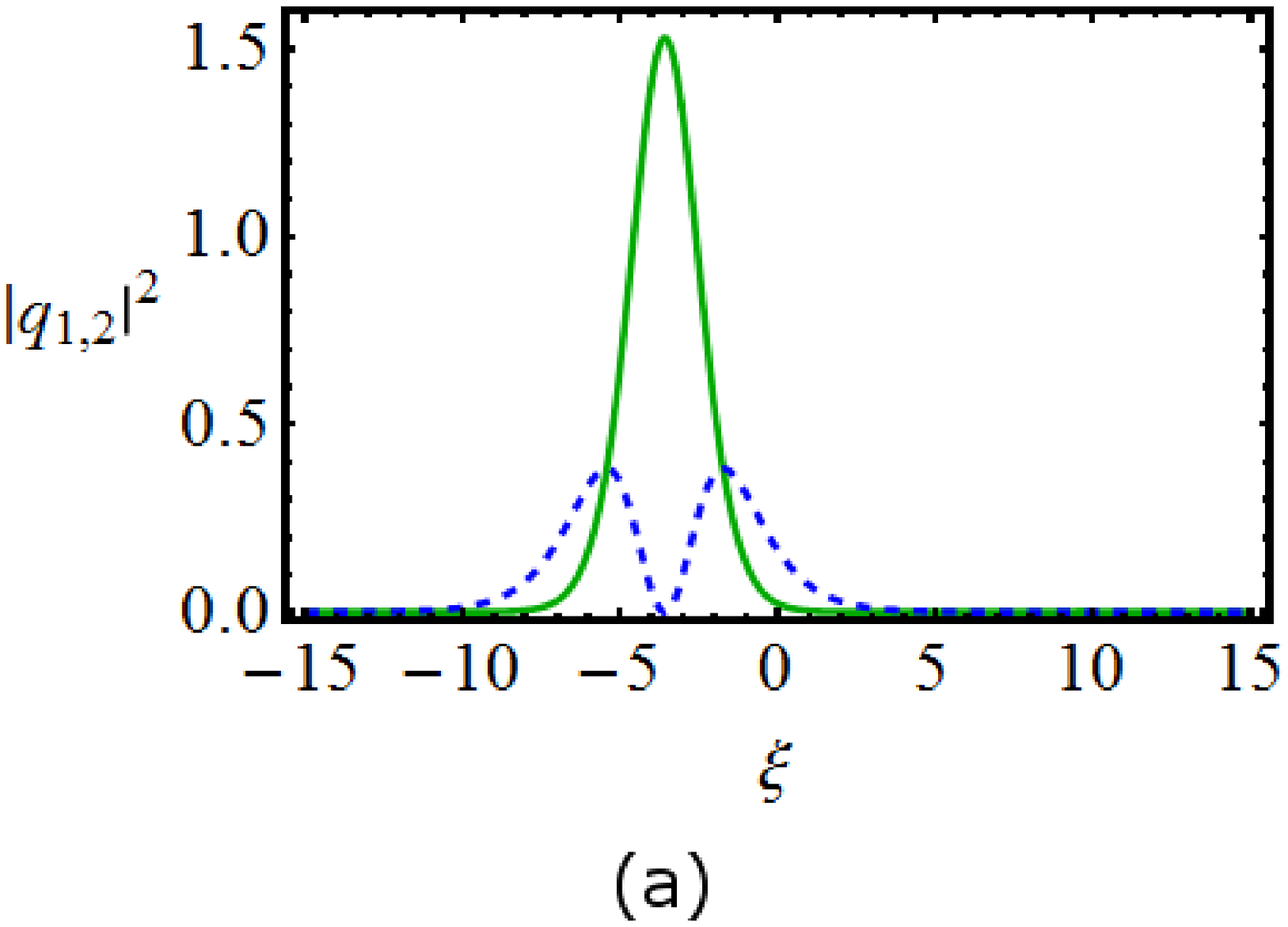}~~~
\includegraphics[width=0.32\linewidth]{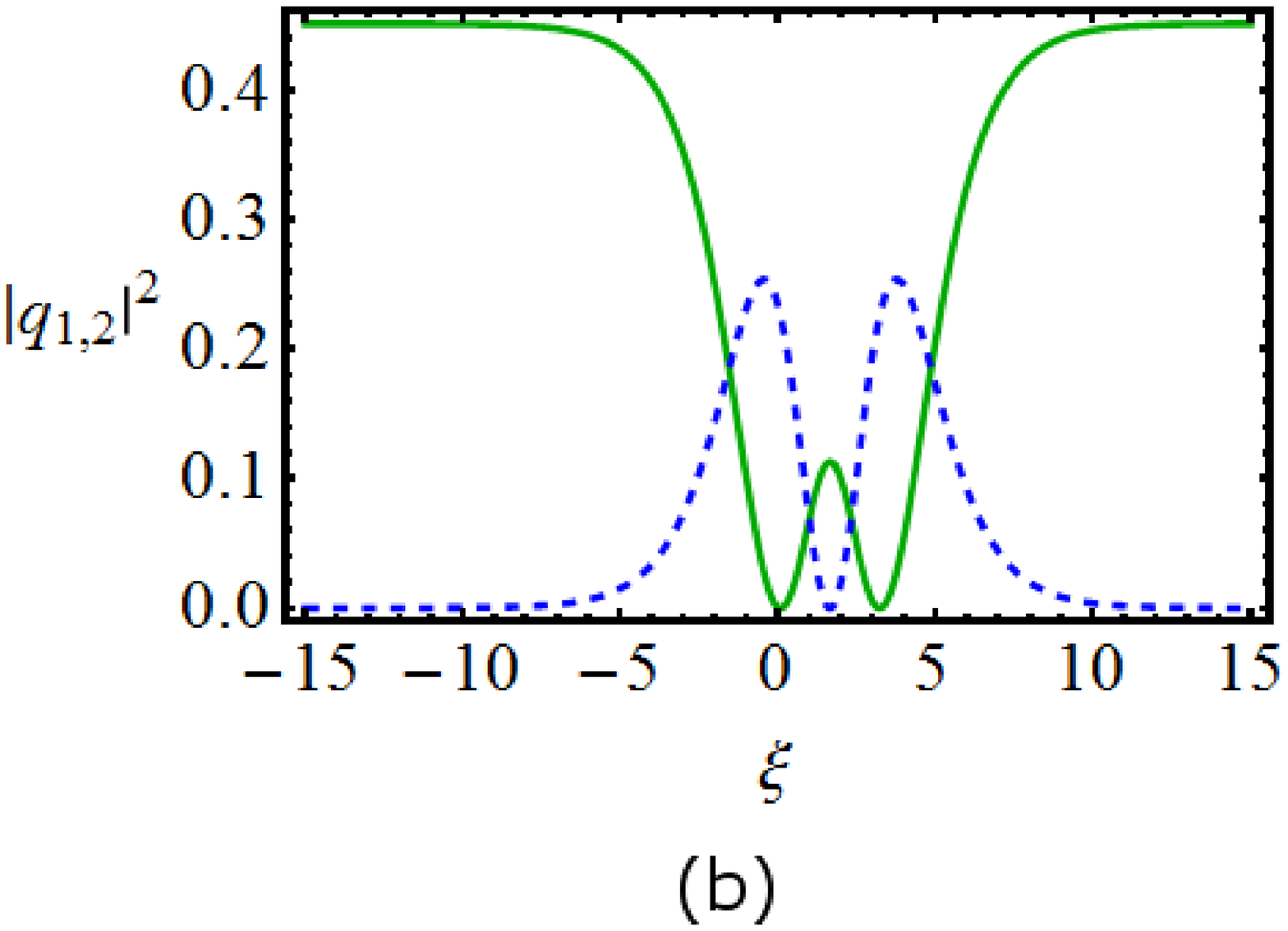}~~~
\includegraphics[width=0.32\linewidth]{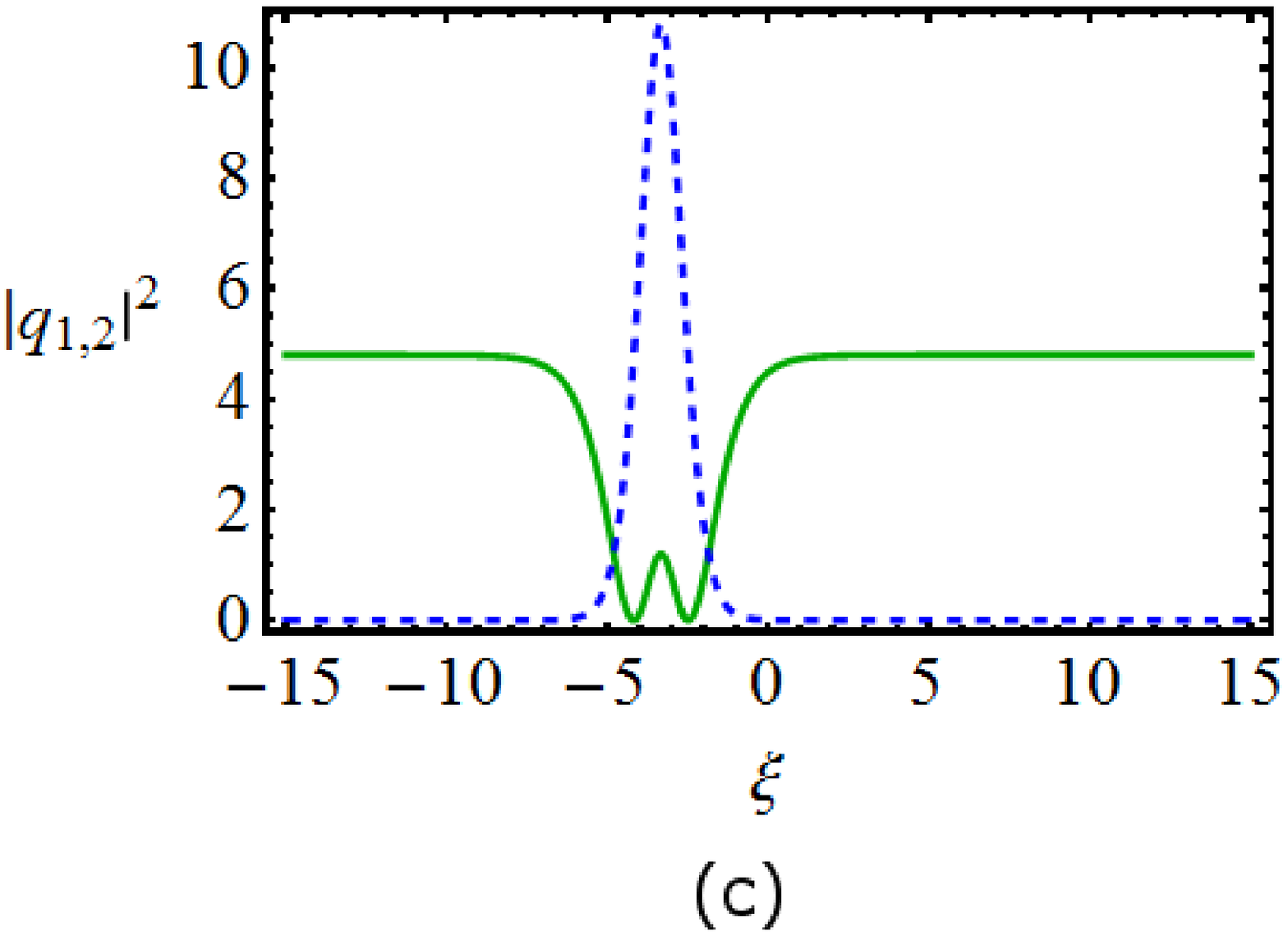}
\end{figure}
\begin{figure}[H]
\caption{The intensity plots of solution 8(b) with $\omega_{2}=0.7$, $k_{2}=0.01$, $\sigma_{1}=1$, $\sigma_{2}=1$, solution 11(b) with $\omega_{2}=-0.9$; $k_{2}=-2$, $\sigma_{1}=-1$, $\sigma_{2}=1$,  solution 13(b) for the choice $\omega_{2}=0.9$, $k_{2}=1$, $\sigma_{1}=-1$,$\sigma_{2}=1$, are shown in first two rows of Figs.~7(a-c) respectively. The third row shows the corresponding two dimensional plots of the $q_1$ (dark green) and $q_{2}$ (dashed blue) components  with $\tau=1.5$. All these figures $\kappa=0.0001$, $\delta_{0}=\delta_{1}=\delta_{2}=1$, $m=1$.}
\end{figure}
\section{Conclusion}
In this paper, we have obtained two distinct sets of elliptic wave solutions of the CNLH system which are expressed in terms of  Lam\'e polynomials of order one and two. The resulting solitary waves in the limit $m=1$ are also discussed in detail. This study reveals the role of nonparaxiality on the  ultra-broad beam nonlinear wave propagation in the CNLH equations (\ref{1}). We have shown that for certain first order solutions the pulse width increases as the nonparaxial parameter is increased while for the remaining solutions it becomes arbitrary. For all the second order solutions the pulse width increases as $\kappa$ increases. We have demonstrated that the speed can be altered by tuning the nonparaxial parameter. This is the most significant  effect of the nonparaxial parameter. Importantly, it has been shown that the effect of nonparaxiality is to remove the phase-locking behaviour of the solutions. In the second order solutions, we have presented novel single and double-hump bright solitary waves, double-hump (W-shape)  dark and bright solitary waves and coupled double-hump dark (W-shape) and standard bright solitary waves. These solitary waves with distinct profiles were found to admit the degenerate property in which the shape of the solution profile remains unaltered by the solution parameters even though these parameters influence the amplitude, pulse width and velocity.  These results will give further impetus into the dynamics of nonlinear travelling waves in the CNLH system and will find application in nonlinear pulse shaping. This procedure can be extended to variants of CNLH system and other CNLS systems by including quintic and power-law nonlinearities. The elliptic solutions as well as hyperbolic solitons obtained here will serve as good platform for further numerical investigation of the CNLH system. Their experimental realization will give further insight into the the nonparaxial effects in optical waveguides. Further investigation of nonparaxial effects in graded index medium is a separate issue and yet to be studied.

\section*{Acknowledgments}
The authors T.K. and K.T. acknowledge the Principal and Management of Bishop Heber College, Tiruchirapalli for the constant support and encouragement. The author A.K. acknowledges DAE, Govt. of India for the financial support through Raja Ramanna Fellowship.


\begin{thebibliography}{9}

\bibitem{whith}
Whitham G B, \textit{Linear and Nonlinear Waves}, John Wiley \& Sons, New York, (1999).

\bibitem{kivs}
Kivshar Y S, Agrawal G P, \textit{Optical Solitons: From Fibers to Photonic Crystals}, Academic Press, San Diego, (2003).

\bibitem{pethic}
Pethick C J, Smith H, \textit{Bose–Einstein Condensation in Dilute Gases}, Cambridge University Press, (2008).

\bibitem{scott}
Scott A C, \textit{Nonlinear Science: Emergence and Dynamics of
Coherent Structures}, Oxford University Press, Oxford, (1999).

\bibitem{david}
Davidson R, \textit{Methods in Nonlinear Plasma Theory}, Academic Press, (1972).

\bibitem{cros}
Crosignani B, Cutolo A, Porto P D, J. Opt. Soc. Am. \textbf{72} (1982) 1136.

\bibitem{polet}
Poletti F, Horak P, J. Opt. Soc. Am. B \textbf{25} (2008) 1645.

\bibitem{men}
Menyuk C R, IEEE J. Quantum Electron. \textbf{25} (1989) 2674.

\bibitem{tk1}
Radhakrishana R, Sahadevan R, Lakshmanan M, Chaos Solitons \& Fractals \textbf{5}, (1995) 2315;
Makha\'nkov V G, Pashaev O K, Theor. Math. Phys. \textbf{53} (1982) 979.

\bibitem{mana}
Manakov S V, Sov. Phys. JETP, \textbf{38} (1974) 248.

\bibitem{radhak}
Radhakrishnan R, Lakshmanan M, Hietarinta J, Phys. Rev. E \textbf{56} (1997) 2213.

\bibitem{shep}
Sheppard A P, Kivshar Y S, Phys. Rev. E \textbf{55} (1997) 4773.

\bibitem{vm}
Vijayajayanthi M, Kanna T, Lakshmanan M, Eur. Phys. J. Spec. Top. \textbf{173} (2009) 57.

\bibitem{laza}
Lazarides N, Tsironis G P,  Phys. Rev. E \textbf{71} (2005) 036614.

\bibitem{ciat}
Ciattoni A, Porto P D, Crosignani B, Yariv A, J. Opt. Soc. Am. B \textbf{17} (2000) 809.

\bibitem{chi}
Chi S, Guo Q, Opt. Lett. \textbf{20} (1995) 1598; Ciattoni A, Crosignani B, Di Porto P, Scheuer J, Yariv A, Opt. Express \textbf{14} (2006) 5517 [and reference there in]; Malomed B A, Marinov K, Pushkarov D I, Shivarova A, Phys. Rev. A \textbf{64} (2001) 023814 [and reference there in].

\bibitem{lax}
Lax M, Louisell W H, Mc Knight W B, Phys. Rev. A \textbf{11} (1975) 1365.

\bibitem{akh}
Akhmediev N, Ankiewicz A, Soto-Crespo J M,  Opt. Lett. \textbf{18} (1993) 411.

\bibitem{fibi}
Fibich G, Phys. Rev. Lett. \textbf{76} (1996) 4356.

\bibitem{cham0}
Chamorro-Posada P, Mc Donald G S,  New G H C, J.
Mod. Opt. \textbf{45} (1998) 1111[and reference there in].

\bibitem{blair}
Blair S, Chaos \textbf{10} (2000) 570.

\bibitem{christ3}
Christian J M, Mc Donald G S, Chamorro-Posada P, Phys. Rev. E \textbf{74} (2006) 066612 [and reference there in].

\bibitem{cros1}
Crosignani B, Yariv A, Mookherjea S, Opt. Lett. \textbf{29} (2004)  1254.

\bibitem{ciat1}
Ciattoni A, Crosignani B, Mookherjea S Yariv A, Opt. Lett. \textbf{30} (2005)  516.

\bibitem{wang}
Wang H, She W, Opt. Express \textbf{14} (2006) 1590.

\bibitem{cham1}
Chamorro-Posada P, Mc Donald G S, Phys. Rev. E \textbf{74} (2006) 036609 [and reference there in].

\bibitem{soy}
Tsoy E N, Akhmediev N, Opt. Commun. \textbf{266} (2006) 660; Si L G, Yang W X, Liu J B, Li J, Yang X, Opt. Express. \textbf{17} (2009) 7771; Ohta Y, Wang D S, Yang J, Stud. Appl. Math. \textbf{127} (2011) 345; Sonnier W J, Christov C I, Discret. Contin. Dyn. S (2009) pages 708-718; Feng B F, J. Phys. A: Math. Theor \textbf{47} (2014) 355203; Wright O C, Appl. Math. Lett. \textbf{16} (2003) 647; Zhao L C, Results Phys. \textbf{2} (2012) 203; Kanna T, Tsoy E N, Akhmediev N, Phys. Lett. A \textbf{330} (2004) 224;
Kanna T, Vijayajayanthi M, Lakshmanan M, Phys. Rev. A \textbf{76} (2007) 013808;
Vijayajayanthi M, Kanna T, Lakshmanan M, Phys. Rev. A \textbf{77} (2008) 013820;
Kanna T, Lakshmanan M, Tchofo Dinda P, Akhmediev N, Phys. Rev. E \textbf{73} (2006) 026604;Yang J, 2010 \textit{Nonlinear Waves in Integrable and Nonintegrable Systems} SIAM.

\bibitem{chow1}
 Chow K W,  Nakkeeran K, Malomed B A, Opt. Commun. \textbf{219} (2003) 251; Hioe F T, Slater J S, J. Phys. A \textbf{35} (2002) 8913; Hioe F T, J. Math. Phys.\textbf{43} (2002) 6325; Hioe F T, Phys. Lett. A \textbf{304} (2002) 30; Chow K W, Lai D W C, Phys. Rev. E \textbf{68} (2003) 017601; Chow K W, Lai D W C, Phys. Rev. E \textbf{65} (2002) 026613;  Florjanczyk M, Tremblay R, Phys. Lett. A \textbf{141} (1989) 34; Petnikova V M,  Shuvalov V V, Vysloukh V A, Phys. Rev. E \textbf{60} (1999) 1009; Wright III O C, Phys. D \textbf{264} (2013) 1-16.

\bibitem{anjan}
Bhrawy A H, Abdelkawy M A, Anjan Biswas, Commun. Nonlinear Sci. Numer. Simulat. \textbf{18} (2013) 915;
Belmonte-Beitia J, P\'erez-Garc\'ia V M, Brazhnyi V, Commun. Nonlinear Sci. Numer. Simulat. \textbf{16} (2011) 158;
Li H, Ma L, Wang K,  Commun. Nonlinear Sci. Numer. Simulat. \textbf{14} (2009) 3296.

 \bibitem{hioe}
 Hioe F T, Slater J S, J. Phys. A \textbf{35} (2002) 8913.

\bibitem{whit}
 Whittaker E T, Watson G N, \textit{A course of modern analysis} Cambridge University Press, London, (1886).

\bibitem{avin}
 Avinash Khare, Kanna T, Tamilselvan K, Phys. Lett. A \textbf{378} (2014) 3093.

 \bibitem{avin1}
 Khare A, Saxena A, Phys. Lett. A \textbf{377} (2013) 2761; Khare A, Saxena A, J. Math. Phys. \textbf{55} (2014)  032701.

\bibitem{Akhm1}
Akhmediev N N, Buryak A V, Soto-Crespo J M, Andersen D R, J. Opt. Soc. Am. B \textbf{12} (1995) 434.

\bibitem{hioe1}
Hioe F T, Phys. Rev. E \textbf{58} (1998) 6700.

\bibitem{xu}
Xu T, Xu X M, Phys. Rev. E \textbf{87}  (2013) 032913.

\end{thebibliography}
\end{document}